\documentclass[11pt]{article}

\pdfoutput=1

\usepackage[T1]{fontenc}
\usepackage[latin9]{inputenc}
\usepackage[a4paper]{geometry}
\usepackage[active]{srcltx}
\usepackage{amsmath}
\usepackage{amssymb}
\usepackage{esint}
\usepackage{ulem}
\usepackage{enumitem}

\makeatletter

\usepackage{textcomp}

\pdfoutput=1 

\usepackage{jheppub}

\newcommand*\xbar[1]{%
  \hbox{%
    \vbox{%
      \hrule height 0.5pt 
      \kern0.3ex
      \hbox{%
        \kern-0.0em
        \ensuremath{#1}%
        \kern-0.0em
      }%
    }%
  }%
}

\usepackage{amsfonts}

\usepackage{stmaryrd}
\usepackage{mathtools}

\usepackage{natbib}
\usepackage{mathrsfs}
\usepackage{verbatim}

\usepackage{bbm}

\newcommand{\be}{\begin{equation}}
\newcommand{\ee}{\end{equation}}
\newcommand{\bea}{\begin{eqnarray}}
\newcommand{\eea}{\end{eqnarray}}

\renewcommand{\d}{\partial}

\title{\boldmath  Logarithmic angle-dependent gauge transformations at null infinity}

\author[a,b,c]{Oscar Fuentealba}
\author[c,d]{and Marc Henneaux}

\affiliation[a]{Instituto de Ciencias Exactas y Naturales (ICEN), Universidad Arturo Prat,\\ Playa Brava 3256, 1111346 Iquique, Chile}
\affiliation[b]{Facultad de Ciencias, Universidad Arturo Prat,\\ Avenida Arturo Prat Chac\'on 2120, 1110939 Iquique, Chile}
\affiliation[c]{Universit\'e Libre de Bruxelles and International Solvay Institutes, \\ ULB-Campus Plaine CP231, B-1050 Brussels, Belgium}
\affiliation[d]{Coll\`ege de France,  Universit\'e PSL, 11 place Marcelin Berthelot, \\ 75005 Paris, France}

\emailAdd{ofuentealba@unap.cl}
\emailAdd{marc.henneaux@ulb.be}

\preprint{}

\abstract{Logarithmic angle-dependent gauge transformations are symmetries of electromagnetism that are canonically conjugate to the standard $\mathcal O(1)$ angle-dependent $u(1)$ transformations.   They were exhibited a few years ago at spatial infinity.   In this paper, we derive their explicit form at null infinity.  We also derive the expression there of the associated "conserved" surface integrals.  To that end, we provide a comprehensive analysis of the behaviour of the electromagnetic vector potential $A_\mu$ in the vicinity of null infinity for generic initial conditions given on a Cauchy hypersurface.  This behaviour is given by a  polylogarithmic expansion involving both gauge-invariant logarithmic terms also present in the field strengths and gauge-variant logarithmic terms with physical content, which we identify.   We show on which explicit terms, and how, do the logarithmic angle-dependent gauge transformations act.  Other results of this paper are a derivation of the 
 matching conditions for the Goldstone boson and for the conserved charges of the angle-dependent $u(1)$ asymptotic symmetries, as well as a clarification of a misconception concerning the non-existence of these angle-dependent $u(1)$ charges in the presence of logarithms at null infinity. 
 We also briefly comment on higher spacetime dimensions.}

\makeatother

\begin{document}
\maketitle \flushbottom

 \newpage{}

\section{Introduction}

\subsection{Logarithmic angle-dependent $u(1)$ gauge transformations}

Following developments first pursued in the context of gravity \cite{Fuentealba:2022xsz}, we showed in the article \cite{Fuentealba:2023rvf} that electromagnetism in $D=4$ spacetime dimensions was invariant under asymptotic symmetries involving gauge transformations growing like $\mathcal O(\log r)$ at spatial infinity.  This was achieved by relaxing in a consistent way (i.e., while keeping the action and the charges well-defined and finite) the boundary conditions on the vector potential. [Gauge transformations that grow like $\mathcal O(r)$ were actually also considered in \cite{Fuentealba:2023rvf} - see also \cite{Campiglia:2016hvg,Seraj:2016jxi} for different approaches - but we shall stick here to $\mathcal O(\log r)$ transformations only.]

One of the interests of enlarging the asymptotic conditions on the vector potential so as to allow $\mathcal O(\log r)$ gauge transformations at infinity is that the generators of these new transformations are canonically conjugate to the generators of the angle-dependent $\mathcal O(1)$ gauge transformations exhibited in \cite{Barnich:2013sxa,He:2014cra,Kapec:2015ena,Strominger:2017zoo}, in the sense that their Poisson brackets yield the identity.  Using general Darboux type theorems \cite{Fuentealba:2023hzq}, one can then provide a definition of the Lorentz generators that is free from the ambiguities due to the fact that angle-dependent $u(1)$ transformations and Lorentz transformations in their original form do not commute \cite{Fuentealba:2023rvf}.

The analysis of  \cite{Fuentealba:2023rvf} was carried out at spatial infinity.  It is the purpose of this paper to describe the logarithmic angle-dependent $u(1)$ gauge transformations at null infinity by identifying their action on the vector potential there. We also give the expression there of the corresponding charges.

The issue is a bit subtle because polylogarithmic terms are known to be omnipresent at null infinity, even in the absence of logarithmic gauge transformations.  One therefore needs to carefully disentangle the logarithms that appear in the field strengths and have nothing to do with gauge transformations from the logarithms present in the relevant gauge transformations.

\subsection{Polylogarithmic terms in the field strengths at null infinity}

In order to disentangle the two types of logarithms, one needs to know the form of the polylogarithmic terms in the field strengths at null infinity.  This has been completely worked out in references \cite{Henneaux:2018gfi,Henneaux:2019yqq}, of which we summarize the results for four spacetime dimensions.  

In $D=4$ Minkowski spacetime, it is natural to assume that the electromagnetic field components $F_{\mu \nu}$  behave
in Minkowskian coordinates $ds^2 = -dt^2 + (dx^i)^2$ as
\be
F_{\mu \nu} = \frac{\xbar F_{\mu \nu}}{r^2} + \mathcal O\left(\frac{1}{r^3}\right) \,, \label{Eq:DecayMink}
\ee
as one goes to  spatial infinity ($t = $const, $\theta =$const, $\varphi = $const, $r \rightarrow \infty$). The asymptotic decay (\ref{Eq:DecayMink}) guarantees the finiteness of the flux at infinity
\be
\oint d^2 S \, n^i  \lambda \, F^{i0} \label{Eq:SurfaceElectricFlux}
\ee
of the electric field multiplied by an arbitrary function $\lambda(x^A)$ of the angles $(x^A)$.   Here, $n^i (x^A)$ is the unit normal to the sphere. Similar magnetix flux expressions also converge.

The surface integral (\ref{Eq:SurfaceElectricFlux}) was shown in \cite{Henneaux:2018gfi} to be the value, expressed at spatial infinity, of the charge-generators of the angle-dependent $u(1)$ transformations exhibited at null infinity in \cite{He:2014cra,Kapec:2015ena,Strominger:2017zoo}.

In polar coordinates, (\ref{Eq:DecayMink}) is equivalent to
\begin{align}
&F_{0r} = \frac{\xbar F_{0r}}{r^2}+  \mathcal O\left(\frac{1}{r^3}\right)\,, \quad 
F_{0 A} = \frac{\xbar F_{0A}}{r}+ \mathcal O\left(\frac{1}{r^2}\right) \, , \label{Eq:DecayMinkPolar0}\\
& F_{r A} = \frac{\xbar F_{rA}}{r}+ \mathcal O\left(\frac{1}{r^2}\right) \,, \quad F_{AB} = \xbar F_{AB} +\mathcal O\left(\frac{1}{r}\right)  \, .
   \label{Eq:DecayMinkPolar1}
\end{align}

If one integrates the Maxwell equations with initial data (\ref{Eq:DecayMink}) all the way to null infinity, one finds that logarithmic terms appear at null infinity even though there is none in (\ref{Eq:DecayMink}).  This is because null infinity corresponds to a Fuchsian singularity of the differential equations that control the dynamics of the electromagnetic field, as in the scalar field case \cite{Fuentealba:2024lll}.

The presence of polylogarithmic terms in the expansion near null infinity is not a surprise since it parallels what happens for gravity \cite{Damour:1985cm,Christodoulou:1993uv,Chrusciel:1993hx,Fried1,Friedrich:1999wk,Friedrich:1999ax,Valiente-Kroon:2002xys,ValienteKroon:2003ix,Kehrberger:2021uvf,Minucci:2022hav,Sen:2024bax,Geiller:2024ryw}.  Logarithms at null infinity were also exhibited for the Maxwell field in  \cite{ValienteKroon:2007bj}, and more recently for $p$-form gauge fields in \cite{Francia:2024hja,Romoli:2024hlc,Manzoni:2025wyr}.

In retarded null coordinates $(u,r,x^A)$
\be
ds^2 = - du^2 - 2 du dr +2 r^2 \xbar \gamma_{AB} dx^A dx^B\,,
\ee
where $\xbar \gamma_{AB}$ is the unit metric on the $2$-sphere, one finds explicitly (see  \cite{Henneaux:2019yqq} and Appendices  {\bf \ref{App:Derivation}} and {\bf \ref{App:FNullInf}})
\begin{align}
F_{ur} &= \frac{\log r}{r^{2}}F_{ur}^{\log}  + \frac{\xbar F_{ur}}{r^2}+ o\left(r^{-2}\right)\,, \\
F_{uA} & =F_{uA}^{(0)}+\frac{\log r}{r}F_{uA}^{\log}+\mathcal{O}\left(r^{-2}\right)\,,\\
F_{rA} & =\frac{1}{r}F_{rA}^{(1)}+\frac{\log r}{r^{2}}F_{rA}^{\log}+\mathcal{O}\left(r^{-3}\right)\,,\\
F_{AB} & = \log r F^{\log}_{AB} + \xbar F_{AB} + o\left(1\right)  \, ,
   \label{Eq:DecayNull}
\end{align}
for some functions $F_{ur}^{\log}$, $\xbar F_{ur}$, $F_{uA}^{(0)}$, $F_{uA}^{\log}$,  $F_{rA}^{(1)}$, $F_{rA}^{\log}$, $F^{\log}_{AB}$ and $\xbar F_{AB}$ of the retarded time $u$ and  the angles  that can be expressed in terms of the initial data.  Not only are polylogarithmic terms present, but they are even the leading terms near null infinity in $F_{ur}$ and $F_{AB}$ for generic initial data of the form (\ref{Eq:DecayMink}).  These leading logarithmic terms do not conflict with finiteness of the charges; their presence simply means that these charges are not given by standard flux integrals at null infinity, as in the scalar case \cite{Fuentealba:2024lll}. 

The leading logarithmic terms are absent if and only if one restricts the leading order of the initial data to be odd under the sphere antipodal map \cite{Henneaux:2018gfi}, i.e., 
\be
F_{\mu \nu} = \frac{\xbar F_{\mu \nu}}{r^2} + \mathcal O\left(\frac{1}{r^3}\right) \, , \quad \xbar F_{\mu \nu} (-n^i) = -\xbar F_{\mu \nu}(n^i)\, . \label{Eq:DecayMinkV2}
\ee
[Information and conventions on the sphere antipodal map are collected in Appendix {\bf \ref{App:Orientation}}.] These are the boundary conditions considered in \cite{Henneaux:1999ct}, inspired by \cite{Regge:1974zd}, and are the boundary conditions explicitly adopted throughout this paper unless otherwise stated.  From the point of view of the action principle, there is actually a need to impose some form of parity conditions on the leading orders of the fields in order to achieve finiteness of action and charges. While the conditions (\ref{Eq:DecayMinkV2}) are the most natural ones because they are compatible with electric and magnetic sources, they are not the only ones.  Other possibilities exist \cite{Henneaux:2018gfi}, which are recalled in Section {\bf \ref{Sec:AsympSpatial}}.

When the asymptotic conditions  (\ref{Eq:DecayMinkV2}) hold at spatial infinity, the behaviour of the electromagnetic field at null infinity is
\be
F_{ur} =  \frac{\xbar F_{ur}}{r^2}+ o\left(r^{-2}\right)\,, \quad 
F_{u A} = \mathcal O \left( 1 \right) \, , \quad
 F_{r A}= \mathcal O\left(\frac{\log r}{r^{2}}\right) \,, \quad F_{AB} =  \mathcal O \left( 1 \right)  \, ,
   \label{Eq:DecayNull2}
\ee
which is, to leading order, the form of the electromagnetic field taken in \cite{He:2014cra,Kapec:2015ena,Strominger:2017zoo,Satishchandran:2019pyc} (which however did not consider subleading logarithmic terms\footnote{The $\mathcal O\left(\frac{\log r}{r^{2}}\right)$ is leading with respect to the $\mathcal O\left(\frac{1}{r^{2}}\right)$ term with which the expansion of $F_{rA}$ starts in \cite{He:2014cra,Kapec:2015ena,Strominger:2017zoo,Satishchandran:2019pyc}, but it is really subleading with respect to the $\mathcal O\left(\frac{1}{r}\right)$ term that is present without parity conditions.}).  Similarly, one has in advanced null coordinates
\be
F_{vr} =  \frac{\xbar F_{vr}}{r^2}+ o\left(r^{-2}\right)\,, \quad 
F_{v A} = \mathcal O \left( 1 \right) \, , \quad
 F_{r A}= \mathcal O\left(\frac{\log r}{r^{2}}\right)  \,, \quad F_{AB} =  \mathcal O \left(1\right)  \, .
   \label{Eq:DecayNull3}
\ee
The parity conditions at spatial infinity not only remove the leading logarithms but also imply the familiar antipodal matching \cite{He:2014cra,Kapec:2015ena,Strominger:2017zoo},
\be
\lim_{ v\rightarrow \infty} \xbar F_{vr}(- x^A) = \lim_{u \rightarrow - \infty}\xbar F_{ur}(x^A)\,, \label{Eq:IntroMatching}
\ee
where $x^A \rightarrow - x^A$ symbolically denotes the $2$-sphere antipodal map in terms of the angles. This was pointed out in \cite{Henneaux:2018gfi,Henneaux:2018hdj}.  In other words, assuming the absence of leading logarithmic terms at null infinity (and hence that the charges are given there by standard flux expressions) implicitly assumes that the parity conditions (\ref{Eq:DecayMinkV2}) hold at spatial infinity,  which implies the matching conditions (\ref{Eq:IntroMatching}).

Because the field strengths involve (subleading) logarithmic terms at null infinity, the gauge potentials will also involve there logarithms, which are not removable by gauge transformations.  It is customary to assume
\be
A_r = \mathcal O\left(\frac{1}{r^2}\right)\, , \quad A_u = \mathcal O\left(\frac{1}{r}\right)\, , \quad A_A = \mathcal O\left(1\right)   \label{Eq:IntroFallOffA}
\ee
(with subleading logarithmic terms)\footnote{One sometimes  imposes the stronger gauge condition $A_r = 0$ but this will not be discussed here as (\ref{Eq:IntroFallOffA}) will be seen to be already too strong.}, which is compatible with (\ref{Eq:DecayNull2}) -- the $\mathcal O\left(\frac{\log r}{r^{2}}\right)$ in $F_{rA}$ comes from the subleading $\mathcal O\left(\frac{\log r}{r}\right)$ term  in $A_A$.  However, in order to reach (\ref{Eq:IntroFallOffA}), one must fix the gauge.  

The main point of our paper is that (\ref{Eq:IntroFallOffA}) involves an improper gauge fixing (in the sense of \cite{Benguria:1976in}).  Explicitly, it freezes the physically-relevant logarithmic angle-dependent $u(1)$ gauge transformations, which are not seen for that reason. We prove that in order to keep the possibility to perform logarithmic gauge transformations, one must relax the asymptotic conditions at null infinity as
\be
A_r = \partial_r \Delta + \mathcal O\left(\frac{1}{r^2}\right)\, , \quad A_u = \partial_u \Delta + \mathcal O\left(\frac{1}{r}\right)\, , \quad A_A = \partial_A \Delta + \mathcal O\left(1\right) \,,  \label{Eq:IntroFallOffLogA}
\ee
with 
\be
\Delta = \frac{\log r}{r}u X(x^A)\,, \label{Eq:DeltaPhi200}
\ee
where $X(x^A)$ is an arbitrary function of the angles (note that $ \partial_A \Delta$ is subleading with respect to $\mathcal O\left(1\right)$).  Because these extra terms are pure gradients, they do not modify the field strengths.  However, they are physically relevant. Both $\mathcal O\left(1\right)$ and $\mathcal O\left(\frac{\log r}{r}\right)$  gauge transformations are improper with non-vanishing charges, which we explicitly write at null infinity.  The latter can be seen only if one allows logarithmic terms in the expansion of the fields.

This is our central result.  We establish it by explicitly integrating the Maxwell equations for the vector potential in the Lorenz gauge from the initial Cauchy hypersurface $t= 0$ (say) to null infinity, determining thereby the asymptotic form (\ref{Eq:IntroFallOffLogA}) from the asymptotic form of the fields at spatial infinity given in \cite{Fuentealba:2023rvf}.  This method provides a wealth of information, including a derivation of the matching conditions for the Goldstone boson of the angle-dependent $u(1)$ symmetry and an understanding of the behaviour of the charges as one goes to null infinity.

Another interesting feature that also emerges from the analysis is that while the parameters of the logarithmic gauge transformations diverge at spatial infinity like $\log r$ and dominate therefore the standard $\mathcal O(1)$ gauge transformations, they become subdominant (but still physically relevant) and behave as $\frac{\log r}{r}$ near null infinity.

\subsection{Organization of the paper}

Our paper is organized as follows.  In Section {\bf \ref{Sec:AsympSpatial}}, we discuss the asymptotic behaviour of the electromagnetic field at spatial infinity and review how logarithmic gauge transformations appear in the expansion of the vector potential at large radial coordinate $r$ on spacelike hyperplanes (e.g., $t=0$).  These are generalizations of the standard strict parity conditions twisted by an $\mathcal O(1)$ gauge transformations.  We also recall that other mathematically consistent parity conditions exist. The next three sections prepare the ground for the introduction of the logarithmic gauge transformations, by providing new useful material on the gauge potential in the absence of these transformations.  In Section {\bf \ref{Sec:AinHyp}}, we integrate the field equations for the vector potential in hyperbolic coordinates, as a first step for getting the vector potential near null infinity.  In Section {\bf \ref{Sec:ConservedCharges}}, we discuss the conserved charges.  One of the aims of this discussion is to stress that both types of charges (Noetherian electric ones and non-Noetherian magnetic ones) remain well defined at null infinity, even for asymptotic conditions that lead to logarithms there.  This clarifies some misconceptions that in the presence of logarithms, these charges would be ill-defined.  We then determine the behaviour of the electromagnetic potential at null infinity (Section {\bf \ref{Sec:VectorANullInf}}).  In Section {\bf \ref{Sec:LogGaugeTransf}} we derive the form of the gauge potentials and of the gauge transformations at null infinity when 
angle-dependent logarithmic $u(1)$ gauge transformations are included.  We also derive the form of the logarithmic conserved charges.  Section {\bf \ref{Sec:MatchingHighD}} is devoted to comments on the generalization to higher spacetime dimensions.  We conclude our analysis in Section {\bf \ref{Sec:Conclusions}} by summarizing our central results and discussing their extension to gravity.  Finally, a few appendices of a more technical or review nature close our paper: Appendix {\bf \ref{App:Orientation}} gives information on the sphere antipodal map, Appendices {\bf \ref{App:Derivation}} and {\bf \ref{App:FNullInf}} review the description of the electromagnetic field strengths in hyperbolic and null coordinates, while the last appendix, Appendix {\bf \ref{App:SphHar}}, recalls general features of spherical harmonics for $p$-forms in $D$ spacetime dimensions.

\section{Asymptotic conditions at spatial infinity}
\label{Sec:AsympSpatial}

\subsection{Standard parity conditions twisted by an $\mathcal O(1)$ gauge transformation}

We rely for our analysis on the formulation of the theory on Cauchy hypersurfaces (which are in particular achronal), taken for definiteness to be the hyperplanes of constant time or their boost-transformed.  Hamiltonian techniques connecting symmetries and charge-generators are direct on Cauchy hypersurfaces. There is indeed a well-defined symplectic structure and unambiguous Poisson brackets, which enable us to use the standard techniques of classical mechanics.

To completely define the theory, one needs to specify the asymptotic conditions that the vector potential components and their conjugate momenta (the electric field components) must statisfy for large $r$, i.e., as one goes to spatial infinity.

One  gradually arrives at the boundary conditions allowing $\mathcal O(\log r)$ gauge transformations -- the subject of our article -- through various intermediate steps.  

One first starts with the asymptotic conditions $A_i =\mathcal O(\frac{1}{r})$, $\pi^i =\mathcal O(\frac{1}{r^2})$ which implies that the field strengths decay as $1/r^2$ at infinity, as it is natural in $4$ spacetime dimensions. Observing then that this obvious asymptotic behaviour of $A_i$ and $\pi^i$  leads to a logarithmically divergent kinetic term $\int d^3x \pi^i \dot{A}_i$ in the action, one next imposes that the leading terms in the asymptotic expansion fulfill the strict parity conditions
$\xbar A_i(-x^A) = \xbar A_i(x^A)$, $\xbar\pi^i(-x^A) = - \xbar \pi^i(x^A)$ (standard parity conditions \cite{Henneaux:1999ct,Regge:1974zd}). 

One then realizes that these boundary conditions, while at first sight natural because they are satisfied by the known solutions and are Lorentz-invariant, suffer from one drawback: they freeze the possibility to perform arbitrary angle-dependent $\mathcal O(1)$ gauge transformations.  In order to incorporate that freedom and thereby eliminate the tension with the null infinity description,  one relaxes the boundary conditions involving strict parity conditions by allowing an $\mathcal O(1)$ gauge transformation of opposite parity in $\xbar A_i$,
\be 
A_r = \frac{\xbar A_r}{r} + \mathcal O\left(\frac{1}{r^2}\right) \, , \qquad A_A = \xbar A_A + \partial _A \xbar \Phi + \mathcal O\left(\frac{1}{r}\right) \, , \label{Eq:ParityA00a}
\ee
with the gauge-invariant momenta remain unchanged,
\be
\pi^r = \xbar \pi^r + \mathcal O\left(\frac{1}{r}\right) \, , \qquad \pi^A = \frac{\xbar \pi^A}{r} + \mathcal O\left(\frac{1}{r^2}\right) \, .\label{Eq:ParityA00b}
\ee
The coefficients of the leading terms are subject to the following parity conditions
\be
\xbar A_r (- x^B) = - \xbar A_r (x^B) \, , \quad \xbar A_A (- x^B) = \xbar A_A (x^B) \, , \quad \xbar \Phi (- x^B) =  \xbar \Phi (x^B)  \, , \label{Eq:ParityA01a}
\ee
\be
\xbar \pi^r (- x^B) =  \xbar \pi^r (x^B) \, , \quad \xbar \pi^A (- x^B) = -\xbar \pi^A (x^B) \,.  \label{Eq:ParityA01ab}
\ee
These are the ``standard parity conditions twisted by a gauge transformation" of \cite{Henneaux:2018gfi}, completed by
\be
A_0 =  \frac{1}{r} \xbar\Psi + \mathcal O\left(\frac{1}{r^2}\right)\, , 
 \quad \xbar \Psi (- x^B) =  - \xbar \Psi (x^B)  \, .
\ee
The boundary conditions imply that $ \xbar F_{AB}$ is odd so that the radial magnetic field $ \epsilon^{AB} \xbar F_{AB}$ is even. 

It is clear that the zero mode of $\xbar \Phi (x^B)$ drops from (\ref{Eq:ParityA00a}) and is therefore pure gauge.  For that reason, it is convenient to set it equal to zero, and this is what we shall do from now on.

The conditions (\ref{Eq:ParityA00a})-(\ref{Eq:ParityA01ab}) are invariant by construction under angle-dependent $\mathcal O(1)$ $u(1)$ gauge transformations and still lead to a consistent (finite, well-defined) Hamiltonian formulation provided one imposes also that the leading $\mathcal O(r^{-3})$ coefficient of Gauss' law be strictly zero, which reads $\partial_A \xbar \pi^A = 0$ \cite{Henneaux:2018gfi}.

\subsection{Non-standard twisted parity conditions}

Alternative consistent boundary conditions, inspired by those of \cite{Henneaux:2018cst}, are possible \cite{Henneaux:2018gfi}.  Although these are not the ones that we have extended to include logarithmic gauge transformations, they provide insightful light on the connection between parity conditions at spatial infinity and presence of leading logarithmic terms at null infinity.  We consider them here for that reason. 

These alternative  boundary conditions are
\be 
A_r = \frac{\xbar A_r}{r} + \mathcal O\left(\frac{1}{r^2}\right) \, , \qquad A_A = \xbar A_A + \partial _A \xbar \Phi + \mathcal O\left(\frac{1}{r}\right) \, ,
\ee
and
\be 
\pi^r = \xbar \pi^r + \mathcal O\left(\frac{1}{r}\right) \, , \qquad \pi^A = \frac{\xbar \pi^A}{r} + \mathcal O\left(\frac{1}{r^2}\right) \, ,
\ee
where $\xbar A_r$,  $\xbar A_A$,  $\xbar \Phi $, $\xbar \pi^r$ and $\xbar \pi^A $ are functions of the angles that obey the parity conditions:
\be
\xbar A_r (- x^B) = - \xbar A_r (x^B) \, , \quad \xbar A_A (- x^B) = -\xbar A_A (x^B) \, , \quad \xbar \Phi (- x^B) =  -\xbar \Phi (x^B)  \, , \label{Eq:ParityA01b}
\ee
\be
\xbar \pi^r (- x^B) =  \xbar \pi^r (x^B) \, , \quad \xbar \pi^A (- x^B) = \xbar \pi^A (x^B) \,  
\ee
($\Rightarrow \xbar F_{AB}$ even) with again the extra condition $\partial_A \xbar \pi^A = 0$ that guarantees that Gauss' law holds to leading order.   One can actually assume $\xbar \Phi = 0$ in this case, because it corresponds to a proper gauge transformation with zero charge \cite{Henneaux:2018gfi}.

While these parity conditions cover the Coulomb solution, they do not include magnetic monopoles which have a field strength $\xbar F_{AB}$ with the opposite parity. 

\subsection{Parity-inverted conditions}

A third set of parity conditions on the leading order of the asymptotic fields is actually consistent, in which one takes $\xbar A_i$ to be odd up to a gauge transformation and $\xbar \pi^i$ to be even.  These parity conditions imply that $\xbar \pi^r$ is odd and does not allow for the Coulomb solution, nor for magnetic monopoles, and so are  physically limited.  They do allow, however, for non-trivial angle-dependent gauge transformations parametrized by an odd function of the angles.  It is thus instructive to explore theoretically the consequences of these inverted parity conditions, if only to understand better the implications of the standard ones.  Furthermore, it is 
natural from the connection interpretation to regard $\xbar A_i$ to have the same parity properties as $\partial_i$, i.e. to be odd in cartesian coordinates.   These conditions were already considered in \cite{Tanzi:2020fmt}.

The inverted parity conditions explicitly read, in spherical coordinates:
\be 
A_r = \frac{\xbar A_r}{r} + \mathcal O\left(\frac{1}{r^2}\right) \, , \qquad A_A = \xbar A_A + \partial _A \xbar \Phi + \mathcal O\left(\frac{1}{r}\right) \, ,
\ee
and
\be 
\pi^r = \xbar \pi^r + \mathcal O\left(\frac{1}{r}\right) \, , \qquad \pi^A = \frac{\xbar \pi^A}{r} + \mathcal O\left(\frac{1}{r^2}\right) \, ,
\ee
where $\xbar A_r$,  $\xbar A_A$,  $\xbar \Phi $, $\xbar \pi^r$ and $\xbar \pi^A $ are functions of the angles that now obey:
\be
\xbar A_r (- x^B) =  \xbar A_r (x^B) \, , \quad \xbar A_A (- x^B) = - \xbar A_A (x^B) \, , \quad \xbar \Phi (- x^B) =  - \xbar \Phi (x^B)  \, , \label{Eq:ParityA01c}
\ee
\be
\xbar \pi^r (- x^B) =  - \xbar \pi^r (x^B) \, , \quad \xbar \pi^A (- x^B) = \xbar \pi^A (x^B) \,.  
\ee

\subsection{Standard parity conditions twisted by a $\mathcal O(\log r)$ gauge transformations}

The standard parity conditions twisted by a gauge transformation (\ref{Eq:ParityA00a})-(\ref{Eq:ParityA01ab}) can be relaxed ("further twisted")
 by adding the  terms induced by a gauge transformation with a gauge parameter that logarithmically grows at spatial infinity.  These lead to the asymptotic conditions 
\cite{Fuentealba:2023rvf}:
\be 
A_r = \frac{\xbar A_r }{r} + \mathcal O\left(\frac{\log r}{r^2}\right) \, , \qquad A_A = \log r \, \partial_A  \Phi_{\textrm{log}} + \xbar A_A + \partial _A \xbar \Phi + \mathcal O\left(\frac{\log r}{r}\right) \, , \label{Eq:ParityA0007Z}
\ee
while the gauge-invariant momenta behave as
\be 
\pi^r = \xbar \pi^r + \mathcal O\left(\frac{1}{r}\right) \, , \qquad \pi^A = \frac{\xbar \pi^A}{r} + \mathcal O\left(\frac{1}{r^2}\right) \, .\label{Eq:ParityA0017Z}
\ee
Here, $\xbar A_r$,  $\xbar A_A$,  $\xbar \Phi_{\log} $, $\xbar \Phi $, $\xbar \pi^r$ and $\xbar \pi^A $ are functions of the angles that obey the parity conditions\footnote{The antipodal map $\vec x^i \rightarrow -\vec x^i$ is symbollically written $x^A \rightarrow - x^A$ when acting on the angles $x^A$ on the sphere, see Appendix {\bf \ref{App:Orientation}} for more information.}:
\begin{align}
&\xbar A_r (- x^B) = - \xbar A_r (x^B) \, , \quad\xbar A_A (- x^B) = \xbar A_A (x^B)   \, , \label{Eq:ParityA017Z} \\
& \Phi_{\textrm{log}}(-x^A) = - \Phi_{\textrm{log}}(x^A)\, , \quad \xbar \Phi (- x^B) =  \xbar \Phi (x^B)  \, , \label{Eq:ParityA017ZBis} \\
&\xbar \pi^r (- x^B) =  \xbar \pi^r (x^B) \, , \quad \xbar \pi^A (- x^B) = -\xbar \pi^A (x^B) \,.  
\end{align}
 One must also impose the condition $\partial_A \xbar \pi^A = 0$ that guarantees that Gauss' law holds to leading order.

The asymptotic conditions for the temporal component are,
\be
A_0 = \frac{\log r}{r} \Psi_{\textrm{log}} + \frac{1}{r} \xbar\Psi + \mathcal O\left(\frac{\log r}{r^2}\right)\, , \label{Eq:ParityA0027Z}
\ee
with 
\be
 \Psi_{\textrm{log}}(-x^A) =  \Psi_{\textrm{log}}(x^A)\, , \quad \xbar \Psi (- x^B) =  - \xbar \Psi (x^B)  \, ,\label{Eq:ParityA0037Z}
\ee
while the conjugate momentum $\pi^0$, constrained to vanish on-shell, can be assumed to decay as fast as desired.  The field $\Psi_{\textrm{log}}(x^A)$ turns out to be canonically conjugate to the field $\xbar \Phi(x^A)$, as the kinetic term (2.25) in the action of \cite{Fuentealba:2023rvf} shows.  It is therefore natural to assume that it has also no zero mode. This condition was actually not imposed in \cite{Fuentealba:2023rvf}, which developed the formalism by keeping the zero modes of both $\xbar \Phi(x^A)$ and $\Psi_{\textrm{log}}(x^A)$, which became dynamical degrees of freedom.  This extension of the formalism is consistent, but for simplicity, we stick here to the condition that these zero modes are equal to zero, which is equivalent to
\be
\oint \sqrt{\xbar \gamma} \, \xbar \Phi = 0 \, , \qquad \oint \sqrt{\xbar \gamma} \, \Psi_{\textrm{log}} = 0 \, .
\ee
Technically, one enforces these conditions by projecting out the zero modes of $\xbar \Phi(x^A)$ and $\Psi_{\textrm{log}}(x^A)$ from the kinetic term (2.25) of the action of \cite{Fuentealba:2023rvf}.  These zero modes  become then pure gauge and can be set to zero.

Understanding the implications at null infinity of the asymptotic conditions (\ref{Eq:ParityA0007Z})-(\ref{Eq:ParityA0037Z}) is the main issue investigated in this paper.
We stress again that the logarithms present in the vector potential on a spacelike hypersurface through a gauge transformation have nothing to do with the logarithms that develop in both the vector potential and the field strengths at null infinity.  They do not affect in particular the form of the field strengths, which are gauge invariant.  The field strengths are free from logarithms on spacelike hypersurfaces.

We also emphasize that the logarithmic gauge transformations dominate the $\mathcal O(1)$ gauge transformations at spatial infinity. As we will see,  this is not the case any more at null infinity.

\section{The vector potential $A_{\mu}$ in hyperbolic coordinates  (in the absence of logarithmic gauge transformations)} \label{Sec:AinHyp}

\subsection{Field equations in hyperbolic coordinates (Lorenz gauge)}

In order to determine the asymptotic form of the vector potential near null infinity for given initial data on a Cauchy hypersurface, we first integrate the equations in hyperbolic coordinates.   As shown in \cite{Ashtekar:1978zz}, hyperbolic coordinates are extremely useful for connecting spatial infinity with null infinity.  Further key insight was given in \cite{Beig:1982ifu}.  

It is an easy generalization to consider arbitrary spacetime dimension $D$, which is what we shall do.  We will then specify to $D=4$.  

The solution for the field strengths has already been determined in \cite{Henneaux:2018gfi,Henneaux:2019yqq} and is recalled in Appendix {\bf \ref{App:Derivation}}. As explained in the introduction, the reason why we want to know explicitly the vector potential, and not just the field strengths, is that the vector potential contains information about ``improper" \cite{Benguria:1976in} (or ``large") gauge transformations, to which the field strength is blind.  In particular, information about the Goldstone boson of the angle-dependent $u(1)$-transformations \cite{He:2014cra,Kapec:2015ena,Strominger:2017zoo}, as well on the logarithmic gauge transformations,   is encoded in $A_\mu$.

We recall that hyperbolic coordinates are defined in the region $r > \vert t \vert  \Leftrightarrow \vert s \vert < 1$ by
\be 
\eta = \sqrt{-t^2 + r^2}, \; \; \; s = \frac{t}{r} 
\ee
(angles $x^A$ unchanged). The inverse coordinate transformation is given by
\begin{equation}
    t = \eta \frac {s}{\sqrt{1-s^2}}\,,\quad  r = \eta \frac
    {1}{\sqrt{1-s^2}}\,. \end{equation}
In hyperbolic coordinates, the Minkowski line element reads
\begin{equation}
g_{\mu\nu}dx^{\mu}dx^{\nu}=d\eta^{2}+\eta^{2}h_{ab}dx^{a}dx^{b}\quad\text{with}\quad(x^{a})=(s, x^{A})\,,
\end{equation}
with
\begin{equation}
h_{ab}dx^{a}dx^{b}=-\frac{1}{(1-s^{2})^{2}}ds^{2}+\frac{\xbar \gamma_{AB}}{1-s^{2}}d x^{A}d x^{B}\,,
\end{equation}
where $\xbar \gamma_{AB}$ is the metric on the unit round ($D-2$)-sphere.

The curvature of the metric $h_{ab}$ on the hyperboloid is
\be
{R^c}_{mab} = \delta^c_a h_{mb} -  \delta^c_b h_{ma}\,,
\ee
so that
\be
[\mathcal{D}_{a}, \mathcal{D}_{b}]v^c = \delta^c_a v_b - \delta^c_b v_a \, , \qquad [\mathcal{D}_{a}, \mathcal{D}_{b}]\theta_c = - h_{cb} \theta_a +  h_{ca} \theta_b\,,
\ee
where $\mathcal{D}_{a}$ is the covariant derivative associated to $h_{ab}$ and $v^c$, $\theta_c$ arbitrary vectors and covectors on the hyperboloid (with $v_a = h_{ab} v^b$).

Other useful facts that we shall repeatedly use are that the hypersurface $t=0$ coincides with $s=0$ and that  on that hypersurface, $r= \eta$ and $\partial_s = r \partial_t$.

We now write down the equations in hyperbolic coordinates.  As observed in \cite{Henneaux:2018gfi}, the boundary conditions on Cauchy hypersurfaces displayed above asymptotically imply the Lorenz gauge.  We shall thus impose $\nabla^{\mu}A_{\mu}=0$.

In hyperbolic coordinates, the Lorenz gauge condition reads
\begin{equation}
\eta^{-D+3}\partial_{\eta}\left(\eta^{D-1}A_{\eta}\right)+\mathcal{D}_{a}A^{a}=0\,.\label{eq:Lorenz-gauge}
\end{equation}
 The Maxwell
equations  reduce then to $\nabla_{\mu}\nabla^{\mu}A^{\nu}=0$ and read
\begin{align}
\eta^{-D+3}\partial_{\eta}\left(\eta^{D-1}\partial_{\eta}A_{\eta}\right)-(D-1)A_{\eta}-2\eta^{-1}\mathcal{D}_{a}A^{a}+\mathcal{D}_{a}\mathcal{D}^{a}A_{\eta} & =0\,,\label{eq:MaxwellE1}\\
\eta^{-D+4}\partial_{\eta}\left[\eta^{D-1}\partial_{\eta}\left(\eta^{-1}A_{a}\right)\right]+\mathcal{D}_{b}\mathcal{D}^{b}A_{a}-A_{a}+2\eta\partial_{a}A_{\eta} & =0\,. \label{eq:MaxwellE2}
\end{align}
By using the Lorenz gauge condition, we can reduce \eqref{eq:MaxwellE1} to an equation for $A_{\eta}$ only
\begin{equation}
\eta^{-D+3}\partial_{\eta}\left(\eta^{D-1}\partial_{\eta}A_{\eta}\right)+2\eta\partial_{\eta}A_{\eta}+(D-1)A_{\eta}+\mathcal{D}_{a}\mathcal{D}^{a}A_{\eta}=0\,.\label{eq:MaxwellE1red}
\end{equation}

In $D$ spacetime dimensions, the field strengths decay at spatial infinity (in Cartesian coordinates) as $\sim r^{-D+2}$ (to leading order) and so one might be tempted to take as asymptotic behaviour for the gauge potentials
\be
A_{\eta}  =\sum_{k \geq 0}\eta^{-D+3 -k}A_{\eta}^{(k)}\,, \qquad
A_{a}  =\sum_{k\geq 0}\eta^{-D+4-k}A_{a}^{(k)}\,. \label{Eq:AsympGP0}
\ee
However, as we pointed out in the introduction, the gauge transformations necessary to achieve (\ref{Eq:AsympGP0}) involve in general an improper gauge symmetry and so this expansion is too restrictive: one must allow terms that decay slowlier but are pure gradients.  In four dimensions, these improper gauge terms involve precisely the logarithmic angle-dependent $u(1)$ gauge transformations, which are the central subject of this paper. 

Our method for handling this question will be to first solve the equations of motion under the conditions  (\ref{Eq:AsympGP0}) and then perform the necessary improper gauge transformations to reach the desirable form. Thus, from now on, we work with the form (\ref{Eq:AsympGP0}) until we explicitly include the improper gauge terms.
In particular, in $D=4$ spacetime dimensions, we start with initial data that fulfill either the standard parity conditions twisted by a gauge transformation, or the non-standard twisted parity conditions, or the inverted parity conditions.  The logarithmic gauge terms, which have been considered only for the standard parity conditions twisted by a gauge transformation (\ref{Eq:ParityA00a})-(\ref{Eq:ParityA01ab}), are  introduced later, in Section {\bf \ref{Sec:LogGaugeTransf}}.

The Lorenz gauge \eqref{eq:Lorenz-gauge} imposes the following conditions on the various terms in the expansions  (\ref{Eq:AsympGP0}),
\be
\mathcal{D}_{a}A^{(k)a}-(k-2)A_{\eta}^{(k)} =0 \,.\label{eq:Lorenz-11V0}
\ee
The $\eta$-component \eqref{eq:MaxwellE1red} of the Maxwell equations becomes
\begin{equation}
\mathcal{D}_{a}\mathcal{D}^{a}A_{\eta}^{(k)}+(k+D-4)(k-2)A_{\eta}^{(k)}=0\,,\label{eq:A(k)_etaV0}
\end{equation}
and  the $a$-components \eqref{eq:MaxwellE2} read
\begin{equation}
\left[\mathcal{D}_{b}\mathcal{D}^{b}+k^{2}-2+(D-4)(k-1)\right]A_{a}^{(k)}+2\partial_{a}A_{\eta}^{(k)}=0\,. \label{eq:Maxwell-11V0}
\end{equation}
While the equations for for $A_\eta^{(k)}$ are homogeneous, those for $A_a^{(k)}$ are inhomogeneous with a source that involves $A_\eta^{(k)}$.

\subsection{Method for finding the general solution}

The strategy for solving the equations of motion consists in first solving the equations for $A_\eta^{(k)}$.   These can again be reduced, as we shall see,   to differential equations of the ultraspherical type.  

The procedure for determining then $A_a^{(k)}$ proceeds differently according to whether $D=4$, $k=0$ or $D-4+k>0$. The first case is more complicated and will be treated separately in Section {\bf \ref{Sec:AmuD=4}}.  The second case will be dealt with now.

The key point is to observe that the definition $F_{a\eta} \equiv E_{a} = \partial_a A_\eta - \partial_\eta A_a$ implies 
\begin{equation}
A_{a}^{(k)}=\frac{1}{k+D-4}\left(-\partial_{a}A_{\eta}^{(k)}+F_{a\eta}^{(k)}\right) \qquad (k +D -4\not= 0)\,.  \label{Eq:AaInTermsOfF}
\end{equation}
One can easily verify that this expression fulfills both (\ref{eq:Lorenz-11V0}) and (\ref{eq:Maxwell-11V0}), given the equation \eqref{eq:A(k)_etaV0} for $A_{\eta}^{(k)}$ and the  equation (\ref{eq:EqforEGen})  fulfilled by $E_{a}^{(k)}$ .
One can view  $-\frac{1}{k+D-4}\partial_{a}A_{\eta}^{(k)}$ as a particular solution of the inhomogeneous system for $A_{a}^{(k)}$ and $\frac{1}{k+D-4}F_{a\eta}^{(k)}$ as the general solution of the homogeneous system.  

It follows that the knowledge of $A_{\eta}^{(k)}$ and of the field strength $F_{a\eta}^{(k)}$, determined in Appendix {\bf \ref{App:Derivation}}, is sufficient to determine $A_a^{(k)}$ when $D-4+k>0$.

\subsection{Higher orders in the $D=4$ expansion - Spacetime dimensions $D>4$}
\label{Subsec:HigherD}

\subsubsection*{Determining $A_\eta^{(k)}$}
We proceed first to solve the equations for $A_{\eta}^{(k)}$,
which reduces to 
\begin{equation}
(1-s^{2})\partial_{s}^{2}A_{\eta}^{(k)}+(D-4)s\partial_{s}A_{\eta}^{(k)}-\xbar D_{A}\xbar D^{A}A_{\eta}^{(k)}-\frac{\left(k+D-4\right)\left(k-2\right)}{1-s^{2}}A_{\eta}^{(k)}=0\,.\label{eq:Ak_eta2D>4}
\end{equation}
We expand $A_{\eta}^{(k)}$ in spherical harmonics,
\begin{equation}
A_{\eta}^{(k)}=(1-s^{2})^{1- \frac{k}{2}}\sum_{l,m}\Theta_{lm}^{(k)}Y_{lm}\qquad\text{with}\qquad\xbar D_{A}\xbar D^{A}Y_{lm}=-l(l+D-3)Y_{lm}\,,
\end{equation}
leading  to the ultraspherical differential equation, 
\begin{equation}
(1-s^{2})\partial_{s}^{2}Y_{n}^{(\lambda)}+(2\lambda-3)s\partial_{s}Y_{n}^{(\lambda)}+(n+1)(n+2\lambda-1)Y_{n}^{(\lambda)}=0\,,   \label{Eq:Ultra000}
\end{equation}
where
\begin{equation}
\lambda=k+\frac{D-5}{2}\qquad\text{and}\qquad n=l-k+1\,.
\end{equation}

The second order differential equation (\ref{Eq:Ultra000}), which reduces to the Legendre differential equation for $\lambda = \frac12$, has been  throroughly discussed in \cite{Ultra} (see also \cite{Henneaux:2018mgn}).
Its general solution is a linear combination of two "basic" solutions denoted $\tilde{P}_{n}^{(\lambda)}(s)$  and $\tilde{Q}_{n}^{(\lambda)}(s)$ and   reads (with $n =l-k+1$), 
\begin{equation}
\Theta_{lm}^{(k)}(s)=\Theta_{lm}^{P(k)}\tilde{P}_{l-k+1}^{\left(k+\frac{D-5}{2}\right)}(s)+\Theta_{lm}^{Q(k)}\tilde{Q}_{l-k+1}^{\left(k+\frac{D-5}{2}\right)}(s)\,.
\end{equation}

One key feature of the basic solutions $\tilde{P}_{n}^{(\lambda)} $ and $\tilde{Q}_{n}^{(\lambda)}$ is \cite{Henneaux:2018mgn}:      
\be 
\tilde{P}_{n}^{(\lambda)}(- s) = (-1)^n \tilde{P}_{n}^{(\lambda)}(s) \, , \qquad \tilde{Q}_{n}^{(\lambda)}(- s) = (-1)^{n+1} \tilde{Q}_{n}^{(\lambda)}(s)\,.  \label{Eq:ParityPQ}
\ee
As stressed in \cite{Troessaert:2017jcm,Henneaux:2018mgn,Fuentealba:2024lll}, these relations control the matching conditions, which relate the fields at $s = \pm 1$.

Other key features, necessary for understanding the limits of the solutions as one goes to null infinity are ($\lambda$ is a half-integer or an integer $\geq 0$):
\begin{itemize}
\item For $n \geq 0$, one has $\tilde{P}_{n}^{(\lambda)}= (1-s^2)^{\lambda - \frac12} P_{n}^{(\lambda)}$ where $P_{n}^{(\lambda)}$ are the ultraspherical (or Gegenbauer) polynomials, while  $\tilde{Q}_{n}^{(\lambda)}=(1-s^2)^{\lambda - \frac12} Q_{n}^{(\lambda)}$ where $Q_{n}^{(\lambda)}$ are the ultraspherical functions of the second kind.  In particular,  $\tilde{P}_{n}^{(\frac12)}=  P_{n}^{(\frac12)}$ are the Legendre polynomials $P_n$, while $\tilde{Q}_{n}^{(\frac12)}=  Q_{n}^{(\frac12)}$ are the Legendre functions of the second kind $Q_n$.  For $\lambda$ a half-integer,  $\tilde{P}_{n}^{(\lambda)}$ is also a polynomial, while for $\lambda$ an integer,  $\tilde{P}_{n}^{(\lambda)}$ is a polynomial multiplied by $\sqrt{1-s^2}$.
\item For $n \geq 0$ and all values of $\lambda$, the $Q$-branch dominates the $P$-branch in the limit $s \rightarrow \pm 1$.   For $\lambda = \frac12$, the polynomials $\tilde P^{(\frac12)}_n = P^{(\frac12)}_n$ tend to finite, non-zero values, while the $\tilde Q^{(\frac12)}_n = Q^{(\frac12)}_n$ diverge logarithmically. For $\lambda > \frac12$, the $\tilde P^{(\lambda)}_n $ (which are polynomials for $\lambda$ half-integer) tend to zero, while the $\tilde Q^{(\lambda)}_n $ tend to finite, non-zero values.
\item For $n <0$, both branches $\tilde P^{(\lambda)}_n $ and $\tilde Q^{(\lambda)}_n $ tend to non-zero finite values.  The $\tilde Q^{(\lambda)}_n $ are polynomials, while the $\tilde P^{(\lambda)}_n $ are also polynomials when $\lambda$ is a half-integer. 
\item Explicit  recurrence formulas for the $\tilde P^{(\lambda)}_n $ and $\tilde Q^{(\lambda)}_n $ are  recalled in Appendix  {\bf \ref{App:Derivation}}.
\end{itemize}

The general solution for $A_{\eta}^{(k)}$ is thus
\begin{equation}
A_{\eta}^{(k)}=(1-s^{2})^{1-\frac{k}{2}}\sum_{l,m}\left[\Theta_{lm}^{P(k)}\tilde{P}_{l-k+1}^{(k+\frac{D-5}{2})}(s)+\Theta_{lm}^{Q(k)}\tilde{Q}_{l-k+1}^{(k+\frac{D-5}{2})}(s)\right]Y_{lm}\,.
\end{equation}
Because of (\ref{Eq:ParityPQ}) and the well-known property $Y_{lm}(-x^A) = (-1)^l Y_{lm}(x^A)$ of the spherical harmonics, the $P$-branch is even under the hyperboloid antipodal map, which is $x^A \rightarrow - x^A$, $s \rightarrow -s$, while the $Q$-branch is odd.  This implies that on the hyperplane $s=0$ ($\Leftrightarrow t=0$), the $P$-branch solution is even under the sphere antipodal map  and its $s$-derivative is odd, while it is the opposite for the $Q$-branch.

Since no parity condition is imposed on $A_{\eta}^{(k)}$ ($k>0$) on $t=0$,
 both the $P$-branch and the $Q$-branch should  be kept.

\subsubsection*{Determining $A_a^{(k)}$}

From (\ref{Eq:AaInTermsOfF}) and the expression of the field strengths given in Appendix {\bf \ref{App:Derivation}}, we get the components $A_a^{(k)}$ ($D-4 + k >0$).   One finds explicitly 
\begin{align}
A_{s}^{(k)} & =-\frac{(1-s^{2})^{-\frac{k}{2}}}{\left(D-4+k\right)}\sum_{l,m}\left[(1-s^{2})\partial_{s}-s\left(2-k\right)\right]\left[\Theta_{lm}^{P(k)}\tilde{P}_{l-k+1}^{\left(k+\frac{D-5}{2}\right)}+\Theta_{lm}^{Q(k)}\tilde{Q}_{l-k+1}^{\left(k+\frac{D-5}{2}\right)}\right]Y_{lm}\nonumber \\
 & \quad+\frac{(1-s^{2})^{-\frac{k}{2}}}{\left(D-4+k\right)}\sum_{l,m}\left[\Xi_{lm}^{P(k)}\tilde{P}_{l-k}^{\left(k+\frac{D-3}{2}\right)}+\Xi_{lm}^{Q(k)}\tilde{Q}_{l-k}^{\left(k+\frac{D-3}{2}\right)}\right]Y_{lm}\,,
\end{align}
\begin{equation}
A_{A}^{(k)T}=(1-s^{2})^{-\frac{k}{2}}\sum_{l>0,m}\left[\alpha_{lm}^{P(k)}\tilde{P}_{l-k}^{\left(k+\frac{D-3}{2}\right)}+\alpha_{lm}^{Q(k)}\tilde{Q}_{l-k}^{\left(k+\frac{D-3}{2}\right)}\right]\Psi_{A,lm}\,,
\end{equation}
\begin{align}
A_{A}^{(k)L} & =-\frac{(1-s^{2})^{1-\frac{k}{2}}}{\left(D-4+k\right)}\sum_{l,m}\left[\Theta_{lm}^{P(k)}\tilde{P}_{l-k+1}^{\left(k+\frac{D-5}{2}\right)}+\Theta_{lm}^{Q(k)}\tilde{Q}_{l-k+1}^{\left(k+\frac{D-5}{2}\right)}\right]\Phi_{A,lm}\nonumber \\
 & \quad-\frac{(1-s^{2})^{-\frac{k}{2}}}{\left(D-4+k\right)}\sum_{l>0,m}\frac{1}{l\left(l+D-3\right)}\nonumber\\
 & \qquad\qquad\left\{ \left[(1-s^{2})\partial_{s}+s\left(D-4+k\right)\right]\left[\Xi_{lm}^{P(k)}\tilde{P}_{l-k}^{\left(k+\frac{D-3}{2}\right)}+\Xi_{lm}^{Q(k)}\tilde{Q}_{l-k}^{\left(k+\frac{D-3}{2}\right)}\right]\Phi_{A,lm}\right\} \,.
\end{align}

\subsection{The vector potential $A_\mu^{(0)}$ in $D=4$ spacetime dimensions}
\label{Sec:AmuD=4}

The leading term  $A_\mu^{(0)}$  in the expansion of the vector potential cannot be obtained by following the above method since $D-4+k$ vanishes then.  While one can still determine $A_\eta^{(0)}$ by solving its homogeneous differential equation, the determination of $A_a^{(0)}$ is more involved.  Furthermore, one must take into account parity conditions specific to $D=4$, $k=0$.

\subsubsection*{Equations in hyperbolic coordinates}

In $D=4$, the Lorenz gauge \eqref{eq:Lorenz-gauge} takes the form
\be
\mathcal{D}_{a}A^{(k)a}-(k-2)A_{\eta}^{(k)} =0 \,.
\ee
The $\eta$-component of the Maxwell equations are
\begin{equation}
\mathcal{D}_{a}\mathcal{D}^{a}A_{\eta}^{(k)}+k(k-2)A_{\eta}^{(k)}=0\,,\quad\text{with}\quad k\geq0\,,\label{eq:A(k)_eta}
\end{equation}
while the $a$-components \eqref{eq:MaxwellE2} read
\be
\left(\mathcal{D}_{b}\mathcal{D}^{b}+k^{2}-2\right)A_{a}^{(k)}+2\partial_{a}A_{\eta}^{(k)}  =0 \,.
\ee

\subsubsection{Solution for $A_{\eta}^{(0)}$}

The equation for $A_{\eta}^{(0)}$ reduces to
\begin{equation}
(1-s^{2})\partial_{s}^{2}A_{\eta}^{(0)}-\xbar D_{A}\xbar D^{A}A_{\eta}^{(0)} = 0\,.\label{eq:Ak_eta2}
\end{equation}
We expand $A_{\eta}^{(0)}$ in spherical harmonics as usual, 
\begin{equation}
A_{\eta}^{(0)}=(1-s^{2})\sum_{l,m}\Theta_{lm}^{(0)}Y_{lm}\qquad\text{with}\qquad\xbar D_{A}\xbar D^{A}Y_{lm}=-l(l+1)Y_{lm}\,.
\end{equation}
This leads to the ultraspherical differential equation, 
\begin{equation}
(1-s^{2})\partial_{s}^{2}Y_{n}^{(\lambda)}+(2\lambda-3)s\partial_{s}Y_{n}^{(\lambda)}+(n+1)(n+2\lambda-1)Y_{n}^{(\lambda)}=0\,.\label{eq:EqY}
\end{equation}
with
\be
\lambda=-\frac{1}{2} \, , \qquad n=l-1\,.
\end{equation}
This case with $\lambda <0$ can be related to the case $\lambda = \frac12$ by observing that the equation (\ref{eq:EqY}), which takes the explicit form
\begin{equation}
(1-s^{2})\partial_{s}^{2}\Theta_{lm}^{(0)}-4s\partial_{s}\Theta_{lm}^{(0)}+\left[l(l+1)-2\right]\Theta_{lm}^{(0)}=0\,,\label{eq:EqTheta}
\end{equation}
coincides with the derivative of  Legendre's differential
equation
\be
(1-s^{2})\partial_{s}^{2}P_{l}-2s\partial_{s}P_{l}+l(l+1)P_{l}=0\,.\label{eq:Lde}
\end{equation}
Hence, the general solution is given by
\begin{eqnarray}
&&
\Theta_{00}^{(0)}(s)= \Theta_{00}^{P(0)}\frac{s}{1-s^2} +\Theta_{00}^{Q(0)}\partial_{s}Q_{0}(s)\,,\label{eq:Theta0-lm0} \\
&&\Theta_{lm}^{(0)}(s)= \Theta_{lm}^{P(0)}\partial_{s}P_{l}(s)+\Theta_{lm}^{Q(0)}\partial_{s}Q_{l}(s)\,, \quad l>0 \label{eq:Theta0-lm}
\end{eqnarray}
where $P_{l}(s)$ are the Legendre polynomials and $Q_{l}(s)$ are the
Legendre functions of the second kind ($P_{l}(s) \equiv P_{l}^{(\frac12)}(s)$,  $Q_{l}(s)\equiv Q_{l}^{(\frac12)}(s)$). ($\partial_s P_0 \equiv 0$ and one must use instead $\frac{s}{1-s^2}$ as non-trivial independent solution.)

Since $\partial_s P_{l}(s)$ is odd under the hyperboloid antipodal map while $\partial_s Q_{l}(s)$ is even, the part of the solution involving $\partial_s P_{l}(s)$ is such that, on $t=0$,  $\xbar A_r$ is odd under the sphere antipodal map and $\xbar \pi^r$ is even, while the part of the solution involving $\partial_s Q_{l}(s)$ enjoys the opposite parity properties.

It follows that both the standard parity conditions twisted by a gauge transformation and the non-standard twisted parity conditions force $\Theta_{lm}^{Q(0)} =0$, i.e., 
\begin{equation}
A_{\eta}^{(0)}=\Theta_{00}^{P(0)} s Y_{00}+ (1-s^{2})\sum_{l>0,m}\Theta_{lm}^{P(0)}\partial_{s}P_{l}(s)Y_{lm}\,. \label{Eq:ExpressionForAeta}
\end{equation}
In contrast, the inverted parity conditions imply $\Theta_{lm}^{P(0)} =0$, so that the solution for $A_\eta^{(0)}$ is then given by
\begin{equation}
A_{\eta}^{(0)}= (1-s^{2})\sum_{l \geq 0,m}\Theta_{lm}^{Q(0)}\partial_{s}Q_{l}(s)Y_{lm}\,. \label{Eq:ExpressionForAetaInverted}
\end{equation}

\subsubsection{Solution for $A_a^{(0)}$}

The system of equations to be solved reads 
\begin{align}
\mathcal{D}_{a}A^{a(0)}+2A_{\eta}^{(0)} & =0\,,\label{eq:Lorenz}\\
\left(\mathcal{D}_{b}\mathcal{D}^{b}-2\right)A_{a}^{(0)}+2\partial_{a}A_{\eta}^{(0)} & =0\,,\label{eq:Maxwell}
\end{align}
which can be explicitly expanded as
\begin{align}
(1-s^{2})\partial_{s}A_{s}^{(0)}-\xbar D_{A}A^{A(0)}-\frac{2}{1-s^{2}}A_{\eta}^{(0)} & =0\,,\label{eq:E1.0}\\
(1-s^{2})\partial_{s}^{2}A_{s}^{(0)}-4s\partial_{s}A_{s}^{(0)}-\xbar D_{A}\xbar D^{A}A_{s}^{(0)}+\frac{2s}{1-s^{2}}\xbar D_{A}A^{A(0)}-\frac{2}{1-s^{2}}\partial_{s}A_{\eta}^{(0)} & =0\,,\label{eq:E2.0}\\
(1-s^{2})\partial_{s}^{2}A_{A}^{(0)}-2s\left(\partial_{s}A_{A}^{(0)}-\partial_{A}A_{s}^{(0)}\right)-\xbar D_{B}\xbar D^{B}A_{A}^{(0)}+A_{A}^{(0)}-\frac{2}{1-s^{2}}\partial_{A}A_{\eta}^{(0)} & =0\,.\label{eq:E3}
\end{align}
There is some redundancy in this system, since the longitudinal part  of (\ref{eq:E3}) is identically satifisfied as a consequence of (\ref{eq:E1.0}) and (\ref{eq:E2.0}).  We thus focus on (\ref{eq:E1.0}), (\ref{eq:E2.0}) and the transverse part of (\ref{eq:E3}).

This  system of inhomogeneous second order linear differential equations for $A_{a}^{(0)}$ can be put in a tractable form as follows. 
The first equation can be solved to give the longitudinal component $A_{A}^{(0)L}$ in terms of $A_{s}^{(0)}$ and $A_{\eta}^{(0)}$,
\be
\xbar D_{A}A^{A(0)} = (1-s^{2})\partial_{s}A_{s}^{(0)}-\frac{2}{1-s^{2}}A_{\eta}^{(0)}  \label{eq:E1} \, .
\ee
Eliminating $A_{A}^{(0)L}$ from (\ref{eq:E2.0}) using this expression yields then a second order linear differential equation for $A_{s}^{(0)}$ with an inhomogeneous term involving $A_{\eta}^{(0)}$ and its derivative with respect to $s$, 
\be
(1-s^{2})\partial_{s}^{2}A_{s}^{(0)}-2s\partial_{s}A_{s}^{(0)}-\xbar D_{A}\xbar D^{A}A_{s}^{(0)}-\frac{4s}{(1-s^{2})^2}A_\eta^{(0)}-\frac{2}{1-s^{2}}\partial_{s}A_{\eta}^{(0)}  =0\, .\label{eq:E2}
\ee
Finally, the transverse part of (\ref{eq:E3}) yields
\begin{equation}
(1-s^{2})\partial_{s}^{2}A_{A}^{(0)T}-2s\partial_{s}A_{A}^{(0)T}-\xbar D_{B}\xbar D^{B}A_{A}^{(0)T}+A_{A}^{(0)T}=0\,,\label{eq:TranserverseEq}
\end{equation}
which is a homogeneous equation for $A_{A}^{(0)T}$.

\subsubsection*{The component $A_s^{(0)}$}
The equation for $A_s^{(0)}$ is inhomogeneous.  Its general solution is given by the sum of the general solution of the homogeneous equation and a particular solution of the inhomogeneous equation.

Let us find first the general solution of the homogeneous truncation of the equation (\ref{eq:E2}), which reads: 
\begin{equation}
(1-s^{2})\partial_{s}^{2}A_{s}^{(0)}-2s\partial_{s}A_{s}^{(0)}-\xbar D_{A}\xbar D^{A}A_{s}^{(0)}=0\,.\label{eq:E2-1}
\end{equation}
We expand the component  $A_{s}^{(0)}$ in spherical
harmonics
\begin{equation}
A_{s}^{(0)}=\sum_{l,m}K_{lm}(s)Y_{lm}\,,
\end{equation}
and get Legendre's differential equation,
\begin{equation}
(1-s^{2})\partial_{s}^{2}K_{lm}-2s\partial_{s}K_{lm}+l(l+1)K_{lm}=0\,.
\end{equation}
Therefore, the homogenous solution for $A_{s}^{(0)}$ is given by
\begin{equation}
A_{s}^{(0)}=\sum_{l,m}\left[K_{lm}^{P}P_{l}(s)+K_{lm}^{Q}Q_{l}(s)\right]Y_{lm} \quad \textrm{(homogeneous solution),}
\end{equation}
where $K_{lm}^{P}$ and $K_{lm}^{Q}$ are integration constants. 

In order to find a particular solution, we expand again $A_{s}^{(0)}$
in spherical harmonics
\begin{equation}
A_{s}^{(0)}=\sum_{l,m}\kappa_{lm}(s)Y_{lm}\,.
\end{equation}

Assume that $A_{\eta}^{(0)}$ is given (\ref{Eq:ExpressionForAeta}), i.e., that the standard parity conditions twisted by a gauge transformation or the non-standard twisted parity conditions hold.  The case of the inverted parity conditions will be considered next.
Then, equation \eqref{eq:E2} becomes\footnote{The equation (\ref{eq:EqP3-2}) can be somewhat simplified by setting $\kappa_{lm} = \partial_s f_{lm}$ but this will not be necessary for getting the solution.} 
\begin{equation}
(1-s^{2})\partial_{s}^{2}\kappa_{lm}-2s\partial_{s}\kappa_{lm}+l(l+1)\kappa_{lm}=2\partial_{s}\Theta_{lm}^{(0)}\,.\label{eq:EqP3-2}
\end{equation}

For $l=0$, the function $\Theta_{lm}^{(0)}$ is equal to $\Theta_{00}^{P(0)}\frac{s}{1-s^2}$ and a particular solution is easily found to be 
\be
\kappa_{00} = \frac{\Theta_{00}^{P(0)}}{1-s^2} \, .
\ee
For $l>0$, the function $\Theta_{lm}^{(0)}$ in $\eqref{eq:Theta0-lm}$ is proportional
to $\partial_{s}P_{l}(s)$, so that the right-hand side of the above equation is
a polynomial of degree $l-2$ in $s$. 
One can then prove that the particular solution $\kappa_{lm}(s)$ can be chosen to be also a polynomial of degree $l-2$,
\be
\kappa_{lm} = \textrm{polynomial of degree $l-2$ in $s$ ($\kappa_{1m}= 0$)} \, .
\ee
Indeed, let us make a polynomial ansatz for the solution
\begin{equation}
\kappa_{lm}(s)=\sum_{n=0}^{l-2}k_{lm}^{(n)}s^{n}\, ,
\end{equation}
and denote the expansion of the Legendre polynomials as
\begin{equation}
P_{l}(s)=\sum_{n=0}^{l}a_{l}^{(n)}s^{n}\,.
\end{equation}
The differential equation becomes
\begin{equation}
\sum_{n=0}^{l-4}(n+2)(n+1)k_{lm}^{(n+2)}s^{n}+\sum_{n=0}^{l-2}\left[l(l+1)-n(n+1)\right]k_{lm}^{(n)}s^{n}=2\Theta_{lm}^{P(0)}\sum_{n=0}^{l-2}(n+2)(n+1)a_{l}^{(n+2)}s^{n}\,.
\end{equation}
We then find that the coefficients of $\kappa_{lm}$ can be obtained
from the following recursion relations:
\begin{align}
& 2(2l-1)k_{lm}^{(l-2)}  =2\Theta_{lm}^{P(0)}l(l-1)a_{l}^{(l)}\,,\\
& 6(l-1)k_{lm}^{(l-3)}  =2\Theta_{lm}^{P(0)}(l-1)(l-2)a_{l}^{(l-1)} = 0\,,\\
& \left[l(l+1)-n(n+1)\right]k_{lm}^{(n)}  \nonumber \\
&\qquad \quad  =2\Theta_{lm}^{P(0)}(n+2)(n+1)a_{l}^{(n+2)} -(n+2)(n+1)k_{lm}^{(n+2)}\,,\qquad n\leq l-4\,,
\end{align}
where we have used that $a_{l}^{(l-1)}$ vanishes because the only powers present in $P_l$ are either all even ($l$ even) or all odd ($l$ odd).
One first determines $k_{lm}^{(l-2)}$ and $k_{lm}^{(l-3)}= 0$ from the first two equations.  One then proceeds successively to determine $k_{lm}^{(l-4)}$, $k_{lm}^{(l-5)}= 0$, $\cdots$ from the last equation, all the way down to $k_{lm}^{(0)}$ (which is zero when $l$ is odd).  The procedure works because the coefficient of $k_{lm}^{(n)}$ in the last equation never vanishes for $n\leq l-4$.

Hence, we have that
\begin{equation}
A_{s}^{(0)}=\sum_{l,m}\left[K_{lm}^{P}P_{l}(s)+K_{lm}^{Q}Q_{l}(s)+\kappa_{lm}(s)\right]Y_{lm}\,,
\end{equation}
where from now on $\kappa_{lm}(s)$ is the particular solution just determined.  Note that $\kappa_{lm}(s)$ inherits from its even source $2\partial_s \Theta_{lm}^{(0)}$ the property
\be
\kappa_{lm}(s) = (-1)^l \kappa_{lm}(s)\,.
\ee

\subsubsection*{The component $A_A^{(0)L}$}
The component $A_A^{(0)L}$ is determined from the Lorenz gauge, which takes the form (\ref{eq:E1}).  For this equation to possess a solution, its right-hand side can have no $l=0$ mode.  Imposing this condition yields
\begin{align}
0 & =\oint d^{2}x\sqrt{\xbar g}\left[(1-s^{2})\partial_{s}A_{s}^{(0)}-\frac{2}{1-s^{2}}A_{\eta}^{(0)}\right]\\
 & =(1-s^{2})\partial_{s}\left(K_{00}^{P}P_{0}(s)+K_{00}^{Q}Q_{0}(s)\right)\,.
\end{align}
The first term vanishes but the second is not zero since
$
\partial_s Q_{0}(s)=\frac{1}{1-s^2}
$.  This imposes
\begin{equation}
K_{00}^{Q}=0\,.
\end{equation}

Expanding $A_{A}^{(0)L}$ in longitudinal spherical harmonics $\Phi_{A,lm}=\partial_{A}Y_{lm}$ (see Appendix {\bf \ref{App:SphHar}}), we directly get
\begin{eqnarray}
A_{A}^{(0)L}&& =-\sum_{l>0,m}\frac{(1-s^{2})}{l(l+1)}\left(K_{lm}^{P}\partial_{s}P_{l}(s)+K_{lm}^{Q}\partial_{s}Q_{l}(s)\right)\Phi_{A,lm} \nonumber \\
&& \qquad  \qquad -\sum_{l>0,m}\frac{1}{l(l+1)}\left[(1-s^{2})\partial_{s}\kappa_{lm}-2\Theta_{lm}^{P(0)}\partial_{s}P_{l}(s)\right]\Phi_{A,lm}\,.
\end{eqnarray}

A direct computation shows that $(A_a^{(0)L}) \equiv (A_s^{(0)},A_A^{(0)L})$ is equal to
\be
A_a^{(0)L} = \partial_a ( \Gamma + \Phi)\,, \label{Eq:L=E+O}
\ee
with
\begin{eqnarray}
&&\Gamma = -\sum_{l>0,m}\frac{(1-s^{2})}{l(l+1)}K_{lm}^{P}\partial_{s}P_{l}(s) Y_{lm} \nonumber \\
&& \qquad  \qquad -\sum_{l>0,m}\frac{1}{l(l+1)}\left[(1-s^{2})\partial_{s}\kappa_{lm}-2\Theta_{lm}^{P(0)}\partial_{s}P_{l}(s)\right]Y_{lm} \, , \\
&& \Phi = -\sum_{l>0,m}\frac{(1-s^{2})}{l(l+1)}K_{lm}^{Q}\partial_{s}Q_{l}(s) Y_{lm}  \,, \\
&& \Gamma(-s, -x^A) = - \Gamma(s, x^A) \, , \qquad \Phi (-s, -x^A) = \Phi (s,x^A) \, , \label{Eq:PhiHyp}
\end{eqnarray}
where we have set the undetermined integration constant in $\Phi$, which drops out from $\partial_a \Phi$, equal to zero.  One can rewrite $\Phi$ in a simpler way as
\be
\Phi = \sum_{l > 0,m}(1-s^{2})C_{lm}^{Q}\partial_{s}Q_{l}(s) Y_{lm} \,, \label{Eq:FormPhi}
\ee
by setting
\be 
 C_{lm}^{Q} = - \frac{1}{l(l+1)}K_{lm}^{Q} \, \quad (l>0) \,.\label{Eq:DefClm}
\ee  

Because $A_a^{(0)L}$ is a pure gradient, it is left arbitrary by the parity conditions, which prescribe the leading part of the vector potential ``up to a gauge".   There is no restriction on the integration constants $K_{lm}^{P}$ and $K_{lm}^{Q}$.  We note, however, that with the parity conditions on the radial components of the fields, $\Gamma$ defines a proper gauge transformation with zero charge so that there is no loss of generality in assuming $K_{lm}^{P} = 0$, while $\Phi$ defines an improper gauge transformation.  This will be shown below when we discuss the charges.

For the inverted parity conditions, the situation is opposite.
The solution for $A_a^{(0)L}$ is again of the form
\be
A_a^{(0)L} = \partial_a ( \Gamma + \Phi) \, , \quad  \Gamma(-s, -x^A) = - \Gamma(s, x^A) \, , \quad \Phi (-s, -x^A) = \Phi (s,x^A) \,,
\ee
with now
\begin{eqnarray}
&&\Gamma = -\sum_{l>0,m}\frac{(1-s^{2})}{l(l+1)}K_{lm}^{P}\partial_{s}P_{l}(s) Y_{lm} \,,  \\
&& \Phi = -\sum_{l>0,m}\frac{(1-s^{2})}{l(l+1)}K_{lm}^{Q}\partial_{s}Q_{l}(s) Y_{lm} + C \nonumber \\
&& \qquad  \qquad -\sum_{l>0,m}\frac{1}{l(l+1)}\left[(1-s^{2})\partial_{s}\kappa'_{lm}-2\Theta_{lm}^{Q(0)}\partial_{s}Q_{l}(s)\right]Y_{lm} \, .
\end{eqnarray}
Here, $\kappa'_{lm}(s)$ is a particular solution of the inhomogeneous equation for $A_s^{(0)}$, which inherits the parity properties of $Q_l(s)$, which we do not need to work out explicitly because it is now $\Phi$ that defines a proper gauge transformation so that only $\Gamma$ is physically relevant.

\subsubsection*{The component $A_A^{(0)T}$}

If we expand $A_{A}^{(0)T}$ in transverse vector
spherical harmonics $\Psi_{A,lm}$,
\begin{equation}
A_{A}^{(0)T}=\sum_{l>0,m}T_{lm}(s)\Psi_{A,lm}\, , 
\end{equation}
Eq. \eqref{eq:TranserverseEq} reduces also to the Legrendre differential equation,
\begin{equation}
(1-s^{2})\partial_{s}^{2}T_{lm}-2s\partial_{s}T_{lm}+l\left(l+1\right)T_{lm}=0\,.
\end{equation}
The solution for $A_{A}^{(0)T}$ is therefore given by
\begin{equation}
A_{A}^{(0)T}=\sum_{l>0,m}\left(T_{lm}^{P}P_{l}(s)+T_{lm}^{Q}Q_{l}(s)\right)\Psi_{A,lm}\,, \label{Eq:ExpressionForAAT}
\end{equation}
where $T_{lm}^{P}$ and $T_{lm}^{Q}$ are integration constants.

With the standard parity conditions twisted by a gauge transformation, $A_{A}^{(0)T}$ must be even, which forces $T_{lm}^{Q}=0$, so that
\begin{equation}
A_{A}^{(0)T}=\sum_{l>0,m}T_{lm}^{P}P_{l}(s)\Psi_{A,lm}\,. \label{Eq:ExpressionForAATStandard}
\end{equation}
For the non-standard twisted parity conditions or the inverted parity conditions, $A_{A}^{(0)T}$ must be odd, which forces $T_{lm}^{P}=0$, so that
\begin{equation}
A_{A}^{(0)T}=\sum_{l>0,m}T_{lm}^{Q}Q_{l}(s)\Psi_{A,lm}\,. \label{Eq:ExpressionForAATInverted}
\end{equation}

\subsubsection*{Connection with the field strength}

\noindent {\it Standard parity conditions twisted by a gauge transformation or non-standard twisted parity conditions}

Putting everything together, we have proved that the general solution to the system of equations for the vector potential in the Lorenz gauge is given by
\begin{align}
A_{\eta}^{(0)} & =\Theta_{00}^{P(0)} s Y_{00} + (1-s^{2})\sum_{l>0,m}\Theta_{lm}^{P(0)}\partial_{s}P_{l}(s)Y_{lm}\,, \label{Eq:ARta007}\\
A_{s}^{(0)} & =\sum_{l,m}\left[K_{lm}^{P}P_{l}(s)+K_{lm}^{Q}Q_{l}(s)+\kappa_{lm}(s)\right]Y_{lm}\,,\\
A_{A}^{(0)L} & =-\sum_{l>0,m}\frac{(1-s^{2})}{l(l+1)}\left[K_{lm}^{P}\partial_{s}P_{l}(s)+K_{lm}^{Q}\partial_{s}Q_{l}(s)\right]\Phi_{A,lm}\nonumber \\
 & \quad-\sum_{l>0,m}\frac{1}{l(l+1)}\left[(1-s^{2})\partial_{s}\kappa_{lm}(s)-2\Theta_{lm}^{P(0)}\partial_{s}P_{l}(s)\right]\Phi_{A,lm}\,,\\
A_{A}^{(0)T} & =\sum_{l>0,m}\left(T_{lm}^{P}P_{l}(s)+T_{lm}^{Q}Q_{l}(s)\right)\Psi_{A,lm}\,,
\end{align}
where $K_{00}^{Q}=0$. 

We can compute the $k=0$ term in the expansion of the field strength from these expressions.   One has
\be
E_a^{(0)} = \partial_a A^{(0)}_\eta \, , \qquad F^{(0)}_{sA} = \partial_s A^{(0)}_A - \partial_A A^{(0)}_s\, , \qquad  F^{(0)}_{AB} =\partial_A A^{(0)}_B - \partial_B A^{(0)}_A
\ee
($\partial_\eta A_a = \mathcal O\left(\eta^{-2}\right)$).  By direct computation, one finds that the expressions derived from the vector potential match those obtained by direct integration of the Maxwell equations for $F_{\mu \nu}$ given in Appendix {\bf \ref{App:Derivation}} provided one makes the following identifications:
\be
\Theta_{00}^{P(0)} = \Xi_{00}^{P(0)} \, , \quad - l(l+1) \Theta_{lm}^{P(0)} = \Xi_{lm}^{P(0)} \quad (l>0) \, ,  \label{Eq:ThetaXi}
\ee
and
\be
T_{lm}^{P} = \alpha_{lm}^{P(0)} \, , \qquad T_{lm}^{Q} = \alpha_{lm}^{Q(0)} \, .
\ee

The first relation follows 
by comparing  (\ref{Eq:ExpForEs}) with the $s$-derivative of (\ref{Eq:ExpressionForAeta}).   It then automatically implies the matching of $E_A^{(0)L}$ given by (\ref{Eq:EAkL}) with $\partial_A A_{\eta}^{(0)}$.   The fact that $E_A^{(0)T}$ is equal to zero from $E_A^{(0)} = \partial_A A^{(0)}_\eta$ matches Eq.(\ref{Eq:EAkT}) with $D=4$ and $k=0$. Similarly, one finds that $F_{sA}^L$ given by $\partial_s A^{(0)L}_A - \partial_A A^{(0)}_s$ also vanishes, in agreement with (\ref{Eq:FsAkL})  with $k=0$. 

The second relation simply follows by comparing $F_{AB}^{(0)}$ given by (\ref{Eq:ExpForFAB}) with $\partial_A A^{(0)T}_B - \partial_B A^{(0)T}_A$ with $A^{(0)T}_A$ given by (\ref{Eq:ExpressionForAAT}).  This automatically ensures that (\ref{Eq:FsAkT})  with $k=0$ matches $\partial_s A^{(0)T}_A$.

Note that the integration constants $K_{lm}^{P}$ and $K_{lm}^{Q}$ drop from the field strength, as they should since they appear in the vector potential only through a gradient.
Note also that in $D=4$ spacetime dimensions, the magnetic term (\ref{Eq:MagneticCharge}) does not derive from a single-valued potential and so, the magnetic charge $g$ is not related to the coefficients $T_{lm}^{P(0)}$ -- it is an independent parameter.

\vspace{.2cm}

\noindent {\it Inverted parity conditions}

For the inverted parity conditions, one must make the following identifications:
\be
 - l(l+1) \Theta_{lm}^{Q(0)} = \Xi_{lm}^{Q(0)}  \, , \qquad
 T_{lm}^{Q} = \alpha_{lm}^{Q(0)} \, .
\ee

\subsection{Comment on the behaviour of the solution as $s \rightarrow \pm 1$}
Because of the generic presence of the ultraspherical functions of the second kind $Q^{(\lambda)}_n$, the vector potential involves logarithms as we take the limit $s \rightarrow \pm 1$.  These logarithms are subleading if the $Q$-branch is absent for $k=0$ but are nevertherless present for $k>0$.  

In order to study more carefully the behaviour of the fields near the critical sets $s \rightarrow \pm 1$ where spatial infinity meet future and past null infinities, we go to Friedrich coordinates \cite{Fried1,Friedrich:1999wk,Friedrich:1999ax}, which cover better that region\footnote{A clear description of the critical sets can be found in \cite{Gasperin:2017apb,Paetz:2018nbd,Gasperin:2021vnm,Mohamed:2021rfg}.}.

We shall not carry this analysis in detail in our paper as we shall go directly to retarded or advanced null coordinates.  We simply point out here that the quantities $A_A A^A$ and $A_w A^w$ ($w = \eta, s$), which are scalars under changes of coordinates in the two-surface spanned by $\eta$ and $s$, go to zero as one takes the limit to the critical sets either coming from spatial infinity ($\rho \rightarrow \infty$ followed by $s \rightarrow \pm 1$), or coming from null infinty ($s \rightarrow \pm 1$ followed by $\rho \rightarrow \infty$).  Here, $\rho$ is the (inverted) Friedrich coordinate $\rho = \eta (1-s^2)^{\frac12}$.

\section{$D=4$ conserved charges}
\label{Sec:ConservedCharges}

Even in the absence of the logarithmic gauge transformations, and independently of the parity conditions imposed on the leading orders of the canonical variables, the electromagnetic field possesses two infinite families of conserved charges given by flux integrals at spatial infinity, one of electric type and the other of magnetic type.  When logarithmic gauge transformations are turned on, further conserved charges arise.

We first exhibit the two infinite families of conserved charges on Cauchy hypersurfaces and turn next to their description in hyperbolic coordinates.   Logarithmic charges are discussed in Section {\bf \ref{Sec:LogGaugeTransf}}.

\subsection{Conserved charges of electric type}
The conserved charges of electric type have a direct Noether interpretation and generate proper and improper gauge transformations. They take the following expression in terms of the canonical variables $(A_i, \pi^i)$, $(A_0, \pi^0)$ \cite{Henneaux:2018gfi}
\be G^e[\epsilon, \mu] = \int d^3 x \left(\epsilon \mathcal G + \mu \pi^0 \right) + \oint d^2 x \left(\xbar \epsilon \xbar \pi^r - \xbar \mu \xbar A_r \sqrt{\xbar \gamma} \right) \, .
\label{Eq:ECharges1}
\ee  
The asymptotic behaviour of the gauge parameters is $\epsilon = \xbar \epsilon+ \frac{\epsilon^{(1)}}{r} + \mathcal O (r^{-2})$ and $\mu = \frac{\xbar \mu}{r} + \mathcal O (r^{-2})$.
The gauge transformations shifts $A_0$ as $\delta_\mu A_0 = \mu$.  The charges are conserved provided $\partial_t \xbar \epsilon = 0$, $\partial_t \xbar \mu = 0$ since the Hamiltonian is invariant under all gauge transformations.

The Hamiltonian formulation of electromagnetism in hyperbolic coordinates proceeds of course in the same way and yields as generator of the gauge transformations 
\be G^e[\epsilon,  \mu'] = \int d^3 x \left(\epsilon \mathcal G +  \mu' \pi^\eta \right) + \oint d^2 x \left( \lambda^{(0)}  \pi^{(0)\eta} -  \mu'^{(0)}  A^{(0)}_\eta \sqrt{\xbar \gamma} \right)
\ee
where, since $A_s = \mathcal O(1)$ as can be seen on the slice $t=0=s$ where $\eta = r$ and $A_s = r A_t$, we have rescaled the gauge parameter $\mu$ under which $A_s$ is shifted by a power of $\eta$, $\mu' = \mathcal O(1)$.  
We have also set
\be 
\epsilon= \lambda^{(0)}+\mathcal{O}\left(\eta^{-1}\right)\, , \qquad
\mu'= \mu'^{(0)}+\mathcal{O}\left(\eta^{-1}\right)\, ,
\ee
 and
\be 
A_{\eta}=\frac{1}{\eta} A^{(0)}_{\eta}+\mathcal{O}\left(\eta^{-2}\right)\, , \qquad
\pi^{\eta}= \pi^{(0)\eta}+\mathcal{O}\left(\eta^{-1}\right)\, .
\ee
This last expansion can be checked from the definition of the conjugate momenta, which yields $\pi^\eta = (\partial_s A_\eta - \partial_\eta A_s) \eta \sqrt{\xbar \gamma}$ (with $\partial_s A_\eta^{(0)} \not=0$ in hyperbolic coordinates).  In particular,
\be
\pi^{(0)\eta} = \partial_s A_\eta^{(0)} \sqrt{\xbar \gamma} \, .
\ee

To make contact with the Maxwell Lagrangian, we eliminate the conjugate momenta through their definition and impose a gauge condition such that $\mu' = \partial_s \epsilon$, i.e., go back to the original, ``unextended" Hamiltonian formulation.  This implies in particular $\mu'^{(0)}= \partial_s \lambda^{(0)}$.  The surface term $Q^e_\lambda$ in the canonical generator of the gauge transformations becomes then
\begin{equation}
Q^e_{\lambda}=\oint \left(\lambda^{(0)}\partial_{s}A_{\eta}^{(0)}-\partial_{s}\lambda^{(0)} A_{\eta}^{(0)}\right)\sqrt{\xbar \gamma}d^{2}x\,, \label{Eq:ECharges2}
\end{equation}
This quantity can equivalently be derived as the Noether charge of the gauge transformations from the action
\begin{equation}
S=-\frac{1}{4}\int d^{4}x\,F^{\mu\nu}F_{\mu\nu}+\intop_{\mathcal{H}}d^{3}x\sqrt{-h}\left(A_{\eta}^{(0)}\mathcal{D}^{a}A_{a}^{(0)}+A_{\eta}^{(0)}{}^{2}\right)\,.
\end{equation}
where the surface term is needed to make the variational principle well-defined \cite{Henneaux:2018gfi,Campiglia:2017mua}.

The connection between expressions in Minkowskian coordinates and in hyperbolic coordinates is easily worked out on the surface $t=0=s$ and is given by the formulas
\begin{align}
A_{\eta}\Big|_{s=0} & =A_{r}=\frac{1}{r}\xbar A_{r}+\frac{1}{r^{2}}A_{r}^{(2)}+\mathcal{O}\left(r^{-3}\right)\,,\\
\partial_{s}A_{\eta}\Big|_{s=0} & =r\partial_{t}A_{r}=\partial_{t}\xbar A_{r}+\frac{1}{r}\partial_{t}A_{r}^{(2)}+\mathcal{O}\left(r^{-2}\right)\nonumber \\
 & =\frac{\xbar\pi^{r}}{r\sqrt{\xbar \gamma}}+\mathcal{O}\left(r^{-2}\right)\,,
\end{align}
where the Hamiltonian equations of the motion in Minkowskian coordinates
\begin{equation}
\partial_{t}\xbar A_{r}=0\,,\qquad\partial_{t}A_{r}^{(2)}=\frac{\xbar\pi^{r}}{\sqrt{\xbar \gamma}}\,,
\end{equation}
have been used. 
Since one has furthermore
\begin{align}
\lambda^{(0)}\Big|_{s=0} & =\xbar\epsilon\,,\qquad\partial_{s}\lambda^{(0)}\Big|_{s=0}=\xbar\mu\,,
\end{align}
one obtains
$ Q^e_{\epsilon,\mu}  = Q^e_{\lambda}$,
as one should.

Now, while the charge is conserved in Minkowskian coordinates if $\partial_t \xbar \epsilon = 0$, $\partial_t \xbar \mu = 0$, it is conserved in hyperbolic coordinates provided
\begin{equation}
(1-s^2)\partial_s^2\lambda^{(0)}-\xbar D_A \xbar D^A \lambda^{(0)}=0\,.
\end{equation}
Both sets of equations are of course equivalent and express, in their respective coordinate systems, that $\Box \epsilon = 0$ to leading order (compare with Eq. (3.5) of \cite{Fuentealba:2024lll}, with $k=-1$).  We recall that the boundary conditions  on the vector potential at infinity implement asymptotically the Lorenz gauge, which is preserved only by gauge transformations fulfilling $\Box \epsilon = 0$.

One way to understand the $s$-dependence of the gauge parameter $\lambda^{(0)}$ is to observe that the Hamiltonian in hyperbolic coordinates is similar to a boost generator and is not invariant under improper gauge transformations, which form a non-trivial representation of the homogeneous Lorentz group (but commute with the translations).

The general solution to the equation for the gauge parameter reads
\begin{equation}
\lambda^{(0)}= \lambda_P+\lambda_Q\,,
\end{equation}
where
\begin{align}
\lambda_P&=\lambda^{P(0)}_{00}s Y_{00}+(1-s^2)\sum_{l>0,m}\lambda^{P(0)}_{lm}\partial_s P_l(s)Y_{lm}\,,\\
\lambda_Q&= \lambda^{Q(0)}_{00}Y_{00}+ (1-s^2)\sum_{l>0,m}\lambda^{Q(0)}_{lm}\partial_s Q_l(s)Y_{lm}\,, \label{Eq:lambdaQ}
\end{align}
where we have isolated the $l=0$ mode.

The leading order of the radial component of the electromagnetic potential is odd
for both the standard parity conditions twisted by a gauge transformation and the non-standard  twisted parity conditions.   It reads (see (\ref{Eq:ExpressionForAeta}))
\begin{equation}
A^{(0)}_{\eta}=\Theta^{P(0)}_{00}s Y_{00}+(1-s^2)\sum_{l>0,m}\Theta^{P(0)}_{lm}\partial_s P_l(s)Y_{lm}\,,
\end{equation}
where, as we have seen, the coefficients $\Theta^{P(0)}_{lm}$ are also the coefficients appearing in the expansion of the radial electric field $E_s^{(0)}$  (see (\ref{Eq:ThetaXi})). 
The charge can then be written as
\begin{equation}
Q^e_{\lambda}=Q^e_P+Q^e_Q\,,
\end{equation}
where
\begin{align}
Q^e_P&=\oint \left(\lambda_P\partial_{s}A_{\eta}^{(0)}-\partial_{s}\lambda_P A_{\eta}^{(0)}\right)\sqrt{\xbar \gamma}d^{2}x\,,\\
Q^e_Q&=\oint \left(\lambda_Q\partial_{s}A_{\eta}^{(0)}-\partial_{s}\lambda_Q A_{\eta}^{(0)}\right)\sqrt{\xbar \gamma}d^{2}x\,.
\end{align}
By direct insertion of the above expressions, we find that $Q^e_P=0$, so that the charge reduces to $Q^e_Q$.    This  is just the translation in hyperbolic coordinates that only the leading even part under of the sphere antipodal map of the gauge parameter on spacelike hypersurfaces and the leading odd part of its  time derivative  contribute to the charges.  These define indeed a gauge parameter that is even under the hyperbolic antipodal map.  The other parity component defines proper gauge transformations.

That $\lambda_P$ drops from the charge justifies the claim made in Section {\bf \ref{Sec:AinHyp}} that $\Gamma$ in (\ref{Eq:L=E+O}) defines a proper gauge transformation with zero charge and that only $\Phi$ is relevant for the physical part of the leading order of $A^L_A$.

In the null infinity limit $s\rightarrow 1$, the expression of the (conserved) charge $Q^e_Q$  becomes
\be
\lim_{s\rightarrow 1} Q^e_Q =    \lambda^{Q(0)}_{00}\Theta^{P(0)}_{00} -\oint \left(\sum_{l>0,m}\lambda^{Q(0)}_{lm}Y_{lm}\right)\left(\sum_{l>0,m}l(l+1)\Theta^{P(0)}_{lm}Y_{lm}\right)\sqrt{\xbar \gamma}d^{2}x\,.
\ee
The first term is the total electric charge multiplied by the $0$-mode of the gauge parameter, the other terms are the charges of the higher harmonics of the angle-dependent $u(1)$-symmetry.

The discussion of the conserved charges associated with improper gauge transformations proceed along exactly the same lines for the  inverted parity conditions.  The Hamiltonian generators and the surface integrals take the same form and so, from the point of view of spatial infinity, handling these different boundary conditions is not a big deal.  

The only difference is that it is now $Q^e_Q$ that vanishes since $A^{(0)}_{\eta}$ has opposite parity properties and reads (see (\ref{Eq:ExpressionForAetaInverted}))
\begin{equation}
A_{\eta}^{(0)}= (1-s^{2})\sum_{l \geq 0,m}\Theta_{lm}^{Q(0)}\partial_{s}Q_{l}(s)Y_{lm}\,. 
\end{equation}
It is therefore the even $\lambda_Q$ that defines proper gauge symmetries for the inverted parity conditions, and the odd $\lambda_P$ that defines improper gauge symmetries.   The charge contains no Coulomb term but generates nevertherless an infinite number of angle-dependent $u(1)$ transformation (the zero mode goes with $\xbar A_r$, not $\xbar \pi^r$).

Even though there is hardly anything to say about the inverted boundary conditions at spatial infinity,  the behaviour as we go to null infinity is worth reporting since $A_\eta$ and $\partial_s A_\eta$ develop logarithms as $s \rightarrow 1$.  One finds in that limit that
the (conserved) charge $Q^e_P$  becomes
\be
\lim_{s\rightarrow 1} Q^e_P =    \lambda^{P(0)}_{00}\Theta^{Q(0)}_{00} -\oint \left(\sum_{l>0,m}l(l+1)\lambda^{P(0)}_{lm}Y_{lm}\right)\left(\sum_{l>0,m}\Theta^{Q(0)}_{lm}Y_{lm}\right)\sqrt{\xbar \gamma}d^{2}x\,.
\ee
In spite of the logarithms in the fields, the charges are finite and well-defined. This is not a surprise since they are conserved, and clearly shows that one cannot use finiteness of the charges as a criterion to privilege one set of asymptotic conditions over the others.  Finiteness of the charges in the limit $s \rightarrow 1$ to the past boundary of future null infinity is always guaranteed if one uses the expressions resulting there upon integration of the Hamiltonian expressions at spatial infinity.

We note that the charges considered here are all expressed in terms of the spherical harmonic coefficients $\Theta^{P(0)}_{lm}$ or $\Theta^{Q(0)}_{lm}$ of the leading term $A_\eta^{(0)}$ in the expansion of the radial vector potential in inverse powers of $\eta$.  These coefficients themselves are related through (\ref{Eq:ThetaXi}) to the leading order of the radial component of the electric field.  Charges associated with the subleading orders $\eta^{-k-1}$ ($k>0$) have been considered in
\cite{Campiglia:2018dyi} and involve the corresponding integration constants $\Xi^{P(k)}_{lm}$ and $\Xi^{Q(k)}_{lm}$.

\subsection{Conserved charges of magnetic type}

The Maxwell equations imply that there are also conserved quantities of magnetic type, which take the form 
\begin{equation}
Q^m_{\epsilon,\mu}=\oint\epsilon^{AB}\left(2\mu\partial_{A}\xbar\pi_{B}-\sqrt{\xbar\gamma} \epsilon \xbar F_{AB}\right)d^{2}x\,, \label{Eq:MagneticCharges1}
\end{equation}
in terms of Hamiltonian quantities on constant $t$ Cauchy hypersurfaces. Here, $\epsilon$ and $\mu$ are arbitrary functions of the angles with the same parity on the sphere as $\epsilon^{AB} \xbar F_{AB}$ and $\epsilon^{AB}\partial_{A}\xbar\pi_{B}$ (the other parity components drop). A straightforward computation shows that these quantities are indeed conserved provided $\dot \mu = \dot \epsilon = 0$.  
When $\epsilon^{AB} \xbar F_{AB}$ is even (standard parity conditions), the second integral gives the magnetic charge when $\epsilon = \epsilon_0$, hence the terminology.

In the electric formulation where there is no magnetic potential, there is no magnetic Gauss law constraint  and one cannot complete the surface integral  (\ref{Eq:MagneticCharges1}) by a bulk term that would make it a well-defined generator, generating a symmetry transformation. Consequently, the surface integral (\ref{Eq:MagneticCharges1}) cannot be viewed as the Noether charge of a symmetry.  This is exactly as in the scalar case \cite{Fuentealba:2024lll,Henneaux:2018mgn,Campiglia:2017dpg,Campiglia:2018see,Francia:2018jtb}.

In hyperbolic coodinates, the charges read
\begin{equation}
Q^m_{\varepsilon}=\oint\epsilon^{AB}\left(\varepsilon\partial_{s}F_{AB}^{(0)}-\partial_{s}\varepsilon F_{AB}^{(0)}-\frac{2s}{1-s^{2}}\varepsilon F_{AB}^{(0)}\right)\sqrt{\xbar\gamma}d^{2}x\,,
\end{equation}
and is conserved provided
\begin{equation}
(1-s^{2})^{2}\partial_{s}^{2}\varepsilon+2s(1-s^{2})\partial_{s}\varepsilon-(1-s^{2})\xbar D_{B}\xbar D^{B}\varepsilon+2\left(1+s^{2}\right)\varepsilon=0\,. \label{Eq:MagneticParameter}
\end{equation}
The connection between the expressions on flat and hyperbolic slices  is obtained as in the electric case by considering $s=0=t$.   The relevant formulas are
\begin{equation}
\partial_{s}F_{AB}\Big|_{s=0}=r\partial_{t}F_{AB}=2\partial_{[A}\xbar\pi_{B]}+\mathcal{O}\left(r^{-1}\right)\,
\end{equation}
and
\begin{equation}
\varepsilon\Big|_{s=0}=\mu\,,\qquad\partial_{s}\varepsilon\Big|_{s=0}=\epsilon\,.
\end{equation}

In order to solve the equation (\ref{Eq:MagneticParameter}), we expand the parameter $\varepsilon$ in spherical harmonics, inserting for convenience a power of $(1-s^2)$,
\begin{equation}
\varepsilon=(1-s^{2})\sum_{l,m}E_{lm}(s)Y_{lm}\,.
\end{equation}
The equation then becomes
\begin{equation}
(1-s^{2})\partial_{s}^{2}E_{lm}-2s\partial_{s}E_{lm}+l(l+1)E_{lm}=0\,,
\end{equation}
which is again the Legendre differential equation. The solution is therefore
given by
\begin{equation}
\varepsilon=\varepsilon_{P}+\varepsilon_{Q}\,,
\end{equation}
where
\begin{align}
\varepsilon_{P} & =(1-s^{2})\sum_{l,m}E_{lm}^{P}P_{l}(s)Y_{lm}\,,\\
\varepsilon_{Q} & =(1-s^{2})\sum_{l,m}E_{lm}^{Q}Q_{l}(s)Y_{lm}\,.
\end{align}

Decomposing similarly the angular components of the field strength in terms
of their $P$ and $Q$ branches
\begin{equation}
F_{AB}^{(0)}=F_{AB}^{(0)P}+F_{AB}^{(0)Q}\,,
\end{equation}
with
\begin{eqnarray}
F_{AB}^{(0)P}&&=g\epsilon_{AB}\sqrt{\xbar\gamma}+\sum_{l>0,m}\alpha_{lm}^{P(0)}P_{l}(s)\Theta_{AB,lm}\,, \\
 F_{AB}^{(0)Q}&&=\sum_{l>0,m}\alpha_{lm}^{Q(0)}Q_{l}(s)\Theta_{AB,lm}\,,
\end{eqnarray}
the charge becomes
\begin{equation}
Q^m_{\varepsilon}=Q^m_{PP}+Q^m_{PQ}+Q^m_{QP}+Q^m_{QQ}\,,
\end{equation}
where
\begin{align}
Q^m_{PP} & =\oint\epsilon^{AB}\left(\varepsilon_{P}\partial_{s}F_{AB}^{(0)P}-\partial_{s}\varepsilon_{P}F_{AB}^{(0)P}-\frac{2s}{1-s^{2}}\varepsilon_{P}F_{AB}^{(0)P}\right)\sqrt{\xbar\gamma}d^{2}x\,,\\
Q^m_{PQ} & =\oint\epsilon^{AB}\left(\varepsilon_{P}\partial_{s}F_{AB}^{(0)Q}-\partial_{s}\varepsilon_{P}F_{AB}^{(0)Q}-\frac{2s}{1-s^{2}}\varepsilon_{P}F_{AB}^{(0)Q}\right)\sqrt{\xbar\gamma}d^{2}x\,,\\
Q^m_{QP} & =\oint\epsilon^{AB}\left(\varepsilon_{Q}\partial_{s}F_{AB}^{(0)P}-\partial_{s}\varepsilon_{Q}F_{AB}^{(0)P}-\frac{2s}{1-s^{2}}\varepsilon_{Q}F_{AB}^{(0)P}\right)\sqrt{\xbar\gamma}d^{2}x\,,\\
Q^m_{QQ} & =\oint\epsilon^{AB}\left(\varepsilon_{Q}\partial_{s}F_{AB}^{(0)Q}-\partial_{s}\varepsilon_{Q}F_{AB}^{(0)Q}-\frac{2s}{1-s^{2}}\varepsilon_{Q}F_{AB}^{(0)Q}\right)\sqrt{\xbar\gamma}d^{2}x\,.
\end{align}

After replacing the functions in the charges, we find that
\begin{equation}
Q^m_{PP}=Q^m_{QQ}=0\,.
\end{equation}
For the standard parity conditions twisted by a gauge transformation, the nonvanishing charge is $Q^m_{QP}$, which
in the $s\rightarrow1$ limit is given by
\begin{equation}
\lim_{s\rightarrow1}Q^m_{QP}=-\oint\epsilon^{AB}\left(\sum_{l,m}E_{lm}^{Q}Y_{lm}\right)\left(g\epsilon_{AB}\sqrt{\xbar\gamma}+\sum_{l>0,m}\alpha_{lm}^{P(0)}\Theta_{AB,lm}\right)\sqrt{\xbar\gamma}d^{2}x\,,
\end{equation}
while for the non-standard twisted parity conditions, as well as for the inverted parity conditions, the nonvanishing charge is $Q^m_{PQ}$, which
in this limit reads
\begin{equation}
\lim_{s\rightarrow1}Q^m_{PQ}=\oint\epsilon^{AB}\left(\sum_{l,m}E_{lm}^{P}Y_{lm}\right)\left(\sum_{l>0,m}\alpha_{lm}^{Q(0)}\Theta_{AB,lm}\right)\sqrt{\xbar\gamma}d^{2}x\,.
\end{equation}

As for the electric charges, there is no problem in taking the null infinity limit $s \rightarrow 1$, even with the non-standard parity conditions for which the magnetic field develop logarithmic terms.

\section{Behaviour near null infinity of $A_\mu$ ($D=4$)}
\label{Sec:VectorANullInf}

\subsection{The vector potential}

We now expand the vector potential near future null infinity, by going to retarded null coordinates.  This is a mechanical process in which one re-expresses the vector potential in $(u, r, x^A)$ coordinates and expand around $s=1$ all functions, including the ultraspherical $\tilde P$'s and $\tilde Q$'s, which bring logs in.  Of course, we   expect the expansion to have in general only limited validity in the vicinity of the past of future null infinity (just as the expansions near spatial infinity should be understood in the sense of \cite{Chrusciel:1993hx}).  In particular the solutions are not expected to be analytic in $u$.  

Retarded null coordinates are related to hyperbolic coordinates as
\begin{equation}
    u = \eta \frac {s-1}{\sqrt{1-s^2}} \, , \quad r = \eta
    \frac{1}{\sqrt{1-s^2}} \quad \Leftrightarrow \quad \eta=\sqrt{-u(2r+u)}\,,\quad s=1+\frac{u}{r}\,,
\end{equation}
($u<0$), which implies
\be
1-s^{2}=-\frac{u(2r+u)}{r^{2}} \, ,
\ee
and 
\be 
A_r = -\frac{u}{\sqrt{-u(2r+u)}}A_{\eta} -\frac{u}{r^{2}}A_{s} \, , \quad A_u = -\frac{\left(u+r\right)}{\sqrt{-u(2r+u)}}A_{\eta}+\frac{1}{r}A_{s}\,.
\ee

We are mostly interested in the behaviour of $A_{A}^{L}$ which contains the information on the Goldstone boson.  We will furthermore consider explicitly only 
the standard parity conditions twisted by a gauge transformation and the non-standard twisted parity conditions, for which $A_\eta^{(0)}$ is given by (\ref{Eq:ARta007}) (in both cases).

After somewhat tedious but conceptually straigtforward computations,  one finds that the null expansions of the potentials read (using $P_l(1) = 1$):
\begin{itemize}
\item Radial component:
\begin{equation}
A_{r}=\frac{1}{r}\Theta_{00}^{P(0)}+\frac{\log r}{r^{2}}A_{r}^{\log}(u,x^{A})+\frac{1}{r^{2}}A_{r}^{(2)}(u,x^{A})+o\left(r^{-2}\right)\,,
\end{equation}
where
\begin{align}
A_{r}^{\log}(u,x^{A}) & =\frac{1}{2}\sum_{l,m}\Theta_{lm}^{Q(1)}Y_{lm}-\frac{u}{2}\sum_{l,m}K_{lm}^{Q}Y_{lm}\,,\\
A_{r}^{(2)}(u,x^{A}) & =\sum_{l>0,m}(-u)\Theta_{lm}^{P(0)}\partial_{s}Y_{lm}+\sum_{l,m}(-u)K_{lm}^{P} Y_{lm}\nonumber \\
 & \quad+\sum_{l>0,m}(-u)\left[K_{lm}^{Q}\left(\frac{1}{2}\left(-\log(-u)+\log2\right)+R_{l}^{(\frac{1}{2})}(1)\right)+\kappa_{lm}(1)\right]Y_{lm}\nonumber \\
 & \quad+\sum_{k=2}\frac{(-2u)^{-k+1}}{k}\left(\sum_{l<k-1,m}\Theta_{lm}^{P(k)}\tilde{P}_{l-k+1}^{(k-\frac{1}{2})}(1)Y_{lm}+\sum_{l,m}\Theta_{lm}^{Q(k)}\tilde{Q}_{l-k+1}^{(k-\frac{1}{2})}(1)Y_{lm}\right)\nonumber \\
 & \quad+\sum_{k=1}\frac{(-2u)^{-k+1}}{2k}\left(\sum_{l<k,m}\Xi_{lm}^{P(k)}\tilde{P}_{l-k}^{(k+\frac{1}{2})}(1)Y_{lm}+\sum_{l,m}\Xi_{lm}^{P(k)}\tilde{Q}_{l-k}^{(k+\frac{1}{2})}(1)Y_{lm}\right)\nonumber \\
 & \quad+\sum_{l,m}\Theta_{lm}^{P(1)}Y_{lm}+\sum_{l,m}\Theta_{lm}^{Q(k)}\left(\frac{1}{2}\left(-\log(-u)+\log2\right)+R_{l}^{(\frac{1}{2})}(1)\right)Y_{lm}\,.
\end{align}
\item $u$-component :
\begin{equation}
A_{u}=\frac{\log r}{r}A_{u}^{\log}(x^{A})+\frac{1}{r}A_{u}^{(1)}(u,x^{A})+o\left(r^{-1}\right)\,,
\end{equation}
where
\begin{align}
A_{u}^{\log}(x^{A}) & =\frac{1}{2}\sum_{l,m}K_{lm}^{Q}Y_{lm}\,,\\
A_{u}^{(1)}(u,x^{A}) & =\Theta_{00}^{P(0)}-\sum_{l>0,m}\Theta_{lm}^{P(0)}\partial_{s}P_{l}(1)Y_{lm}+\sum_{l,m}K_{lm}^{P}Y_{lm}\nonumber \\
 & \quad+\sum_{l>0,m}\left[K_{lm}^{Q}\left(\frac{1}{2}\left(-\log(-u)+\log2\right)+R_{l}^{(\frac{1}{2})}(1)\right)+\kappa_{lm}(1)\right]Y_{lm}\nonumber \\
 & \quad-\sum_{k=2}\frac{2(k+1)}{k}(-2u)^{-k}\left(\sum_{l<k-1,m}\tilde{P}_{l-k+1}^{(k-\frac{1}{2})}(1)Y_{lm}+\sum_{l,m}\Theta_{lm}^{Q(k)}\tilde{Q}_{l-k+1}^{(k-\frac{1}{2})}(1)Y_{lm}\right)\nonumber \\
 & \quad+\sum_{k=1}\frac{(-2u)^{-k}}{k}\left(\sum_{l<k,m}\Xi_{lm}^{P(k)}\tilde{P}_{l-k}^{(k+\frac{1}{2})}(1)Y_{lm}+\sum_{l,m}\Xi_{lm}^{Q(k)}\tilde{Q}_{l-k}^{(k+\frac{1}{2})}(1)Y_{lm}\right)\,.
\end{align}
\item Angular component (transverse):
\begin{equation}
A_{A}^{T}=\log rA_{A}^{\log T}(x^{A})+A_{A}^{(0)T}(u,x^{A})+o\left(1\right)\,,
\end{equation}
where
\begin{align}
A_{A}^{\log T}(x^{A}) & =\frac{1}{2}\sum_{l>0,m}T_{lm}^{Q}P_{l}(1)\Psi_{A,lm}\,,\\
A_{A}^{(0)T}(u,x^{A}) & =\sum_{l>0,m}\left[T_{lm}^{P}+\frac{1}{2}\left(-\log(-u)+\log2\right)+R_{l}^{(\frac{1}{2})}(1)\right]\Psi_{A,lm}\nonumber \\
 & \quad+\sum_{k=1}(-2u)^{-k}\left(\sum_{l<k,m}\alpha_{lm}^{P(k)}\tilde{P}_{l-k}^{(k+\frac{1}{2})}(1)\Psi_{A,lm}+\sum_{l>0,m}\alpha_{lm}^{Q(k)}\tilde{Q}_{l-k}^{(k+\frac{1}{2})}(1)\Psi_{A,lm}\right)\,.
\end{align}
\item Angular component (longitudinal):
\begin{equation}
A_{A}^{L}=A_{A}^{(0)L}(u,x^{A})+o(1)\,,
\end{equation}
where
\begin{align}
A_{A}^{(0)L}(u,x^{A}) & =\sum_{l>0,m}\frac{1}{l(l+1)}\left[-K_{lm}^{Q}+2\Theta_{lm}^{P(0)}\partial_{s}P_{l}(1)\right]\Phi_{A,lm}\nonumber \\
 & \quad-\sum_{k=2}\sum_{l<k-1,m}\frac{(-2u)^{-k}}{k}\Theta_{lm}^{P(k)}\tilde{P}_{l-k+1}^{(k-\frac{1}{2})}(1)\Phi_{A,lm}\nonumber \\
 & \quad-\sum_{k=1}\frac{(-2u)^{-k}}{l(l+1)}\left(\sum_{l<k,m}\Xi_{lm}^{P(k)}\tilde{P}_{l-k}^{(k+\frac{1}{2})}(1)\Phi_{A,lm}+\sum_{l,m}\Xi_{lm}^{Q(k)}\tilde{Q}_{l-k}^{(k+\frac{1}{2})}(1)\Phi_{A,lm}\right)\,.
\end{align}
\end{itemize}

The first striking feature of these expansions near null infinity is the appearance of polylogarithmic terms.
The logarithms are brought in by the $Q$-branch of the solutions and appear at null infinity even though there is no logarithm in the initial data.   

A second striking property of the vector potential that we just derived at null infinity is that, except for the angular component $A_A^L$, it does not obey the boundary conditions usually assumed there \cite{Strominger:2017zoo}, namely,
$
A_r = \mathcal O\left(\frac{1}{r^2}\right)$, $ A_u = \mathcal O\left(\frac{1}{r}\right)$, $A_A = \mathcal O\left(1\right)$.
One gets instead $A_r = \mathcal O\left(\frac{1}{r}\right)$ and $ A_u = \mathcal O\left(\frac{\log r}{r}\right)$. 

However, the terms by which $A_\mu$ fail to fulfill these conditions can be removed by a gauge transformation, which, as we shall argue in Section {\bf \ref{Sec:LogGaugeTransf}}, is proper and does not have any physical impact.  Indeed, if we perform the logarithmic gauge transformation with gauge parameter
\begin{equation}
\varepsilon=- \log r \, \Theta_{00}^{P(0)} +\frac{\log r}{r}\varepsilon^{\log}(u,x^{A})\, , \quad \varepsilon^{\log}(u,x^{A}) = \frac{1}{2}\sum_{l,m}\Theta_{lm}^{Q(1)}Y_{lm}-\frac{u}{2}\sum_{l,m}K_{lm}^{Q}Y_{lm}\,,\label{eq:Corrective}
\end{equation}
we get
\be
A_{r}+\partial_{r}\varepsilon  =\frac{1}{r^{2}}\left(A_{r}^{(2)}+\varepsilon^{\log}\right)+o\left(r^{-2}\right)\,.
\ee
and
\be
A_{u}+\partial_{u}\varepsilon = \frac{1}{r} A_{u}^{(1)}(u,x^{A})\,.
\ee
This gauge transformation does not change the leading behaviour of $A_A^L$, which is still
\begin{equation}
A_{A}^{L}+\partial_{A}\varepsilon=\xbar A_{A}^{L}+o(1)\,.
\end{equation}
Explicitly,
\be
\xbar A_{A}^{L} = \partial_A (\Gamma  + \Phi) \, , \qquad \Gamma + \Phi =   \xbar \Upsilon+ \mathcal O\left(\frac{1}{u}\right)\,,
\ee
with
\begin{eqnarray}
&&\xbar \Upsilon = \lim_{u \rightarrow - \infty} (\Gamma + \Phi) =   \xbar \Gamma + \xbar \Phi\,, \\
&& \xbar \Gamma =  \sum_{l>0,m}\frac{2}{l(l+1)}\Theta_{lm}^{P(0)}\partial_{s}P_{l}(1) Y_{lm}\,,\\
&&  \xbar \Phi = \sum_{l> 0,m}C_{lm}^{Q}Y_{lm}  \label{Eq:DefPhiNull}
\end{eqnarray}
(with $C_{lm}^{Q}$ defined in terms of $K^{Q}_{lm}$ in (\ref{Eq:DefClm})).

Concerning $A^T_A$, there is an unremovable logarithm when $T_{lm}^{Q} \equiv \alpha^{Q(0)}_{lm} \not=0$.   This term cannot be removed because it is present in the field strengths (see Appendix {\bf \ref{App:FNullInf}}).  It is then even the leading contribution.

\subsubsection*{Standard parity conditions twisted by a gauge transformation}

We now impose the parity conditions on the leading orders of the initial data and consider first the standard parity conditions twisted by a gauge transformation.
In that case, one must take $T^Q_{lm} \equiv \alpha^{Q(0)}_{lm}= 0$ in order for $\xbar A^T = A_A^{(0)T}$ to be even under the antipodal map and $\xbar \pi^{A,T}=  F_{sA}^T \sin \theta$ to be odd.

This implies that the leading logarithm in $A_A^{(0)T}$ at null infinity is absent.  The potential at null infinity then fulfills all the boundary conditions usually assumed there \cite{Strominger:2017zoo}, including those for $A^T_A$,
\be
A_r = \mathcal O\left(\frac{1}{r^2}\right) \, , \qquad A_u = \mathcal O\left(\frac{1}{r}\right) \, , \qquad A_A = \mathcal O\left(1\right) \, . \label{Eq:AAtNInfty}
\ee

Assuming the absence of leading logarithmic terms at null infinity is thus equivalent to adopting the ``standard" parity conditions on the initial data.

\subsubsection*{Non-standard twisted parity conditions}
In contrast, the non-standard twisted parity conditions keep the leading logarithmic term in $A^T_A$, which therefore fails to fulfill the boundary conditions usually assumed at null infinity.   We stress again that it is quite striking that a seemingly innocent change of parity conditions at spatial infinity on the coefficients of the same leading order of the fields leads to radically different behaviours at null infinity.

The inverted parity conditions are even worse in the sense that they lead to dominant unremovable logarithms in the other components as well.

\subsection{Matching with charges at null infinity}

In the case of the standard parity conditions twisted by a gauge transformation,  the charges of electric type (\ref{Eq:ECharges1}) or (\ref{Eq:ECharges2}) can be easily rewritten as surface integrals at null infinity involving $F_{ur}$ in the limit $u \rightarrow - \infty$ (or of $F_{vr}$ in the limit $v \rightarrow  \infty$),
\be
Q^e_\lambda = \lim_{u \rightarrow - \infty} \oint \xbar \lambda (x^A) \xbar F_{ur} \sqrt{\xbar \gamma}
\ee
where $\xbar \lambda$ is the gauge parameter at null infinity (see (\ref{Eq:Fur4D2}) and recall the connection (\ref{Eq:ThetaXi}) between $\Theta_{lm}^{P(0)}$ and $ \Xi_{lm}^{P(0)}$).  One can view this expression as the flux at null infinity of the electric field, which indeed decays as $1/r^2$ there, weighted by the angle-dependent function $\xbar \lambda$.  
A direct symplectic interpretation of the charges as generators of angle-dependent gauge transformations has been provided in \cite{He:2014cra,Kapec:2015ena,Strominger:2017zoo} following null infinity techniques developed in \cite{Ashtekar:1987tt}.  

As also derived in \cite{He:2014cra,Kapec:2015ena,Strominger:2017zoo}, the null infinity approach needs the antipodal matching condition
$
\lim_{v \rightarrow \infty}\xbar F_{vr}(-x^A) = \lim_{u \rightarrow -\infty}\xbar F_{ur}(x^A)$,
which is precisely the relation (\ref{Eq:Match0}) obtained in Appendix {\bf \ref{App:FNullInf}} by integration to null infinity of the initial data.  
Null infinity and spatial infinity results therefore perfectly match.

The same features also hold for the angle-dependent charges of magnetic type, but there, in order to have a symplectic interpretation, one needs to introduce dual potentials \cite{Strominger:2015bla,Henneaux:2020nxi}.

The situation is more complicated with the other parity conditions.  To illustrate the point, let us consider the inverted parity conditions, where both $F_{ur}$ and $F_{AB}$ develop leading logarithms at null infinity.  Nothing special occurs at spatial infinity, but the presence of leading logarithms at null infinity implies major changes there.  First, the angle-dependent charges are given by the coefficient of the $\frac{\log r}{r^2}$ term in $F_{ur}$, not by the coefficient of the $\frac{1}{r^2}$ term.  This implies in particular that it is not given by a standard flux at null infinity (while it still  is at spatial infinity).  Using the standard flux expression at null infinity and concluding not only that the charges are ill-defined on the grounds that the field does not decay sufficiently rapidly but also that the boundary conditions should accordingly be rejected would thus be incorrect since this is not the appropriate expression for the charges.   We expect that it should be possible to derive the correct expression for the charges at null infinity in the presence of leading logarithms directly by null infinity symplectic techniques from their action on the fields, but there are subtleties (in addition to the usual ones) due to the presence of logarithms in the symplectic structure itself.

Another change is that the coefficients of the leading terms in the asymptotic expansion near null infinity obey 
matching conditions with the opposite sign, as expressed by (\ref{Eq:OppositeMatching}) - and as in the scalar field case \cite{Fuentealba:2024lll}.

\subsection{Matching conditions for the Goldstone field}
\label{SubSec:Goldstone}

In order to complete the comparison between past null infinity, spatial infinity and future null infinity, it remains to derive the matching conditions for the Goldstone field.

The Goldstone field is the physical part $\Phi$ of the longitudinal component $A_A^L$ of the vector potential and is of order $\mathcal O(1)$. Since $\Phi$, which involves $\partial_s Q_l(s) Y_{lm}(x^A)$, is even under the hyperboloid antipodal map, we get the matching relation
\be
\lim_{u \rightarrow - \infty} \Phi (u, x^A) = \lim_{v \rightarrow  \infty} \Phi (v, -x^A)\,,
\ee
in complete agreement with \cite{He:2014cra,Kapec:2015ena,Strominger:2017zoo}.  

This is for both the standard parity conditions twisted by a gauge transformation and the non-standard twisted parity conditions.  For the inverted parity conditions, the Goldstone field develops a logarithm and the coefficient of the leading logarithm is odd.

It is interesting to point out that $\Phi$ and the gauge parameter $\lambda_Q$ obey exactly the same equations of motion and are given by identical expressions (compare (\ref{Eq:FormPhi}) with (\ref{Eq:lambdaQ})).  This is of course as it should be and provides a useful way to determine the Goldstone field without having to solve for the vector potential.   One imposes strict parity conditions for which $\Phi=0$ and then turn on $\Phi$ by making a gauge transformation.  The Goldstone field is the parameter of that gauge transformation and obeys the wave equation, which preserves the Lorenz gauge condition (to leading order).

\section{Angle-dependent logarithmic $u(1)$ gauge transformations ($D=4$)}
\label{Sec:LogGaugeTransf}

\subsection{Angle-dependent logarithmic $u(1)$ gauge transformations and charges}
We turn now to the boundary conditions (\ref{Eq:ParityA0007Z})-(\ref{Eq:ParityA0037Z}) that allow for angle-dependent logarithmic $u(1)$ gauge transformations.  By construction, these transformations simply amount to shifting $\Phi_{\log}$ and $\Psi_{\log}$,
\be
\Phi_{\log} \rightarrow \Phi_{\log} + \epsilon_{\log} \, , \qquad \Psi_{\log} \rightarrow \Psi_{\log} + \mu_{\log}\, ,
\ee
where $\epsilon_{\log} $ is odd while $\mu_{\log} $ is even and has no zero mode.   Using symplectic methods at spatial infinity, the corresponding charge-generators have been shown in \cite{Fuentealba:2023rvf} to be the next terms $\xbar \Phi$ and $\xbar \Psi$ in the expansion of the potential (which match with the Goldstone field of \cite{He:2014cra,Kapec:2015ena,Strominger:2017zoo} at null infinity).

Rewritten in hyperbolic coordinates following the rules of Section {\bf \ref{Sec:ConservedCharges}}, these charges associated with the angle-dependent logarithmic gauge transformations $\delta A_\mu = \partial_\mu \lambda$ with
\be
\lambda = \log \eta \lambda_{\textrm{log}}  + \mathcal O\left(1\right) \label{Eq:LogarithmicGauge}
\ee
read, on spacelike hyperplanes (e.g., $s=0$)
\be
Q_{\lambda_{\log}}\Big|_{s=0}  =\oint d^{2}x\sqrt{\xbar g}\left(\epsilon_{\log}\xbar\Psi-\mu_{\log}\xbar\Phi\right)\,,
\ee
or 
\be
Q_{\lambda_{\log}}\Big|_{s=0}  =\oint d^{2}x\sqrt{\xbar g}\left(\lambda_{\log}\partial_{s}\Phi-\partial_{s}\lambda_{\log} \, \Phi\right)_{s=0}\,,
\ee
using the relations
\begin{equation}
\Phi\Big|_{s=0}=\xbar\Phi\,,\qquad\partial_{s}\Phi\Big|_{s=0}=\xbar\Psi\,,
\end{equation}
and
\be
\lambda_{\log}\Big|_{s=0}  =\epsilon_{\log}\,,\qquad\partial_{s}\lambda_{\log}\Big|_{s=0}=\mu_{\log}\,.
\ee

We see that only the odd part of $\epsilon_{\log}$ and the even part of $\mu_{\log}$ contribute to the charge since $\xbar\Psi$ is odd and $\xbar\Phi$ is even.  We can therefore argue that the even part of $\epsilon_{\log}$ and the odd part of $\mu_{\log}$ define proper gauge symmetries that do not change the physical state of the system\footnote{Things are slightly more subtle because logarithmic gauge transformations with even $\epsilon_{\log}$ and odd $\mu_{\log}$, while having zero charge, do change the asymptotic conditions by shifting $\Phi_{\log}$ and $\Psi_{\log}$ as $\Phi_{\log} \rightarrow  \Phi_{\log} + \epsilon_{\log} $,  $\Psi_{\log} \rightarrow  \Psi_{\log} + \mu_{\log} $, which is not permissible in the formulation of \cite{Fuentealba:2023rvf} where these variables obey strict parity conditions.  Nevertherless, one can allow $\Phi_{\log}$ and $\Psi_{\log}$ not to have any definite parity provided one systematically substracts the even part of $\Phi_{\log}$ and the odd part of $\Psi_{\log}$ in the canonical action of \cite{Fuentealba:2023rvf}.  These opposite parity components are then pure gauge that can be shifted arbitrarily because they do not appear in the action.  The statement that the even part of $\epsilon_{\log}$ and the odd part of $\mu_{\log}$ define proper gauge symmetries is then fully correct.}.

The charges associated with the $\mathcal O(1)$ angle-dependent gauge transformations $\delta A_\mu = \partial_\mu \xbar \lambda $, $\xbar \lambda = \mathcal O (1)$, get an extra contribution from $\Psi_{\log}$ \cite{Fuentealba:2023rvf} and are given by
\begin{align}
Q_{\xbar\lambda}\Big|_{s=0} & =\oint d^{2}x\left[\xbar\epsilon\left(\xbar\pi^{r}+\sqrt{\xbar \gamma}\,\Psi_{\log}\right)-\sqrt{\xbar \gamma}\,\xbar\mu\,\xbar A_{r}\right]\,,
\end{align}
with 
\be
\xbar\lambda\Big|_{s=0} =\xbar\epsilon\,,\qquad\partial_{s}\xbar\lambda\Big|_{s=0}=\xbar\mu\,.
\ee
This expression can be rewritten as 
\begin{align}
Q_{\xbar\lambda}\Big|_{s=0} & =\oint d^{2}x\sqrt{\xbar \gamma}\,\left(\xbar\lambda\partial_{s}A_{\eta}^{(0)}-\partial_{s}\xbar\lambda A^{(0)}_{\eta}\right)\,,
\end{align}
using the Hamiltonian equations of the motion 
\begin{equation}
\partial_{t}\xbar A_{r}=0\,,\qquad\partial_{t}A_{r}^{(2)}=\frac{1}{\sqrt{\xbar \gamma}}\left(\xbar\pi^{r}+\sqrt{\xbar \gamma}\,\Psi_{\log}\right)\,,
\end{equation}
with
\begin{align}
A_{\eta}\Big|_{s=0} & =A_{r}=\frac{1}{r}\xbar A_{r}+\frac{1}{r^{2}}A_{r}^{(2)}+\mathcal{O}\left(r^{-3}\right)\,,\\
\partial_{s}A_{\eta}\Big|_{s=0} & =r\partial_{t}A_{r}=\partial_{t}\xbar A_{r}+\frac{1}{r}\partial_{t}A_{r}^{(2)}+\mathcal{O}\left(r^{-2}\right)\nonumber \\
 & =\frac{1}{r\sqrt{\xbar \gamma}}\left(\xbar\pi^{r}+\sqrt{\xbar \gamma}\,\Psi_{\log}\right)+\mathcal{O}\left(r^{-2}\right)\,.
\end{align}
($A_{\eta} = \frac{A_{\eta}^{(0)}}{r} + \cdots$).

The total charge generating the complete gauge transformation
\begin{equation}
\lambda=\log\eta\,\lambda_{\log}+\xbar\lambda+\mathcal{O}\left(\eta^{-1}\right)\,,
\end{equation}
is obtained by taking the sum and is given by
\begin{equation}
Q_{\lambda}\Big|_{s=0}=\oint d^{2}x\sqrt{\xbar \gamma}\left[\xbar\lambda\partial_{s} A_{\eta}^{(0)}-\partial_{s}\xbar\lambda \, A_{\eta}^{(0)}+\lambda_{\log}\partial_{s}\Phi-\partial_{s}\lambda_{\log}\Phi\right] \label{Eq:LogChargeOns=0}
\end{equation}
(modulo weakly vanishing bulk terms).

\subsection{Integration of the gauge parameter in hyperbolic coordinates}

To determine the form of the gauge potential at null infinity, we use the observation made at the end of Subsection {\bf \ref{SubSec:Goldstone}}.  That is, we perform an angle-dependent logarithmic $u(1)$ gauge transformation on the known solutions determined above which did not include any such logarithmic gauge transformation term.   Thus we add to the previous solutions the gauge variation due to (\ref{Eq:LogarithmicGauge}) written at null infinity.  

To find that gauge variation, we need to integrate the wave equation $\Box \lambda = 0$ for the gauge parameter $\lambda$ all the way to null infinity.  This is in order to preserve asymptotically the Lorenz gauge condition, known to hold in Minkowskian coordinates with our asymptotic conditions. This yields the equations
\begin{align}
\mathcal{D}_{a}\mathcal{D}^{a}\lambda_{\log} & =0\,, \label{eq:Symm01}\\
\mathcal{D}_{a}\mathcal{D}^{a}\xbar\lambda+2\lambda_{\log} & =0\,.\label{eq:Symm1}
\end{align}
These equations show in particular that in order to preserve the Lorenz gauge,  a logarithmic gauge transformation must necessarily be accompanied by a non-vanishing $\mathcal O(1)$ gauge transformation.

The wave equation for $\lambda$ guarantees that the charges
\begin{equation}
Q_{\lambda}=\oint d^{2}x\sqrt{\xbar \gamma}\left[\xbar\lambda\partial_{s} A_{\eta}^{(0)}-\partial_{s}\xbar\lambda \, A_{\eta}^{(0)}+\lambda_{\log}\partial_{s}\Phi-\partial_{s}\lambda_{\log}\Phi\right]\,,
\end{equation}
which reduce to (\ref{Eq:LogChargeOns=0}) on $s=0$, are conserved  since the equations for $A_{\eta}^{(0)}$ and $\Phi$ are ($\mathcal{D}_{a}A^{a(0)}= \mathcal{D}_{a}\mathcal{D}^{a}\Phi$),
\begin{align}
\mathcal{D}_{a}\mathcal{D}^{a}\Phi+2 A_{\eta}^{(0)} & =0\,,\\
\mathcal{D}_{a}\mathcal{D}^{a}\Phi_{\log} & =0\,.
\end{align}

Because the equations (\ref{eq:Symm01}) and (\ref{eq:Symm1}) lead again to ultraspherical differential equations which are by now standard, we just report the form of their solutions.
\begin{itemize}
\item The solution for $\lambda_{\log}$ is generically (i.e., without parity conditions) given by
$$
\lambda_{\log}=(1-s^{2})\sum_{l,m}\left[\omega_{lm}^{P}\partial_{s}P_{l}(s)+\omega_{lm}^{Q}\partial_{s}Q_{l}(s)\right]Y_{lm}\, ,
$$
(with the $l=0$ term of the $P$-branch understood to be $\omega_{00}^{P} s Y_{00}$)
but parity conditions rule out the Q-branch, i.e., $\omega_{lm}^{Q}=0$.
Thus, $\lambda_{\log}$ contains only the $P$-branch:
$$
\lambda_{\log}=\omega_{00}^{P} s Y_{00} + (1-s^{2})\sum_{l>0,m}\omega_{lm}^{P}\partial_{s}P_{l}(s)Y_{lm}\,.
$$
Furthermore, since $\partial_s \lambda_{\log} \vert_{s=0}$ coincides with $\mu_{\log}$, which has no zero mode, one must set $\omega_{00}^{P} = 0$, so that
\begin{equation}
\lambda_{\log}= (1-s^{2})\sum_{l>0,m}\omega_{lm}^{P}\partial_{s}P_{l}(s)Y_{lm}\,.
\end{equation}

\item The solution for $\xbar\lambda$ generically reads
\begin{equation}
\xbar\lambda=\sum_{l,m}\left\{ (1-s^{2})\left[\lambda_{lm}^{P}\partial_{s}P_{l}(s)+\lambda_{lm}^{Q}\partial_{s}Q_{l}(s)\right]+\tau_{lm}(s)\right\} Y_{lm}\,,
\end{equation}
where $\tau_{lm}(s)$ is a polynomial of degree $l-1$. The odd
part of the homogeneous solution of $\xbar\lambda\Big|_{s=0}$, which generates
proper gauge transformations, can be put it to zero. This rules
out the P-branch ($\lambda_{lm}^{P}=0$). The polynomial $\tau_{lm}$ is a particular solution of the inhomogeneous equation (\ref{eq:Symm1}) 
and satisfies therefore the following differential equation
\begin{equation}
(1-s^{2})\partial_{s}^{2}\tau_{lm}+l(l+1)\tau_{lm}-2\omega_{lm}^{P}\partial_{s}P_{l}=0\,.\label{eq:kappa-equation}
\end{equation}
We can take $\tau_{lm}$ to have the parity of $\partial_{s}P_{l}$, i.e., 
\begin{equation}
\tau_{lm}(-s)=-(-1)^{l}\tau_{lm}(s) \, .
\end{equation}
 Hence, the solution for $\xbar\lambda$ reads
\begin{equation}
\xbar\lambda=\sum_{l,m}\left[(1-s^{2})\lambda_{lm}^{Q}\partial_{s}Q_{l}(s)+\tau_{lm}(s)\right]Y_{lm}\,.\label{eq:xbar lambda}
\end{equation}
The first term in this expression 
is even and generates improper gauge transformations, while the second
term is odd and defines a proper gauge transformation with no
contribution to the charges.
\end{itemize}
Putting everything together, we thus have
\begin{align}
\lambda & =\log\eta\left[ (1-s^{2})\sum_{l>0,m}\omega_{lm}^{P}\partial_{s}P_{l}(s)Y_{lm}\right]\nonumber \\ &\quad+\sum_{l,m}\left[(1-s^{2})\lambda_{lm}^{Q}\partial_{s}Q_{l}(s)+\tau_{lm}(s)\right]Y_{lm}
 + o(1)\,.
\end{align}

\subsection{Gauge potential at null infinity}

We now express the gauge parameter at null infinity, which is direct since it is a scalar.  We  need only to express $\eta$ and $s$ in terms of retarded Bondi coordinates\footnote{One can see that the logarithmic term $\log \eta \,  \lambda_{\log}$ has a somewhat pathological behaviour at the critical sets by rewriting it in Friedrich coordinates, in which it reads $\left(\log \rho  - \frac12 \log(1-s^2)\right)(1-s^2)\sum_{l,m}\omega_{lm}^{P}\partial_{s}P_{l}(s)Y_{lm}$. This expression vanishes at null infinity ($s \rightarrow 1$) but blows up if one takes first the limit $\rho \rightarrow \infty$.  This is not a surprise since $\log \eta \,  \lambda_{\log}$ logarithmically blows up (in a fully controlled way \cite{Fuentealba:2023rvf}) at spatial infinity.}. One gets by direct substitution
\begin{equation}
\lambda=\lambda^{(0)}+\frac{\log r}{r}\lambda_{\log}^{(1)}+\mathcal{O}\left(r^{-1}\right)\,,  \label{Eq:LogNullInf}
\end{equation}
where
\begin{align}
\lambda^{(0)} & =\sum_{l,m}\left[\lambda_{lm}^{Q}+\tau_{lm}(1)\right]Y_{lm}\,,\\
\lambda_{\log}^{(1)} & =-\frac{1}{2}u\sum_{l>0,m}l(l+1)\omega_{lm}^{P}Y_{lm}-\frac{1}{2}u\sum_{l,m}l(l+1)\lambda_{lm}^{Q}Y_{lm}+\frac{1}{2}\sum_{l,m}\gamma_{lm}^{Q(0)}Y_{lm}\,.
\end{align}
Here we have explicitly added the $\frac{1}{\eta}$-contribution to $\lambda = \log \eta \, \lambda_{\log} + \xbar \lambda + \frac{\lambda^{(2)}}{\eta}$,  
$$
\frac{\lambda^{(2)}}{\eta} = \frac{(1-s^2)^\frac12}{\eta}\sum_{l,m}\left[\gamma_{lm}^{P(0)}\tilde{P}_{l}^{(\frac{1}{2})}(s)+\gamma_{lm}^{Q(0)}\tilde{Q}_{l}^{(\frac{1}{2})}(s)\right]Y_{lm}\,,
$$
since its $Q$-branch is of order $\mathcal O\left(\frac{\log r}{r}\right)$ at null infinity.

A remarkable feature of the logarithmic $u(1)$ gauge transformations, parametrized by $\omega_{lm}^P$,  is that they become subdominant at null infinity with respect to the standard $\mathcal O(1)$ gauge transformations, even though they are the leading terms at spatial infinity.   This is because they involve the $P$-branch while the $\mathcal O(1)$ terms involve the $Q$-branch.

 In fact, the logarithmic $u(1)$ gauge transformations are exactly of the  same order as the $\varepsilon^{\textrm{log}}$ gauge transformation needed in (\ref{eq:Corrective}) to eliminate the logarithmic terms by which the gauge potential failed to fulfill the standard boundary conditions at null infinity.  Although of the same order, however, they originate from terms with opposite antipodal parities and which give a non-zero contribution to the charges.  So, contrary to the proper gauge parameter $\varepsilon^{\textrm{log}}$ , they define improper gauge transformations and must be kept.   

The boundary conditions at null infinity incorporating the logarithmic $u(1)$ gauge transformations are obtained by adding the effect of the gauge transformation (\ref{Eq:LogNullInf}) at null infinity.  Since the $\mathcal O(1)$ gauge transformation have already been included in the asymptotic form of the potential, we focus only on the (relevant part of the) new logarithmic contribution, which we parametrize as in (\ref{Eq:LogNullInf}) with $\lambda^{(1)}_{\log}  \rightarrow \Phi^{(1)}_{\log} $,
where 
\be
 \Phi^{(1)}_{\log} = u X(x^A) \, .
\ee
One finds explicitly
\be
A_r = - \,\frac{\log r}{r^2} \Phi^{(1)}_{\log} +\mathcal O \left(\frac{1}{r^2}\right) , \quad A_u = \frac{\log r}{r}X + \mathcal O \left(\frac{1}{r}\right)\, \quad A_A = \mathcal O (1)\, .  \label{Eq:LogFormNull}
\ee
The effect of the logarithmic gauge transformations is thus to introduce physically relevant logarithmic terms of the form (\ref{Eq:LogFormNull}) in the asymptotic behaviour of the fields at null infinity.  Even though occurring through gradients, these terms must be kept because they cannot be removed by proper gauge transformations.  One might  view $ \Phi^{(1)}_{\log}$ as a  logarithmic Goldstone field. The field $\Delta$ of the introduction (Eq. (\ref{Eq:DeltaPhi200})) is equal to $ \frac{\log {r}}{r}\Phi^{(1)}_{\log}$.

\subsection{Direct integration of the vector potential in hyperbolic coordinates}

Rather than performing a logarithmic gauge transformation on the electromagnetic potentials derived in Section {\bf \ref{Sec:AinHyp}}, which did not have any logarithmic gauge transformation term in their initial data, one can directly integrate the equations of motion with initial conditions that have already in them the logarithmic gauge transformation terms.

It is of course obvious that the two methods give the same final result, as we now briefly verify.  We thus assume that the asymptotic behavior of the gauge field in hyperbolic coordinates
is given by
\begin{align}
A_{\eta} & =\sum_{k=0}\eta^{-(k+1)}A_{\eta}^{(k)}\,,\\
A_{a} & =\log\eta\,\partial_{a}\Phi_{\log}+\sum_{k=0}\eta^{-k}A_{a}^{(k)}\,,
\end{align}
where $\Phi_{\log}(x^{a})$ is a function on the unit hyperboloid $\mathcal{H}$, such that
\be
 \Phi_{\log}(s=0, x^{A}) = \Phi^{\textrm{Hamiltonian}}_{\log}(x^{A}) \, , \qquad  \partial_s \Phi_{\log}(s=0, x^{A}) = \Psi^{\textrm{Hamiltonian}}_{\log}(x^{A}) \, .
\ee
(the Hamiltonian quantities are the ones that appear in the initial data (\ref{Eq:ParityA0007Z})-(\ref{Eq:ParityA0037Z})).  We absorb the contribution of  $\Phi_{\log}(x^{a})$ to $A_{\eta}^{(0)}$ in a redefinition of  $A_{\eta}^{(0)}$, so that the expansion of  $A_{\eta}$ is unchanged.

The Lorenz gauge $\nabla_{\mu}A^{\mu}=0$ 
reads in hyperbolic coordinates
\begin{equation}
\eta^{-1}\partial_{\eta}\left(\eta^{3}A_{\eta}\right)+\mathcal{D}_{a}A^{a}=0\,,\label{eq:Lorenz-gaugeV2}
\end{equation}
and imposes the following conditions
\begin{align}
\mathcal{D}_{a}\mathcal{D}^{a}\Phi_{\log} & =0\,,\\
\mathcal{D}_{a}A^{(k)a}-(k-2)A_{\eta}^{(k)} & =0\,,\quad\text{with}\quad k\geq 0\,.
\end{align}

The Maxwell equations in this gauge reduce to $\nabla_{\mu}\nabla^{\mu}A^{\nu}=0$,
with
\begin{itemize}
\item $\eta$-component:
\begin{equation}
\mathcal{D}_{a}\mathcal{D}^{a}A_{\eta}^{(k)}+k(k-2)A_{\eta}^{(k)}=0\,,\quad\text{with}\quad k\geq 0\,.
\end{equation}
\item $a$-component:
\begin{align}
\left(\mathcal{D}_{b}\mathcal{D}^{b}-2\right)\left(\partial_{a}\Phi_{\log}\right) & =0\,,\\
\left(\mathcal{D}_{b}\mathcal{D}^{b}+k^{2}-2\right)A_{a}^{(k)}+2\partial_{a}A_{\eta}^{(k)} & =0\,,\quad\text{with}\quad k\geq0\,.
\end{align}
\end{itemize}

The equations for $A_\mu^{(k)}$ are unchanged and have thus the same solutions. The equation for $\Phi_{\log}$ is the same as for $\lambda_{\log}$.  
The solutions for the case $k=0$ and $\Phi_{\log}$ are thus given by (taking into account the parity conditions and the fact that $\Phi_{\log}$ has no zero mode)
\begin{align}
\Phi_{\log} & =(1-s^{2})\sum_{l>0,m}\Phi_{lm}^{P}\partial_{s}P_{l}(s)Y_{lm}\,,\\
A_{\eta}^{(0)} & =\Theta_{00}^{P(0)}sY_{00}+(1-s^{2})\sum_{l>0,m}\Theta_{lm}^{P(0)}\partial_{s}P_{l}(s)Y_{lm}\,,\\
A_{s}^{(0)} & =\sum_{l,m}\left[K_{lm}^{P}P_{l}(s)+K_{lm}^{Q}Q_{l}(s)+\kappa_{lm}(s)\right]Y_{lm}\,,\\
A_{A}^{(0)L} & =-\sum_{l,m}\frac{(1-s^{2})}{l(l+1)}\left[K_{lm}^{P}\partial_{s}P_{l}(s)+K_{lm}^{Q}\partial_{s}Q_{l}(s)\right]\Phi_{A,lm}\nonumber \\
 & \quad-\sum_{l,m}\frac{1}{l(l+1)}\left[(1-s^{2})\partial_{s}\kappa_{lm}(s)-2\Theta_{lm}^{P(0)}\partial_{s}P_{l}(s)\right]\Phi_{A,lm}\,,\\
A_{A}^{(0)T} & =\sum_{l,m}\left(T_{lm}^{P}P_{l}(s)+T_{lm}^{Q}Q_{l}(s)\right)\Psi_{A,lm}\,,
\end{align}
where
\begin{equation}
(1-s^{2})\partial_{s}^{2}\kappa_{lm}-2s\partial_{s}\kappa_{lm}+l(l+1)\kappa_{lm}=2\Theta_{lm}^{P(0)}\partial_{s}P_{l}(s)\,.
\end{equation}

The fact that $\Phi_{\log}$ and $\lambda_{\log}$ are given by identical expressions confirms that one can perform the logarithmic gauge transformation before or after integrating the equations, without changing the final result.  In the present case, this is actually obvious and a bit of a tautology, but in more complicated systems, one approach might be faster than the other.

It is interesting to work out the connection between the field strengths and the vector potential.  The field strengths are insensitive to the logarithmic gauge transformations and are unchanged, but the relation (\ref{Eq:ThetaXi}) between the coefficients $\Xi_{lm}^{P(0)}$ and $\Theta_{lm}^{P(0)}$ is modified because we have absorbed the $\Phi_{\log}$ contribution to $A_\eta$ in a redefinition of $A_\eta^{(0)}$.  To see this, it is sufficient to focus on the component $E_{s}=F_{s\eta}$.  We find
\begin{align}
F_{s\eta} & =\frac{1}{\eta}\partial_{s}\left(A_{\eta}^{(0)}-\Phi_{\log}\right)+\mathcal{O}\left(\eta^{-2}\right)\nonumber \\
 & =\frac{1}{\eta}\left[\left(\Theta_{00}^{P(0)}-\Phi_{00}^{P}\right)Y_{00}-\sum_{l>0,m}l(l+1)\left(\Theta_{lm}^{P(0)}-\Phi_{lm}^{P}\right)P_{l}(s)Y_{lm}\right]+\mathcal{O}\left(\eta^{-2}\right)\,,
\end{align}
Then,
\begin{equation}
\Xi_{00}^{P(0)}=\Theta_{00}^{P(0)}-\Phi_{00}^{P}\quad,\quad\qquad\Xi_{lm}^{P(0)}=-l(l+1)\left(\Theta_{lm}^{P(0)}-\Phi_{lm}^{P}\right)\quad\left(l>0\right)\,.\label{eq:Xi}
\end{equation}

\subsection{Logarithmic gauge transformations at null infinity}

The boundary field $X$ appearing at null infinity can be expressed in terms of $\Phi_{lm}^{P}$ as
\be
X =-\frac{1}{2}\sum_{l>0,m}l(l+1)\Phi_{lm}^{P}Y_{lm}\,.
\ee 

By construction, logarithmic gauge transformations just shift the boundary field $\Phi^{(1)}_{\log}$, 
\be
 \Phi^{(1)}_{\log} \rightarrow  \Phi^{(1)}_{\log}  + \lambda^{(1)}_{\log} \, .
\ee
In particular, they  shift the coefficients $\Phi_{lm}^{P}$ appearing in $X$ as
\be
\Phi_{lm}^{P} \rightarrow \Phi_{lm}^{P}+ \omega_{lm}^{P} \, .
\ee
 
It would be interesting to explore whether there is a logarithmic memory effect of the type of \cite{Bieri:2013hqa,Strominger:2014pwa} related to the new boundary field.  It would also be interesting to study directly at null infinity these more general boundary conditions, along the lines of \cite{Geiller:2022vto}.   

We also note that because the logarithmic Goldstone field originates from $\partial_s P_l$, which is odd under the hyperboloid antipodal map, it obeys odd matching conditions, opposite to those of the standard Goldstone field.

\subsection{Charges at null infinity}
We close this section by expressing the logarithmic $u(1)$ charges in the limit $s \rightarrow \pm 1$ as one goes to null infinity. 

First, we express them in terms of the integration constants $\omega^P_{mn}$ and $\lambda^Q_{lm}$ characterizing the gauge parameters $\lambda_{\log}$ and $\xbar \lambda$, as well as the integration constants $C^Q_{mn}$ and $\Theta^P_{lm}$ characterizing the Goldstone field $\Phi$ and the radial component of the vector potential and its time derivatives (see (\ref{Eq:FormPhi}) and (\ref{Eq:ExpressionForAeta})).  We get
\begin{eqnarray}
Q_{\lambda}&&=\sum_{l>0,m}(-1)^{m}l(l+1)(\omega_{lm}^{P}C_{l,-m}^{Q}-\lambda_{lm}^Q \Theta^P_{l,-m}) \nonumber \\
&& \qquad \qquad \times \left[P_{l}(s)(1-s^{2})\partial_{s}Q_{l}(s)-Q_{l}(s)(1-s^{2})\partial_{s}P_{l}(s)\right] \nonumber \\
&& \qquad \qquad \qquad \qquad + \lambda_{00}^Q \Theta^P_{00} 
\end{eqnarray}
where we have explicitly performed the integral over the angles and used the orthogonality relation of the spherical harmonics,
\begin{equation}
\oint d^{2}x\sqrt{\xbar \gamma}\,Y_{l,m}Y_{l',m'}=(-1)^{m}\delta_{l,l'}\delta_{m,-m'}\,.
\end{equation}

In the limit $s \rightarrow 1$, this charge becomes
\be
Q_{\lambda}\Big|_{s=1}  =\sum_{l>0,m}(-1)^{m}l(l+1)(\omega_{lm}^{P}C_{l,-m}^{Q}-\lambda_{lm}^{Q}\Theta_{l,-m}^{P}) + \lambda_{00}^Q \Theta^P_{00} \,,
\ee
We get the same expression in the limit $s \rightarrow -1$
\be
Q_{\lambda}\Big|_{s=-1}  =\sum_{l>0,m}(-1)^{m}l(l+1)(\omega_{lm}^{P}C_{l,-m}^{Q}-\lambda_{lm}^{Q}\Theta_{l,-m}^{P}) + \lambda_{00}^Q \Theta^P_{00} \,,
\ee
as can be seen by using the identities
\begin{equation}
P_{l}(-1)=(-1)^{l}\,,\qquad \lim _{s\rightarrow -1}(1-s^{2})\partial_{s}Q_{l}(s)=(-1)^{l}\,.
\end{equation}
Thus,
\be
Q_{\lambda}\Big|_{s=1}  =Q_{\lambda}\Big|_{s=-1}\,,\\
\ee
as it should since these charges are conserved.

This analysis confirms that nothing very special occurs to the charges as we go to null infinity.  These remain finite and well-defined.   Furthermore, they give the "initial data" ($u \rightarrow - \infty$) for  the charges at null infinity (which are generically not conserved there due to the fluxes).

It is interesting to point out that one may express the charge associated with standard $\mathcal O(1)$  gauge transformations as a modified flux integral of the field strength $F_{ur}$. One gets
\begin{equation}
Q_{\lambda^{(0)}}=\oint d^{2}x\sqrt{\xbar\gamma}\,\lambda^{(0)}\left(\xbar F_{ur}+2X\right)\,,  \label{Eq:ChargeO(1)NullInf}
\end{equation}
with an extra term proportional to $X$, in addition to the expression $\oint d^{2}x\sqrt{\xbar\gamma}\,\lambda^{(0)}\xbar F_{ur}$ valid prior to the extension of the formalism to include logarithmic gauge transformations.  We recall that $\xbar F_{ur}$ is the coefficient of the $1/r^2$ in the expansion of $F_{ur}$ near null infinity (see (\ref{Eq:FurSimpl}) and (\ref{Eq:Fur4D2})).  Indeed, one gets
\begin{align}
&\oint d^{2}x\sqrt{\xbar\gamma}\,\lambda^{(0)}\left(\xbar F_{ur}+2X\right) \nonumber\\
&\quad  =\oint d^{2}x\sqrt{\xbar\gamma}\,\left(\sum_{l,m}\lambda_{lm}^{Q}Y_{lm}\right)\left(\Xi_{00}^{P(0)}Y_{00}+\sum_{l>0,m}\Xi_{lm}^{P(0)}Y_{lm}+\Phi_{00}^{P}Y_{00}-\sum_{l,m}l(l+1)\Phi_{lm}^{P}Y_{lm}\right)\nonumber \\
 & \quad =\oint d^{2}x\sqrt{\xbar\gamma}\,\left(\sum_{l,m}\lambda_{lm}^{Q}Y_{lm}\right)\left(\Theta_{00}^{P(0)}Y_{00}-\sum_{l>0,m}l(l+1)\Theta_{lm}^{P(0)}Y_{lm}\right)\, , \nonumber 
\end{align}
and thus
\be
Q_{\lambda^{(0)}}= \lambda_{00}^{Q}\Theta_{00}^{P(0)}-\sum_{l>0,m}(-1)^{m}l(l+1)\lambda_{lm}^{Q}\Theta_{l,-m} \, .
\ee

Similarly, the logarithmic charge is given by the surface integral
\be
Q_{\lambda^{(1)}_{\log}} = -\frac{2}{u}\oint d^{2}x\sqrt{\xbar\gamma}\,\lambda_{\log}^{(1)}\xbar\Phi
\ee
at null infinity (where one keeps only the improper gauge part of $\lambda_{\log}^{(1)}$ involving $\omega_{lm}^{P}$), with $\xbar \Phi$ given by (\ref{Eq:DefPhiNull}), since this integral reproduces the desired expression,
\be
Q_{\lambda_{\log}^{(1)}} = \sum_{l>0,m}(-1)^{m}l(l+1)\omega_{lm}^{P}C_{l,-m}^{Q} \, .
\ee

The additional term $2 X$ in (\ref{Eq:ChargeO(1)NullInf}), which is shifted under logarithmic gauge transformations, is necessary to ensure that logarithmic charges and $\mathcal O(1)$ charges,
are canonically conjugate, $\delta_{\lambda^{(1)}_{\log}} Q_{\lambda^{(0)}} \sim \oint d^{2}x\sqrt{\xbar\gamma}\, \lambda^{(1)}_{\log}\lambda^{(0)}$ since $F_{ur}$, which is gauge invariant, fulfills $\delta_{\lambda^{(1)}_{\log}} F_{ur} = 0$ and is thus insufficient by itself to make the charges conjugate.

While the charges at null infinity match the spatial infinity conserved charges in the limit $u \rightarrow - \infty$, they are not conserved along null infinity due to the fluxes.

It would be interesting to derive the expression for the charges at null infinity from first "null infinity principles", based on the induced symplectic structure there, which would presumably include now a contribution from the asymptotic form of $A_r$.  This is beyond the scope of our paper.  We just stress again that $\Psi_{\log}$ and $\Phi_{\log}$, which are present in the generator of $\mathcal O(1)$ gauge transformations through $A_r$, play an important role since these are shifted by the logarithmic gauge transformations.

As a final note, we observe that we could have carried the analysis by keeping the zero mode of $\xbar \Phi$ and $\Psi_{\log}$.  The gauge parameter $\lambda_{\log}$ acquires then a zero mode, which appears in the same way as the zero mode of $A_\eta^{(0)}$ since both obey the same equation.  One finds explicitly in Bondi coordinates
\begin{equation}
\lambda=\log r\,\lambda_{\log}^{(0)}+\lambda^{(0)}+\frac{\log r}{r}\lambda_{\log}^{(1)}+\mathcal{O}\left(r^{-1}\right)
\end{equation}
where
\begin{align}
\lambda_{\log}^{(0)} & =\frac{1}{2}\omega_{00}^{P}Y_{00}\\
\lambda^{(0)} & =\frac{1}{2}\log(-2u)\omega_{00}^{P}Y_{00}+\sum_{l,m}\left[\lambda_{lm}^{Q}+\tau_{lm}(1)\right]Y_{lm}\\
\lambda_{\log}^{(1)} & =\frac{u}{2}\left[\omega_{00}^{P}Y_{00}-\sum_{l>0,m}l(l+1)\omega_{lm}^{P}Y_{lm}\right]-\frac{u}{2}\sum_{l,m}l(l+1)\lambda_{lm}^{Q}Y_{lm}+\frac{1}{2}\sum_{l,m}\gamma_{lm}^{Q(0)}Y_{lm} \, .
\end{align}
These expressions reduce to the previous formulas when $\omega_{00}^{P}=0$.  The zero mode of  $\lambda_{\log}$ is the dominant term. The corresponding gauge transformation induces a $\mathcal O(r^{-1})$ term in $A_r$ and a $\mathcal O(1)$ term in $A_u$.  These gauge contributions are, however, proper gauge terms that can be removed since the zero modes of $\xbar \Phi$ and $\Psi_{\log}$ are pure gauge. [If one "gives life" to these zero modes by sticking to the untruncated kinetic term of \cite{Fuentealba:2023rvf}, these are not pure gauge and the logarithmic charge acquires a term involving $\omega_{00}^{P}$ and the zero mode $C$ of $\Phi$, 
\be
Q_{\lambda_{\log}^{(1)}}  =\omega_{00}^{P}C-\sum_{l>0,m}(-1)^{m}l(l+1)\omega_{lm}^{P}C_{l,-m}^{Q} \, . \quad ]
\ee

\section{Comments on  higher dimensions}
\label{Sec:MatchingHighD}

The discussion of the improper gauge symmetries of electromagnetism and the form that they take at null infinity is direct in higher dimensions given the already existing results \cite{Henneaux:2019yqq} and the procedure outlined above for handling them.

The idea is indeed simple.  First one starts with configurations of the electromagnetic potential that do not allow for the possibility of performing improper gauge symmetries, i.e., which decay at spatial infinity as in (\ref{Eq:AsympGP0}).  The corresponding solution has been determined in Subsection {\bf \ref{Subsec:HigherD}} in terms of the field strengths, which are given in \cite{Henneaux:2019yqq}.  
The behaviour of the electromagnetic field near null infinity is also given in \cite{Henneaux:2019yqq} and recalled in Appendices {\bf \ref{App:Derivation}} and {\bf \ref{App:FNullInf}}, implying 
 matching conditions of mixed type for the field strengths, as discussed in \cite{Fuentealba:2024lll,Henneaux:2019yqq}.

The second step is to perform an improper gauge transformation on the initial data (\ref{Eq:AsympGP0}). What explicit form this improper gauge symmetry should take depend on what type of asymptotic hehaviour one ultimately allows.  At least two different versions exist on the market \cite{Henneaux:2019yqq,Fuentealba:2023huv}.  But whatever these are, the procedure always consists in solving the wave equation for the parameter of the improper gauge transformations, which preserve indeed the Lorenz gauge. The scalar wave equation in higher dimensions has been studied in \cite{Fuentealba:2024lll,Henneaux:2018mgn} and we can thus just bring to the electromagnetic case the results of these references.   We leave that task to the reader. There is no logarithmic term to be worried about, because, in higher dimensions, these become  subleading gauge transformations already at spatial infinity (they appear at leading order in $D=4$ spacetime dimensions because the roots of the relevant indicial equation coincide \cite{Fuentealba:2024lll}). 

\section{Conclusions}
\label{Sec:Conclusions}

In this paper, we have exhibited the form of the logarithmic angle-dependent $u(1)$ transformations at null infinity.  Perhaps the main message that our analysis conveys is that these transformations can be seen at null infinity only if the condition $A_r = 0$ is relaxed to the weaker form $A_r = \mathcal O \left(\frac{\log r}{r^2}\right)$.  It is true that $A_r$ can always be written as $\partial_r f$ for some $f$, but the gauge transformation defined by $f$ is in general improper. Performing it  to set $A_r =0$ would truncate the theory.  Similar considerations apply to higher dimensions where setting $A_r=0$ at null infinity would remove subleading (but physically relevant) gauge transformations. 

Setting $A_r=0$ is not a problem if one is interested only in the $\mathcal O(1)$ angle-dependent gauge transformations, since these do not act on $A_r$.  But allowing $A_r \not=0$ is critical in order to exhibit the  logarithmic symmetry at null infinity.

Also to be emphasized is that our derivation of the Goldstone boson and of the charges at null infinity are local in $u$.  They are obtained by matching the relevant quantities in the limit $u \rightarrow - \infty$ with the Hamiltonian expressions.   Our construction involves no integration over $u$, or averages between $u = - \infty$ and $u = \infty$.

Our analysis assumed the background spacetime to be Minkowski space. We plan to extend our work to gravity and to provide a description of the logarithmic supertranlations \cite{Fuentealba:2022xsz} at null infinity.  The matching of the gravitational Goldstone boson was already derived in \cite{Fuentealba:2023syb}. This  would also yield a description of the asymptotic properties of the electromagnetic field in curved, asymptotically flat backgrounds.

Finally, we have studied in depth the expression of the conserved charges as one goes to null infinity.  At spatial infinity, the charges takes always a simple uniform flux expression, reflecting the simple asymptotic behaviour of the fields at large spatial distances, independently of the parity conditions that the leading orders of the fields would satisfy.  In spite of the fact that different parities lead to very different behaviours as one goes to null infinity, with one branch developing dominant logarithmic terms, the conserved charges remain well defined and finite -- they are just not determined in that case by the coefficient of $1/r^2$ term in the expansion of the fields but instead by the coefficient of the $\log r/r^2$ term.

\section*{Acknowledgments}

It is a pleasure to thank C\'edric Troessaert for discussions in the early stages of this work. O.F. is grateful to the Coll\`ege de France and the Universit\'e Libre de Bruxelles for kind hospitality while this article was completed.  This work was partially supported by  FNRS-Belgium (convention IISN 4.4503.15), as well as by funds from the Solvay Family.  The research of O.F. is partially supported by ANID through the Fondecyt grant N$^\circ$ 11251195, and by the Vicerrector\'{i}a de Investigaci\'on e Innovaci\'on of the Universidad Arturo Prat through an UNAP Consolida grant.

\appendix

\section*{Appendices}

\section{Notes and conventions on the sphere antipodal map}
\label{App:Orientation}

The purpose of this appendix is to clarify some conventions concerning the way we handle parity properties of tensors on the sphere that are convenient but might be misleading.

We consider for definiteness the unit $2$-sphere in $3$-dimensional Euclidean space with coordinates $x^i$ ($D=4$ Minkowski space).

The antipodal map is the transformation $S: x^i \rightarrow - x^i$ and is its own inverse ($S = S^{-1}$).  In spherical coordinates, it reads
\be
S: \quad r \rightarrow r \, , \qquad \theta \rightarrow \pi - \theta \, , \qquad \varphi \rightarrow \varphi + \pi \, .
\ee
The $2$-plane $x^3= 0$ and the $2$-sphere $r=1$ are both invariant under the antipodal map.  However, while the antipodal map preserves the orientation of the $2$-plane ($\frac{\partial}{\partial x^1} \rightarrow -\frac{\partial}{\partial x^1}$, $\frac{\partial}{\partial x^2} \rightarrow -\frac{\partial}{\partial x^2}$), it does not preserve the orientation of the $2$-sphere ($\frac{\partial}{\partial \theta} \rightarrow -\frac{\partial}{\partial \theta}$, $\frac{\partial}{\partial \varphi} \rightarrow \frac{\partial}{\partial \varphi}$).  This is an important difference. 

This difference in the orientation-behaviour of the antipodal map (in the ($D-2$)-planes versus the ($D-2$)-sphere) exists in all dimensions.

A tensor field $T$ is even under the antipodal map if $T^* S = S$, where $T^* S$ is the image of $T$ under the antipodal map.  It is odd if $T^* S = -S$.  For example, an odd $2$-form $F$  possesses cartesian components $F_{ij}$ which are odd functions of $x^k$.  This means that $F_{r \varphi}$ is an odd function, but that $F_{r \theta}$ and $F_{\theta \varphi}$ are even. The integral $\oint_{S_2} F = \oint_{S_2} F_{\theta \varphi} d\theta \wedge d\varphi$ can be different from zero, while the same sphere integral vanishes for an even $2$-form, for which $ F_{\theta \varphi}$ is odd.

In order not to have to distinguish between $\theta$ and $\varphi$ in computations dealing with parity properties under the antipodal map, it has become customary to collectively denote the coordinates on the sphere as $x^A$ ($A=1,2$) and to symbolically write the antipodal map as $x^A \rightarrow - x^A$. This is very useful in determining whether a function constructed out of the fields and their derivatives is even or odd. But it must be stessed that there is no coordinate system on the sphere (defined on an open set homeomorphic to $\mathbb{R}^2$) that covers simultaneously a region and its antipodal image and in which the antipodal map takes this form, since the $2$-dimensional change of coordinates $x^A \rightarrow - x^A$ does not change the orientation, while the antipodal map does.

For instance, the coordinates $(x^1, x^2)$ are good coordinates on the upper hemisphere ($x^3 >0$) or the lower hemisphere ($x^3 <0$), between which there is no overlap (these coordinates fail on the equator). The antipodal map does take the form $x^1 \rightarrow - x^1$, $x^2 \rightarrow - x^2$ but it also exchanges the two hemispheres.   As we have also seen, if the coordinates $(x^1, x^2)$ have positive orientation in the upper hemisphere, they have negative orientation in the lower hemisphere.

This means that some care must be taken when integrating $2$-forms over the sphere.  One may hastily conclude that the integral over the sphere of $F_{AB} dx^A \wedge dx^B$ vanishes if the components $F_{AB}$ are odd on the sphere (i.e., ${F_{AB}}_{\textrm{lower}}(-x^A) = - {F_{AB}}_{\textrm{upper}}(x^A)$), because the contribution from the lower hemisphere cancels the contribution from the upper hemisphere but actually, it adds up since the orientations are different.  And indeed, as we have pointed out above, $F_{\theta \varphi}$ is then even.

\section{The electromagnetic field $F_{\mu \nu}$ in hyperbolic coordinates}
\label{App:Derivation}

This section reviews the general solution of Maxwell's equations for the field strengths $F_{\mu \nu}$ in hyperbolic coordinates, performed in the references \cite{Henneaux:2018gfi,Henneaux:2019yqq}.

\subsection{Maxwell's equations}

In hyperbolic coordinates, the free Maxwell equations $\nabla_{\mu}F^{\mu\nu}=0$ become
\begin{align}
\mathcal{D}_{a}E^{a} & =0\,, \qquad
\eta^{D-5}\mathcal{D}^{b}F_{ba}-\partial_{\eta}\left(\eta^{D-3}E_{a}\right)  =0\,,\label{eq:MaxwellE1-1}
\end{align}
where $E_{a}=F_{a\eta}$ and $\mathcal{D}^{a}=h^{ab}\mathcal{D}_{b}$.
The Bianchi identities read
\begin{align}
\partial_{\eta}F_{ab} & =\partial_{b}E_{a}-\partial_{a}E_{b}\,, \qquad 
\partial_{[a}F_{bc]}  =0\,.\label{eq:BianchiDGen1}
\end{align}

Expanding the gauge-invariant field strengths as 
\begin{align}
    E_b &=  \sum_{k\geq 0} \eta^{-D + 3 - k} E_b^{(k)} \, ,  \qquad
    F_{ab} = \sum_{k \geq 0} \eta^{-D + 4 - k} F_{ab}^{(k)}, 
    \label{eq:ExpHyp1}
   \end{align}
we find that the equations to be solved decouple order by order and read
\begin{gather}
\mathcal D^a E^{(k)}_a = 0,\label{eq:MaxwellHyp1}\\
     \mathcal D^b F^{(k)}_{ba} + k E^{(k)}_a = 0,  \label{eq:MaxwellHyp2} \\
     (D-4+k) F_{ab}^{(k)} = \d_a E^{(k)}_b - \d_b E^{(k)}_a,
\label{eq:MaxwellHyp3} \\
    \d_{[a} F^{(k)}_{bc]} = 0.  \label{eq:MaxwellHyp4}
\end{gather}
The $s=0$ hypersurface coincides with the $t=0$ hyperplane, on which $\eta$ reduces to $r$ and the above asymptotic developments coincide with the standard ones ($F_{\mu \nu} \sim 1/r^{D-4}$ in Minkowskian coordinates).  One key interest of hyperbolic coordinates is that Lorentz-invariant equations decouple order by order in a $1/\eta$ expansion \cite{Beig:1982ifu}.

By taking the divergence of the Bianchi identities and using the Maxwell equations, one easily derives
\begin{align}
\mathcal{D}_{a}\mathcal{D}^{a}E_{b}^{(k)}+[k^{2}+ (D-4) k -(D-2)]E_{b}^{(k)} & =0\,,\label{eq:EqforEGen}\\
\mathcal{D}_{a}\mathcal{D}^{a}F_{bc}^{(k)}+[k^{2} + (D-4) k -2(D-3)]F_{bc}^{(k)} & =0\,.\label{eq:EqforFGen}
\end{align}

From the first equation and the divergence-free condition on $E_{b}^{(k)}$, one can derive second order evolution equations that involve only the $s$-component of $E^{(k)}_a$ (the radial electric field), one for each $k$: 
\begin{eqnarray}
    &&  (1-s^2) \d_s^2 E^{(k)}_s + (D-6)s \d_s  E^{(k)}_s + (D-4) E^{(k)}_s - \xbar \triangle E^{(k)}_s \nonumber \qquad \qquad \\
   && \qquad \qquad  \qquad \qquad - (1-s^2)^{-1}k (D-4+k) E^{(k)}_s=0 \, , \label{eq:EvolEs}
\end{eqnarray}
where $\xbar \triangle$ is the (connection) Laplacian on the $(D-2)$-sphere, $\xbar \triangle = \xbar D_A \xbar D^A$.  Here and below, $\xbar D_A$ is the covariant derivative with respect to the sphere metric $\xbar \gamma_{AB}$ and the angle indices are raised and lowered with $\xbar \gamma^{AB}$ and $\xbar \gamma_{AB}$, respectively.  Simarly, the second equation together with the curl-free condition on $F_{ab}$ imply second order evolution equations for $F_{AB}^{(k)}$, 
\begin{eqnarray}
   &&  (1-s^2) \d_s^2 F^{(k)}_{AB} + (D-6) s \d_s  F^{(k)}_{AB}+ 2 (D-4)  F^{(k)}_{AB} \nonumber \qquad \qquad 
    \\ && \qquad \qquad \qquad - \xbar \triangle  F^{(k)}_{AB} - (1-s^2)^{-1} k(D-4+k)  F^{(k)}_{AB}= 0 \, . \label{eq:EvolFAB}
\end{eqnarray}

\subsection{Method for finding the general solution}
\label{SubSec:Method1}

Splitting the components of the fields into $E^{(k)}_s$, $E^{(k)}_A$, $F_{sA}^{(k)}$ and $F_{AB}^{(k)}$ into space ($x^A$) and time ($s$), we can decompose the Maxwell equations and the Bianchi identities as
\begin{eqnarray}
&& \xbar D^A E^{(k)}_A = (1-s^2) \partial_s E^{(k)}_s + (D-4) s E^{(k)}_s  \, , \label{eq:MaxwellDec1}\\
&& (1-s^2)\xbar D^B F^{(k)}_{Bs} + k E^{(k)}_s =0 \, , \label{eq:MaxwellDec2} \\
&& -(1-s^2)^2 \partial_s F_{sA}^{(k)}-(D-6) (1-s^2) s F_{sA}^{(k)}  + (1-s^2)\xbar D^B F^{(k)}_{BA} + k E^{(k)}_A =0 \, , \label{eq:MaxwellDec3}\\
&& (D-4+k) F_{sB}^{(k)} = \d_s E^{(k)}_B - \d_B E^{(k)}_s  \, , \label{eq:MaxwellDec4}\\
&& (D-4+k) F_{AB}^{(k)} = \d_A E^{(k)}_B - \d_B E^{(k)}_A \, , \label{eq:MaxwellDec5}\\
&&  \d_{s} F^{(k)}_{BC} + \d_{B} F^{(k)}_{Cs} + \d_{C} F^{(k)}_{sB} = 0 \, , \label{eq:MaxwellDec6} \\
&&  \d_{[A} F^{(k)}_{BC]} = 0 \, . \label{eq:MaxwellDec7} 
\end{eqnarray}

Suppose that solutions $E^{(k)}_s$ and $F^{(k)}_{AB}$ of (\ref{eq:EvolEs}) and (\ref{eq:EvolFAB}) are given.  Suppose furthermore that the $2$-forms $F^{(k)}_{AB}$ are exact on the $(D-2)$-sphere,  which is compatible with the evolution equation (\ref{eq:EvolFAB})\footnote{The closedness condition is dictated by (\ref{eq:MaxwellDec5}) for $D>4$ or $k>0$ and is equivalent to (\ref{eq:MaxwellDec7}) for $D>4$ since $H^2(S^{D-2})$ then vanishes.  In $D=4$, we can allow a closed term $\epsilon_{AB} \sqrt{\xbar \gamma} H^{(0)}$ in $F_{AB}^{(0)}$, which must be $s$-independent by (\ref{eq:MaxwellDec6}).}.

Then, the equations (\ref{eq:MaxwellDec1}) and (\ref{eq:MaxwellDec5}) give the divergence and the curl of $E^{(k)}_A$ in terms of $E^{(k)}_s$ and $F^{(k)}_{AB}$, while the equations (\ref{eq:MaxwellDec2}) and (\ref{eq:MaxwellDec6})  give the divergence and the curl of $ F^{(k)}_{Bs}$ in terms of $E^{(k)}_s$ and $F^{(k)}_{AB}$.  

Knowing the divergence and the curl of a $1$-form $\omega^1$ on the $(D-2)$-sphere ($D\geq 4$),
$$
\xbar d \omega^1 = \psi^2 \, , \qquad \xbar d \, ^* \! \omega^1 = \chi^{D-2} 
$$
with $\psi^2$ and $\chi^{D-2}$ given,  determines that $1$-form uniquely provided the necessary integrability conditions $\xbar d \psi^2 = 0$ and $\oint \chi^{D-2}=0$ are fulfilled.  The first condition evidently holds in our case since $F_{AB}^{(k)}$ and $\partial_s F_{AB}^{(k)}$ are exact $2$-forms (or drop from these equations for $D=4$, $k=0$), while the second condition imposes
\be 
\oint d^{D-2} x \sqrt{\xbar \gamma} \Big[(1-s^2) \partial_s E^{(k)}_s + (D-4) s E^{(k)}_s \Big] =0 \,, \label{eq:MaxwellInteg1}
\ee
and  
\be 
k \oint d^{D-2} x \sqrt{\xbar \gamma}  E^{(k)}_s = 0\,. \label{eq:MaxwellInteg2}
\ee
This kills the zero mode of $E_s^{(k)}$ except for $k=0$ \cite{Henneaux:2019yqq} (see also below).

The expression for $ F^{(k)}_{Bs}$ obtained from (\ref{eq:MaxwellDec2}) and (\ref{eq:MaxwellDec6}) is compatible with (\ref{eq:MaxwellDec4}) (same curl and divergence) if and only if $E^{(k)}_s$ fulfills the differential equation (\ref{eq:EvolEs}).  
In the same way, the expression for  $E^{(k)}_A$ obtained from (\ref{eq:MaxwellDec1}) and (\ref{eq:MaxwellDec5}) is compatible with (\ref{eq:MaxwellDec3}) (same curl and divergence) if and only if $F^{(k)}_{AB}$ fulfills the differential equation (\ref{eq:EvolFAB})

We can thus proceed as follows in order to solve the Maxwell equations for $F_{\mu \nu}$  at order $k$.  
\begin{itemize}
\item Step 1: Find the most general solution of the second order linear differential equations (\ref{eq:EvolEs}) and (\ref{eq:EvolFAB}), subject to the conditions (\ref{eq:MaxwellInteg1}),  (\ref{eq:MaxwellInteg2}) and (\ref{eq:MaxwellDec7}).  As we shall review below, this can be achieved because these equations can be rewritten in the form of ultraspherical differential equations \cite{Henneaux:2019yqq}.
\item Step 2: Determine then $E^{(k)}_A$ from (\ref{eq:MaxwellDec1}) and (\ref{eq:MaxwellDec5}) as well as $ F^{(k)}_{Bs}$  from (\ref{eq:MaxwellDec2}) and (\ref{eq:MaxwellDec6}).
\end{itemize}
When $D-4 +k >0$, we can circumvent the determination of $ F^{(k)}_{Bs}$ from its divergence and its gradient by using directly  (\ref{eq:MaxwellDec4}).

\subsection{Explicit solution}

\subsubsection{$E_s$ and $F_{AB}$}

Following the procedure just described, we first solve the equations (\ref{eq:EvolEs}) and (\ref{eq:EvolFAB}) for $E_s$ and $F_{AB}$, with the condition $\partial_{[C} F_{AB]} = 0$.

To that end, we expand the fields in terms of the appropriate spherical harmonics (see Appendix {\bf \ref{App:SphHar}} for explicit definitions)
\begin{align}
E_{s}^{(k)}&=(1-s^{2})^{-\frac{k}{2}}\sum_{l,m}\Xi_{lm}^{(k)}(s)Y_{lm}\,, \label{Eq:ExpForEs}\\
F_{AB}^{(k)}&=(1-s^{2})^{-\frac{k}{2}}\sum_{l,m}\alpha_{lm}^{(k)}(s)\Theta_{AB,lm}\,, \label{Eq:ExpForFAB}
\end{align}
where the summation over $m$ for given $l$ spans all corresponding degeneracy values, which is in general not the same for zero forms and exact $2$-forms  (see Appendix {\bf \ref{App:SphHar}}).
We then find that the coefficients $\Xi_{lm}$ and $\alpha_{lm}^{(k)}(s)$ should obey the following second order linear differential equation
\begin{equation}
(1-s^{2})\partial_{s}^{2}Y_{n}^{(\lambda)}+(2\lambda-3)s\partial_{s}Y_{n}^{(\lambda)}+(n+1)(n+2\lambda-1)Y_{n}^{(\lambda)}=0\,,  \label{Eq:Ultra1}
\end{equation}
where 
\begin{equation}
\lambda=k+\frac{D-3}{2}\qquad\text{and\ensuremath{\qquad n=l-k}}\,.
\end{equation}
The relevant values of $\lambda$ are positive half-integers (even spacetime dimensions) and positive integers (odd spacetime dimensions).
In $4$ spacetime dimensions, the extra term 
\be \label{Eq:MagneticCharge}
H^{(0)}_{AB} = \epsilon_{AB} \sqrt{\xbar \gamma} H^{(0)}\,,
\ee 
which we expand as
\be
H^{(0)} =  \sum_{l,m}\Upsilon_{lm}^{(0)}(s)Y_{lm}
\ee
should be added to $F_{AB}^{(0)}$, which might indeed be closed without being exact.  The coefficients $\Upsilon_{lm}^{(k)}(s)$ obey also (the $D=4$ version of) (\ref{Eq:Ultra1}).

The general solution of the differential equation (\ref{Eq:Ultra1}) is explicitly reviewed in the papers \cite{Henneaux:2018mgn} and \cite{Henneaux:2019yqq}, based on the book   \cite{Ultra}.  It is given by
\begin{equation}
Y_{n}^{(\lambda)}(s)=A\tilde{P}_{n}^{(\lambda)}+B\tilde{Q}_{n}^{(\lambda)}\,,
\end{equation}
where:
\begin{enumerate}[label=(\roman*)]
\item For $n \geq 0$,
\begin{equation}
\tilde{P}_{n}^{(\lambda)}=(1-s^{2})^{\lambda-\frac{1}{2}}P_{n}^{(\lambda)}\,,\qquad\tilde{Q}_{n}^{(\lambda)}=(1-s^{2})^{\lambda-\frac{1}{2}}Q_{n}^{(\lambda)}\,,
\end{equation}
where $P_{n}^{(\lambda)}$ stands for the ultraspherical (or Gegenbauer) polynomial parametrized by $\lambda$ and $n$, while 
 $Q_{n}^{(\lambda)}$ is the corresponding ultraspherical function of the second kind.  Both sets fulfill the same recurrence relation
\be
n T_n^{(\lambda)}(s) = 2 (n + \lambda -1) s T^{(\lambda)}_{n-1}(s) - (n+2 \lambda - 2) T^{(\lambda)}_{n-2}(s)
\ee
but with different starting points, explicitly,
\be
P_0^{(\lambda)}(s) = 1 \, , \qquad P_1^{(\lambda)}(s) = 2 \lambda s \, ,
\ee
for the $P$-branch and
\begin{equation}
 Q_{0}^{(\lambda)}(s)=\int_{0}^{s}\left(1-x^{2}\right)^{-\lambda-\frac{1}{2}}dx\, , \qquad Q_1^{(\lambda)}(s) = 2 \lambda s Q_0^{(\lambda)}(s) - (1-s^2)^{- \lambda + \frac12} \, ,
\end{equation}
for the $Q$-branch.  This yields
\be
Q_{n}^{(\lambda)}(s)=P_{n}^{(\lambda)}(s)Q_{0}^{(\lambda)}(s)+R_{n}^{(\lambda)}(s)(1-s^{2})^{-\lambda+\frac{1}{2}}\, ,
\ee
where $R_{n}^{(\lambda)}(s)$ are polynomials of degree $n-1$.
\item For $n<0$, the two branches $\tilde{P}_{n}^{(\lambda)}$ and $\tilde{Q}_{n}^{(\lambda)}$ are again both determined by the same recurrence relation, which now reads 
\be
(2 \lambda + n - 1) Z_{n-1}^{(\lambda)} = 2 (n + \lambda) s Z^{(\lambda)}_{n}(s) - (n+1) Z^{(\lambda)}_{n+2}(s)\,,
\ee  
with respective starting points
\be
\tilde{P}_{-1}^{(\lambda)}(s) = \int_0^s (1-x^2)^{\lambda - \frac32} dx \, , \qquad \tilde{P}_{-1}^{(\lambda)} = s \tilde{P}_{-1}^{(\lambda)}(s) + \frac{1}{2(\lambda -1)} (1-x^2)^{\lambda - \frac12}\,, 
\ee
and
\be
\tilde{Q}_{-1}^{(\lambda)}(s) = 1 \, , \qquad \tilde{Q}_{-2}^{(\lambda)} = s  \, .
\ee
It is now the $Q$-branch that is polynomial, while the $P$-branch is polynomial only in even spacetime dimensions.
\end{enumerate}
Note that for $\lambda = \frac12$, which occurs only when $D=4$ and $k=0$, the $P$-solutions are the Legendre polynomials $P_l \equiv P^{(\frac12)}_l$ while the $Q$-solutions are the Legendre functions of the second kind $Q_l \equiv Q^{(\frac12)}_l$.

We note in this context useful relations controlling the behaviour of the of the Legendre polynomials and functions of the second kind near $s=1$, 
\begin{align}
\partial_{s}P_{l}(s) & =\frac{1}{2}l(l+1)+\mathcal{O}(1-s)\,,\\
\partial_{s}Q_{l}(s) & =\partial_{s}\left[\frac{1}{2}P_{l}(s)\log\left(\frac{1+s}{1-s}\right)+\tilde{Q}_{l}(s)\right]\nonumber \\
 & =\frac{1}{2\left(1-s\right)}+\frac{1}{4}\left[l(l+1)\log\left(\frac{2}{1-s}\right)+4\partial_{s}\tilde{Q}_{l}(1)+1-l(l+1)\right]+o(1-s)\nonumber \,.
\end{align}

Putting everything together, we get  for $E_s^{(k)}$ and $F_{AB}^{(k)}$:
\begin{align}
E_{s}^{(k)} & =(1-s^{2})^{-\frac{k}{2}}\sum_{l,m}\left[\Xi_{lm}^{P(k)}\tilde{P}_{l-k}^{(\lambda)}(s)+\Xi_{lm}^{Q(k)}\tilde{Q}_{l-k}^{(\lambda)}(s)\right]Y_{lm}\,, \label{Eq:ExpForEs2}\\
F_{AB}^{(k)} & =(1-s^{2})^{-\frac{k}{2}}\sum_{l>0,m}\left[\alpha_{lm}^{P(k)}\tilde{P}_{l-k}^{(\lambda)}(s)+\alpha_{lm}^{Q(k)}\tilde{Q}_{l-k}^{(\lambda)}(s)\right]\Theta_{AB,lm}\,. \label{Eq:ExpForFAB2}
\end{align}
We stress that the summation over $m$ for given $l$ spans all corresponding degeneracy values, which is in general not the same for zero forms and exact $2$-forms.  Also, we could include a $l=0$ term in (\ref{Eq:ExpForFAB}) since $\Theta_{AB,lm}$ identically vanishes for $l=0$.

The zero mode conditions (\ref{eq:MaxwellInteg1}) and (\ref{eq:MaxwellInteg2}) are easily seen to imply \cite{Henneaux:2019yqq}\footnote{The condition (\ref{eq:MaxwellInteg2}) yields $\Xi_{00}^{P(k)}\tilde{P}_{-k}^{(\lambda)}(s)+\Xi_{00}^{Q(k)}\tilde{Q}_{l-k}^{(\lambda)}(s)$ for $k \not=0$ and thus $\Xi_{00}^{P(k)}=0 =\Xi_{00}^{Q(k)}$ ($k \not=0$) since $\tilde{P}_{-k}^{(\lambda)}(s)$ and $\tilde{Q}_{-k}^{(\lambda)}(s)$ are independent functions of $s$.  One then gets from (\ref{eq:MaxwellInteg2}) the condition $(1-s^2) [\Xi_{00}^{P(0)}\partial_s \tilde{P}_{0}^{(\lambda)}(s) + \Xi_{00}^{Q(0)}\partial_s \tilde{Q}_{0}^{(\lambda)}(s) ]+(D-4) s [\Xi_{00}^{P(0)} \tilde{P}_{0}^{(\lambda)}(s) + \Xi_{00}^{Q(0)} \tilde{Q}_{0}^{(\lambda)}(s) ] =0 $, which is identically fulfilled by $\tilde{P}_{0}^{(\lambda)}(s)$ but not by $\tilde{Q}_{0}^{(\lambda)}(s) $, implying $\Xi_{00}^{Q(0)}= 0$ with $\Xi_{00}^{P(0)}$ unrestricted.}
\be
\Xi_{00}^{P(k)} = \Xi_{00}^{P(0)} \delta^k_0 \, , \qquad \Xi_{lm}^{Q(k)} = 0 \, .
\ee

Finally, in $D=4$, the extra term $H^{(0)}$ in $F_{AB}^{(0)}$ is given by 
\be
H^{(0)}= g \, ,
\ee
once one imposes the extra condition $\partial_s F_{AB}^{(0)} = 0$.  The integration constant $g$ is the magnetic charge.

\subsubsection{Other components of the electromagnetic field}

By direct manipulations following the method of Subsection {\bf \ref{SubSec:Method1}}, one easily derives the explicit expressions of the other components of the electromagnetic field, which are \cite{Henneaux:2019yqq},
\begin{align}
E_{A}^{(k)L} & =-(1-s^{2})^{-\frac{k}{2}}\sum_{l>0,m}\frac{1}{l(l+D-3)}   \nonumber \\
& \qquad \quad  \Big\{\left[(1-s^{2})\partial_{s}+(D-4 +k)s\right]\left[\Xi_{lm}^{P(k)}\tilde{P}_{n}^{(\lambda)}(s)+\Xi_{lm}^{Q(k)}\tilde{Q}_{n}^{(\lambda)}(s)\right]\Phi_{A,lm}\Big\}\,,\label{Eq:EAkL}\\
E_{A}^{(k)T} & =(D-4 + k)(1-s^{2})^{-\frac{k}{2}}\sum_{l>0,m}\left[\alpha_{lm}^{P(k)}\tilde{P}_{l-k}^{(\lambda)}(s)+\alpha_{lm}^{Q(k)}\tilde{Q}_{l-k}^{(\lambda)}(s)\right]\Psi_{A,lm}\,, \label{Eq:EAkT}\\
F_{sA}^{(k)L} & =-k(1-s^{2})^{-\frac{k}{2}-1}\sum_{l>0,m}\frac{1}{l(l+D-3)}\nonumber \\
& \qquad \qquad \qquad \qquad \qquad \qquad \qquad  \Big\{\left[\Xi_{lm}^{P(k)}\tilde{P}_{l-k}^{(\lambda)}(s)+\Xi_{lm}^{Q(k)}\tilde{Q}_{l-k}^{(\lambda)}(s)\right]\Phi_{A,lm}\Big\}\,,\label{Eq:FsAkL}\\
F_{sA}^{(k)T} & =(1-s^{2})^{-\frac{k}{2}-1} \nonumber \\
& \qquad \qquad \qquad \sum_{l,m}\left[(1-s^{2})\partial_{s}+ks\right]\left[\alpha_{lm}^{P(k)}\tilde{P}_{l-k}^{(\lambda)}(s)+\alpha_{lm}^{Q(k)}\tilde{Q}_{l-k}^{(\lambda)}(s)\right]\Psi_{A,lm}\,.\label{Eq:FsAkT}
\end{align}

\section{Behaviour near null infinity and matching conditions for $F_{\mu \nu}$ ($D=4$)}
\label{App:FNullInf}

We first derive the matching conditions for the field strength $F_{\mu \nu}$.  In order to achieve this task,  we need to expand the solution $F_{\mu \nu}(\eta, s, x^A)$ in the vicinity of null infinity.  This was done in \cite{Henneaux:2018gfi,Henneaux:2019yqq} for $E_{s}$ by relying on the method of \cite{Fried1,Friedrich:1999wk,Friedrich:1999ax}.   Only $E_s$ was considered in \cite{Henneaux:2018gfi,Henneaux:2019yqq} because $E_s$ is the only component of $F_{\mu \nu}$ relevant to the charges of electric type.  We report in this section this result for $D=4$ spacetime dimensions.  We also derive the matching conditions for the radial component of the magnetic field, relevant to the charges of magnetic type.

\subsection{General form of the electromagnetic field}

We compute the electromagnetic field at null infinity, first without imposing any kind of parity conditions.

One has
\begin{equation}
    F_{ur} = \frac{1-s^2}{\eta} E_s \, , \quad F_{uA}=\frac{r+u}{\sqrt{-u(2r+u)}}E_{A}+\frac{1}{r}F_{sA}\, , \quad F_{rA}=\frac{u}{\sqrt{-u(2r+u)}}E_{A}-\frac{u}{r^{2}}F_{sA}\,.
\end{equation}

Expressing the solution for $F_{ur}(\eta, s, x^A)$ and $F_{AB}(\eta, s, x^A)$ in null coordinates and expanding around $s=1$ then yields, without assuming parity conditions,
\begin{itemize}
\item $F_{ur}$:
\begin{equation}
F_{ur}=\frac{\log r}{r^{2}}F_{ur}^{\log}+\frac{1}{r^{2}}\xbar F_{ur}+o\left(r^{-2}\right)\,,
\end{equation}
where 
\begin{align}
F_{ur}^{\log} & = \frac12 \sum_{l,m}\Xi_{lm}^{Q(0)} Y_{lm}\,,\\
\xbar F_{ur} & =\sum_{l,m}\left[\Xi_{lm}^{P(0)}+\Xi_{lm}^{Q(0)}\left(\frac{1}{2}\left(-\log(-u)+\log2\right)+R_{l}^{(\frac{1}{2})}(1)\right)\right]Y_{lm}\nonumber \\
 & \quad+\sum_{k>0}\sum_{l,m}\left(-2u\right)^{-k}\left[\Xi_{lm}^{P(k)}\tilde P_{l-k}^{(k+\frac{1}{2})}(1)+\Xi_{lm}^{Q(k)}\tilde{Q}_{l-k}^{(k+\frac{1}{2})}(1)\right]Y_{lm}\,.  \label{Eq:Fur4D}
\end{align}
where we have used $P_{l}^{(\frac{1}{2})}(1) = 1$.   Note that $\tilde P_{l-k}^{(k+\frac{1}{2})}(1)=0$ whenever $l-k \geq 0$, so that only the $\tilde P$-terms with $l<k$ appear in  (\ref{Eq:Fur4D}).  A similar observation applies to the sums below.
\item $F_{uA}$:
\begin{equation}
F_{uA} =F_{uA}^{(0)}+\frac{\log r}{r}F_{uA}^{\log}+\mathcal{O}\left(r^{-2}\right)\,,
\end{equation}
with
\begin{align}
F_{uA}^{(0)} & =-\sum_{l>0,m}\frac{(-2u^{-1})}{l(l+1)}\Xi_{lm}^{Q(0)}\Phi_{A,lm}-\sum_{k=1}\sum_{0<l<k,m}2k\frac{(-2u)^{-(k+1)}}{l(l+1)}\Xi_{lm}^{P(k)}\tilde{P}_{l-k}^{(k+\frac{1}{2})}(1)\Phi_{A,lm}\nonumber \\
 & \quad-\sum_{k=1}\sum_{l>0,m}2k\frac{(-2u)^{-(k+1)}}{l(l+1)}\Xi_{lm}^{Q(k)}\tilde{Q}_{l-k}^{(k+\frac{1}{2})}(1)\Phi_{A,lm}\nonumber \\
 & \quad+\sum_{l,m}(-2u)^{-1}\alpha_{lm}^{Q(0)}\Psi_{A,lm}+\sum_{k=1}\sum_{0<l<k,m}2k\frac{(-2u)^{-(k+1)}}{l(l+1)}\alpha_{lm}^{P(k)}\tilde{P}_{l-k}^{(k+\frac{1}{2})}(1)\Psi_{A,lm}\nonumber \\
 & \quad+\sum_{k=1}\sum_{l>0,m}2k(-2u)^{-(k+1)}\alpha_{lm}^{Q(k)}\tilde{Q}_{l-k}^{(k+\frac{1}{2})}(1)\Psi_{A,lm}\,,\\
 F_{uA}^{\log}&=-\sum_{l>0,m}\frac{1}{4}\Xi_{lm}^{Q(0)}\Phi_{A,lm}+\sum_{l,m}\frac{l(l+1)}{4}\alpha_{lm}^{Q(0)}\Psi_{A,lm}\,.
\end{align}
\item $F_{rA}$:
\begin{equation}
F_{rA} =\frac{1}{r} F_{rA}^{(1)}+\frac{\log r}{r^{2}}F_{rA}^{\log}+\mathcal{O}\left(r^{-3}\right)\,,
\end{equation}
with
\begin{align}
F_{rA}^{(1)}&=\sum_{l>0,m}\frac{1}{2l(l+1)}\Xi_{lm}^{Q(0)}\Phi_{A,lm}+\sum_{l,m}\frac{1}{2}\alpha_{lm}^{Q(0)}\Psi_{A,lm}\,,\\
F_{rA}^{\log} & =-\sum_{l>0,m}\frac{u}{4}\Xi_{lm}^{Q(0)}\Phi_{A,lm}-\sum_{l>0,m}\frac{1}{4l(l+1)}\Xi_{lm}^{Q(1)}P_{l-1}^{(3/2)}(1)\Phi_{A,lm}\nonumber \\
 & \quad-\sum_{l,m}\frac{ul(l+1)}{4}\alpha_{lm}^{Q(0)}\Psi_{A,lm}+\sum_{l,m}\frac{1}{4}\alpha_{lm}^{Q(1)}P_{l-1}^{(3/2)}(1)\Psi_{A,lm}\,.
\end{align}
\item $F_{AB}$:
\begin{equation}
F_{AB}=\log r F_{AB}^{\log}+\xbar F_{AB}+o(1)\,,
\end{equation}
with
\begin{align}
F_{AB}^{\log} & =\frac{1}{2}\sum_{l>0,m}\alpha_{lm}^{Q(0)}\Theta_{AB,lm}\,,\\
\xbar F_{AB} & =\sum_{l>0,m}\left[\alpha_{lm}^{P(0)}+\alpha_{lm}^{Q(0)}\left(\frac{1}{2}\left(-\log(-u)+\log2\right)+R_{l}^{(\frac{1}{2})}(1)\right)\right]\Theta_{AB,lm}\\
 & +\sum_{k\geq 1}\sum_{l>0,m}\left(-2u\right)^{-k}\left[\alpha_{lm}^{P(k)}\tilde{P}_{l-k}^{(k+\frac{1}{2})}(1) + \alpha_{lm}^{Q(k)}\tilde{Q}_{l-k}^{(k+\frac{1}{2})}(1)\right]\Theta_{AB,lm} \nonumber \\
& + g \epsilon_{AB} \sqrt{\xbar \gamma}  \qquad \textrm{(in $4$ dimensions)} 
\end{align}
\end{itemize}

The key feature of these expansions near null infinity is the appearance of polylogarithmic terms.
The logarithms are brought in by the $Q$-branch of the solutions and appear at null infinity even though there is no logarithm in the initial data. The logarithms are even dominant if $\Xi_{lm}^{Q(0)}\not=0$ or $\alpha_{lm}^{Q(0)}\not=0$.

We now impose the parity conditions.

\subsection{Matching in the case of standard parity conditions twisted by a gauge transformation}

The imposition of the  standard parity conditions implies that we must take the $P$-branch both for $E_s^{(0)}$ in (\ref{Eq:ExpForEs2}) and for $F_{AB}^{(0)}$ in (\ref{Eq:ExpForFAB2}) ($\Xi_{lm}^{Q(0)}= 0 = \alpha_{lm}^{Q(0)}$)  \cite{Fuentealba:2024lll,Henneaux:2018gfi,Henneaux:2018hdj,Henneaux:2019yqq}.    This eliminates the leading logarithms and the asymptotic expressions of the field simplifies to
\begin{equation}
F_{ur}=\frac{1}{r^{2}}\xbar F_{ur}+o\left(r^{-2}\right)\,, \label{Eq:FurSimpl}
\end{equation}
with
\be
\xbar F_{ur}  =\sum_{l,m} \Xi_{lm}^{P(0)}Y_{lm} + \sum_{k>0}\sum_{l,m}\left(-2u\right)^{-k}\left[\Xi_{lm}^{P(k)}\tilde P_{l-k}^{(k+\frac{1}{2})}(1) Y_{lm}+\Xi_{lm}^{Q(k)}\tilde{Q}_{l-k}^{(k+\frac{1}{2})}(1)\right]\,,  \label{Eq:Fur4D2}
\ee
and
\begin{equation}
F_{AB}=\xbar F_{AB}+o(1)\,,
\end{equation}
with
\begin{align}
\xbar F_{AB} & =\sum_{l>0,m}\alpha_{lm}^{P(0)}\Theta_{AB,lm} \nonumber \\
 & +\sum_{k\geq 1}\sum_{l,m}\left(-2u\right)^{-k}\left[\alpha_{lm}^{P(k)}\tilde{P}_{l-k}^{(k+\frac{1}{2})}(1) + \alpha_{lm}^{Q(k)}\tilde{Q}_{l-k}^{(k+\frac{1}{2})}(1)\right]\Theta_{AB,lm} \, , \nonumber \\
& +   g \epsilon_{AB} \sqrt{\xbar \gamma} \, 
\end{align}
with the obvious simplifications for the other components completely determined by $E_s$ and $F_{AB}$.

This calls for two comments:
\begin{itemize}
\item These expansions coincide to leading order with the expansion usually assumed at null infinity \cite{Strominger:2017zoo,Satishchandran:2019pyc}. This means that these references implicitly assume that the initial data fulfill the above parity conditions to leading order.  It should be stressed that there are generically, however, subleading logarithmic terms unless one imposes  $\Xi_{lm}^{Q(k)}= 0 = \alpha_{lm}^{Q(k)}$ for all $k$'s to eliminate all the $Q$-branches -- something that may be argued to be artificial and in any case unnecessary. 
\item Because the leading term $\xbar F_{ur}$ in $F_{ur}$ involves only the $P$-branch as we take the limit to the past of future null infinity, which is even under the hyperboloid antipodal map (sphere antipodal map accompanied with hyperbolic time reversal $s \rightarrow - s$ for which $P_l(-1) = (-1)^l$), it obeys the matching condition of \cite{He:2014cra,Kapec:2015ena,Strominger:2017zoo},
\be
\lim_{v \rightarrow \infty}\xbar F_{vr}(-x^A) = \lim_{u \rightarrow -\infty}\xbar F_{ur}(x^A) = \sum_{l,m} \Xi_{lm}^{P(0)}Y_{lm} \, , \label{Eq:Match0}
\ee
since the $k> 0$ terms disappear in the limit.
Note that we also have
\be
\lim_{v \rightarrow \infty}\xbar F_{AB}(-x^C) = -\lim_{u \rightarrow - \infty} \xbar F_{AB}(x^C) = - \sum_{l>0,m}\alpha_{lm}^{P(0)}\Theta_{AB,lm} - g \epsilon_{AB} \sqrt{\xbar \gamma}\, .\label{Eq:Match2}
\ee
\end{itemize} 
Some authors sometimes assume $\lim_{u \rightarrow \infty}\xbar F_{AB}(-x^C) = 0$, i.e., $\xbar A_A$ is a pure gradient in that limit, but this is not needed here. 

The derivation of the matching conditions (\ref{Eq:Match0}) and (\ref{Eq:Match2}) proceeds by expanding the expressions for the fields near past null infinity, in advance Bondi coordinates $(v,r,x^A)$ and making then the comparison.  The key relations are the parity properties of the $P$ and the $Q$ branches, namely, $P_l(-s) = (-1)^l P_l(s)$ and $Q_l(-s) = - (-1)^l Q_l(s)$, which exchanges the past of future null infinity ($s=1$) with the future of past null infinity ($s=-1$) \cite{Henneaux:2018gfi,Henneaux:2018hdj}.  This is exactly as in the case of the scalar field explained in detail  \cite{Fuentealba:2024lll}.

In four spacetime dimensions, one can define the radial magnetic field
\be
\xbar {\mathcal B} = \frac 12 \frac{\epsilon^{AB}}{\sqrt{\xbar \gamma}} \xbar F_{AB}
\ee
and obtain a matching analogous to that of the radial electric field $\xbar F_{ur}$
\be
\lim_{v \rightarrow \infty}\xbar {\mathcal B}(-x^C) = \lim_{u \rightarrow - \infty} \xbar {\mathcal B}(x^C) = g Y_{00} + \sum_{l>0,m} l (l+1)\alpha_{lm}^{P(0)} Y_{lm}\, .
\ee

\subsection{Matching in the case of non-standard twisted parity conditions}
With the second set of parity conditions, $A_{A}^{(0)T}$ must be odd, which forces $T_{lm}^{P}=0$.

These parity conditions leave the radial electric field unchanged, so that (\ref{Eq:FurSimpl}), (\ref{Eq:Fur4D2}) and (\ref{Eq:Match0}) still hold.  However, the angular components acquire a dominant logarithmic term,
\begin{equation}
F_{AB}=\log r F_{AB}^{\log}+\mathcal O (1)\,,
\end{equation}
with
\be
F_{AB}^{\log}  =\frac{1}{2}\sum_{l,m}\alpha_{lm}^{Q(0)}\Theta_{AB,lm}\,.
\ee
In this case, the leading logarithmic term survives.  Since they come with the $Q$-branch which obeys the opposite parity  properties $Q_l(-s) = - Q_l(s)$, one gets 
\be
\lim_{v \rightarrow \infty} F_{AB}^{\log}(-x^C) = \lim_{u \rightarrow - \infty}  F_{AB}^{\log}(x^C) =  \sum_{l>0,m}\alpha_{lm}^{Q(0)}\Theta_{AB,lm}\, ,
\ee
\be
\lim_{v \rightarrow \infty} {\mathcal B}^{\log}(-x^C) = -\lim_{u \rightarrow - \infty}  {\mathcal B}^{\log}(x^C) =  -\sum_{l>0,m} l (l+1)\alpha_{lm}^{Q(0)} Y_{lm}\, .
\ee
There is an extra minus sign in the matching of the leading component of $F_{AB}$ (now logarithmic) with respect to the standard case.

It we take the completely inverted parity conditions, one must also pick the $Q$-branch for the radial component of the electric field, leading to 
\begin{equation}
F_{ur}=\frac{\log r}{r^{2}}F_{ur}^{\log}+ \mathcal O\left(1\right)\,, \quad F_{ur}^{\log}  = \frac12 \sum_{l,m}\Xi_{lm}^{Q(0)} Y_{lm}\,, 
\end{equation}
and
\be
\lim_{v \rightarrow \infty}\xbar F_{vr}^{\log}(-x^A) = - \lim_{u \rightarrow -\infty}\xbar F_{ur}^{\log}(x^A) \, . \label{Eq:OppositeMatching}
\ee

\section{Spherical harmonics in higher dimensions
\label{App:SphHar}}
\subsection{Functions}
The functions $f(x^A)$ on the ($D-2)$-sphere can be expanded in terms of eigenfunctions of the Laplacian $\xbar \triangle$, which are called spherical harmonics and denoted $Y_l$ ($l= 0, 1, 2, \cdots$).  The eigenvalues are equal to $- l (l+D-3)$ (i.e., $- l (l+1)$ in $4$ spacetime dimensions),
\be
\xbar \triangle Y_{l} =- l (l+D-3) Y_{l}
\ee
 and are degenerate.  The multiplicity of the eigenvalue corresponding to $l$ is 
\be
 \frac{2l+ D-3}{l} \begin{pmatrix} l + D-4 \\ l-1 \end{pmatrix}.  \label{Eq:Degeneracy0}
\ee
For the $2$-sphere ($D=4$), this is equal to $2 l + 1$ and the eigenfunctions are distinguished by the ``magnetic quantum number $m$" that takes integer values ranging from $-l$ to $l$.  We shall still denote the degeneracy index by $m$ in higher dimensions. 

We will use both notations $Y_l$ and $Y_{lm}$ for the spherical harmonics, depending on whether we want to be explicit about the degeneracy or not (due to the rotational symmetry, all values of $m$ play the same role in our analysis).
In the $(l,m)$ notations, we thus have
\be
\xbar \triangle Y_{lm} =- l (l+D-3) Y_{lm}  
\ee
with $l$ a non-negative integer and $m$ taking (\ref{Eq:Degeneracy0}) different values.

\subsection{Vector fields}

We can also expand vectors tangent to the sphere in terms of the eigenfunctions of the vector-Laplacian.  Using the equivalent form language, any $1$-form can be decomposed as the sum of an exact $1$-form (``longitudinal vector") and a co-exact $1$-form (``transverse vector") without extra contribution since $H^1(\xbar d)(S_{D-2})=0$ ($D \geq 4$).  Explicitly,
\be
v_A = v_A^L + v_A^T \, , \qquad v_A^L = \xbar D_A \Phi \, , \qquad v_A^T = \xbar D^B \Psi_{AB} \, , \qquad \Psi_{AB} = - \Psi_{BA} \, .
\ee

The eigenvalues of the Laplacian  for the longitudinal $1$-forms are parametrized by a non-negative integer $q = 0, 1, 2, \cdots$ and given explicitly by \cite{Iwasaki,Copeland:1984qk}
\be
- (q+1)(q+D-2) + D-3\label{Eq:EingenLong0}
\ee
with degeneracy\footnote{For comparison with \cite{Iwasaki}, we stress that the ``Laplacian" is here the ``connection Laplacian" $\xbar \triangle = \xbar D_B \xbar D^B$. The Laplace-De Rham operator $\hat \triangle = \xbar d \, \delta + \delta \, \xbar d$ considered in \cite{Iwasaki} is related to it (for $1$-forms on the $S_{D-2}$-sphere) as $- \hat \triangle \omega_A = \xbar \triangle - (D-3) \omega_A \Leftrightarrow \xbar \triangle \omega_A= - \hat \triangle \omega_A + (D-3) \omega_A$. Here, the codifferential  $\delta $ is the adjoint of the exterior derivative. }
\be
\frac{(q+D-2)! (2q + D -1)}{q! (D-3)! (q + D-2)(q+1)} \, .\label{Eq:DegenLong0}
\ee
Similarly, the eigenvalues of the Laplacian for the transverse $1$-forms are also parametrized by a non-negative integer $q = 0, 1, 2, \cdots$ and given explicitly by \cite{Iwasaki}
\be
- (q+2)(q+D-3) +D-3   \label{Eq:EingenTrans0}
\ee
with degeneracy
\be
\frac{(q+D-2)! (2q + D -1)}{q! (D-4)! (q + D-3)(q+2)} \, . \label{Eq:DegenTrans0}
\ee

It is useful to define the longitudinal spherical harmonics
\be
\Phi_{A,lm} = \partial_A Y_{lm}
\ee
($l>0$) which obey
\be
\xbar \triangle \Phi_{A,lm} = - [l (l+D-3) - D+3] \Phi_{A,lm}
\ee
(note the misprint in formula (B.3) of \cite{Henneaux:2019yqq} where $d$ should be viewed as the spatial dimension $D-1$ - and not as the spacetime dimension as it is everywhere else in that article).  This formula matches (\ref{Eq:EingenLong0}) if we set $l=q+1$ ($l =1,2, \cdots >0$).  The degeneracy is given by (\ref{Eq:DegenLong0}) and coincides with (\ref{Eq:Degeneracy0}) as it should.

In four spacetime dimensions, one can construct a basis of transverse spherical harmonics $\{\Psi_{A, lm}\}$ as follows,
\be
\Psi_{A, lm} = {\epsilon_A}^B \partial_B Y_{lm} \, ,
\ee
($l>0$). Note that $\Psi_{A, lm}$ has the same parity as $Y_{lm}$ because both $\partial_A$ and $\epsilon_{AB}$ are odd (see previous Appendix {\bf \ref{App:Orientation}} for $\epsilon_{AB}$).  One has
\be
\xbar \triangle \Psi_{A,lm} = - [l (l+1) - 1] \Psi_{A,lm} \, ,
\ee
which is in agreement with (\ref{Eq:EingenTrans0}) if we set again $l=q+1$ together with $D=4$.  The degeneracy is equal to $2l+1$, also in agreement with  (\ref{Eq:DegenTrans0}). 
In higher dimensions, we denote in the same way the transverse spherical harmonics $ \Psi_{A, lm}$.  Their eigenvalues are given by (\ref{Eq:EingenTrans0})
\be
\xbar \triangle \Psi_{A,lm} = - [l (l+D-3) - 1] \Psi_{A,lm}\,,
\ee
($l = q+1$) but the degeneracy index runs now over (\ref{Eq:DegenTrans0}) values, a number different from (\ref{Eq:DegenLong0}) for $D>4$ (we use nevertherless the same letter $m$ because there is no risk of confusion and furthermore, the explicit degeneracy will not be critical in the formulas of interest to our study).

Finally, we define the spherical harmonics $\Theta_{AB, lm}$ for exact $2$-forms from the transverse spherical harmonics through
\be
\Theta_{AB, lm} = \partial_A \Psi_{B, lm} - \partial_B \Psi_{A, lm} \, .
\ee
They are in same number as the $\Psi_{A,lm}$'s and fulfill
\be
\xbar \triangle \Theta_{AB, lm} = - [l (l+D-3) - D+4] \Theta_{AB, lm} \, .
\ee

Finally, we collect the transformation properties of the spherical harmonics under the antipodal map $\vec x \rightarrow - \vec x$.  These are
\begin{eqnarray}
&& Y_{lm}(-\vec x) = (-1)^l Y_{lm}(\vec x)\,, \\
&& \Phi_{A,lm}(-\vec x) = - (-1)^l \Phi_{A,lm}(\vec x)\,, \\
&& \Psi_{A, lm}(-\vec x) = (-1)^l \Psi_{A, lm}(\vec x)\,, \\
&&\Theta_{AB, lm}(-\vec x) = -(-1)^l \Theta_{AB, lm}(\vec x) \,.
\end{eqnarray}
These follow from the known parity properties of the spherical harmonics, taking into account that both $\partial_A$ and $\epsilon_{AB}$ are odd under the antipodal map.



\begin{thebibliography}{99}

\bibitem{Fuentealba:2022xsz}
O.~Fuentealba, M.~Henneaux and C.~Troessaert,
``Logarithmic supertranslations and supertranslation-invariant Lorentz charges,''
JHEP \textbf{02} (2023), 248
doi:10.1007/JHEP02(2023)248
[arXiv:2211.10941 [hep-th]].

\bibitem{Fuentealba:2023rvf}
O.~Fuentealba, M.~Henneaux and C.~Troessaert,
``A note on the asymptotic symmetries of electromagnetism,''
JHEP \textbf{03} (2023), 073
doi:10.1007/JHEP03(2023)073
[arXiv:2301.05989 [hep-th]].

\bibitem{Campiglia:2016hvg}
M.~Campiglia and A.~Laddha,
``Subleading soft photons and large gauge transformations,''
JHEP \textbf{11} (2016), 012
doi:10.1007/JHEP11(2016)012
[arXiv:1605.09677 [hep-th]].

\bibitem{Seraj:2016jxi}
A.~Seraj,
``Multipole charge conservation and implications on electromagnetic radiation,''
JHEP \textbf{06} (2017), 080
doi:10.1007/JHEP06(2017)080
[arXiv:1610.02870 [hep-th]].

\bibitem{Barnich:2013sxa}
G.~Barnich and P.~H.~Lambert,
``Einstein-Yang-Mills theory: Asymptotic symmetries,''
Phys. Rev. D \textbf{88} (2013), 103006
doi:10.1103/PhysRevD.88.103006
[arXiv:1310.2698 [hep-th]].

\bibitem{He:2014cra}
  T.~He, P.~Mitra, A.~P.~Porfyriadis and A.~Strominger,
  ``New Symmetries of Massless QED,''
  JHEP {\bf 1410} (2014) 112
  [arXiv:1407.3789 [hep-th]].

\bibitem{Kapec:2015ena}
  D.~Kapec, M.~Pate and A.~Strominger,
  ``New Symmetries of QED,''
  arXiv:1506.02906 [hep-th].

\bibitem{Strominger:2017zoo}
  A.~Strominger,
  ``Lectures on the Infrared Structure of Gravity and Gauge Theory,''
  arXiv:1703.05448 [hep-th].

\bibitem{Fuentealba:2023hzq}
O.~Fuentealba and M.~Henneaux,
``Simplifying (super-)BMS algebras,''
JHEP \textbf{11} (2023), 108
doi:10.1007/JHEP11(2023)108
[arXiv:2309.07600 [hep-th]].

 \bibitem{Henneaux:2018gfi}
  M.~Henneaux and C.~Troessaert, ``Asymptotic symmetries of electromagnetism at spatial infinity,'' JHEP \textbf{05} (2018), 137 doi:10.1007/JHEP05(2018)137 [arXiv:1803.10194 [hep-th]].

\bibitem{Henneaux:2019yqq}
M.~Henneaux and C.~Troessaert,
``Asymptotic structure of electromagnetism in higher spacetime dimensions,''
Phys. Rev. D \textbf{99} (2019) no.12, 125006
doi:10.1103/PhysRevD.99.125006
[arXiv:1903.04437 [hep-th]].



\bibitem{Fuentealba:2024lll}
O.~Fuentealba and M.~Henneaux,
``Logarithmic matching between past infinity and future infinity: The massless scalar field in Minkowski space,''
JHEP \textbf{03} (2025), 081
doi:10.1007/JHEP03(2025)081
[arXiv:2412.05088 [gr-qc]].



 
\bibitem{Damour:1985cm}
T.~Damour,
``Analytical calculations of gravitational radiation,'' In {\it Proceedings of the Fourth Marcel
Grossmann Meeting}, pages 365-392, Elsevier Science Publishers (Amsterdam: 1986),
PRINT-86-0221 (MEUDON).

\bibitem{Christodoulou:1993uv}
D.~Christodoulou and S.~Klainerman,
``The Global nonlinear stability of the Minkowski space,''
Princeton University Press (Princeton: 1993)

\bibitem{Chrusciel:1993hx}
P.~T.~Chrusciel, M.~A.~H.~MacCallum and D.~B.~Singleton,
``Gravitational waves in general relativity: 14. Bondi expansions and the polyhomogeneity of Scri,''
[arXiv:gr-qc/9305021 [gr-qc]].

   \bibitem{Fried1}
  H. Friedrich, ``Gravitational fields near space-like and null infinity,'' J. Geom. Phys. {\bf 24} (1998) 83-163.
  
\bibitem{Friedrich:1999wk}
  H.~Friedrich and J.~Kannar,
  ``Bondi type systems near space - like infinity and the calculation of the NP constants,''
  J.\ Math.\ Phys.\  {\bf 41} (2000) 2195
  [gr-qc/9910077].
  
\bibitem{Friedrich:1999ax}
  H.~Friedrich and J.~Kannar,
  ``Calculating asymptotic quantities near space - like and null infinity from Cauchy data,''
  Annalen Phys.\  {\bf 9} (2000) 321
  [gr-qc/9911103].

\bibitem{Valiente-Kroon:2002xys}
J.~A.~Valiente-Kroon,
``A New class of obstructions to the smoothness of null infinity,''
Commun. Math. Phys. \textbf{244} (2004), 133-156
doi:10.1007/s00220-003-0967-5
[arXiv:gr-qc/0211024 [gr-qc]].

\bibitem{ValienteKroon:2003ix}
J.~A.~Valiente Kroon,
``Does asymptotic simplicity allow for radiation near spatial infinity?,''
Commun. Math. Phys. \textbf{251} (2004), 211-234
doi:10.1007/s00220-004-1174-8
[arXiv:gr-qc/0309016 [gr-qc]].

\bibitem{Kehrberger:2021uvf}
L.~M.~A.~Kehrberger,
``The Case Against Smooth Null Infinity I: Heuristics and Counter-Examples,''
Annales Henri Poincare \textbf{23} (2022) no.3, 829-921
doi:10.1007/s00023-021-01108-2
[arXiv:2105.08079 [gr-qc]].

\bibitem{Minucci:2022hav}
M.~Minucci, R.~Panosso Macedo and J.~A.~V.~Kroon,
``The Maxwell-scalar field system near spatial infinity,''
J. Math. Phys. \textbf{63} (2022) no.8, 082501
doi:10.1063/5.0104602
[arXiv:2206.04366 [gr-qc]].

\bibitem{Sen:2024bax}
A.~Sen,
``Gravitational Wave Tails from Soft Theorem: A Short Review,''
[arXiv:2408.08851 [hep-th]].

\bibitem{Geiller:2024ryw}
M.~Geiller, A.~Laddha and C.~Zwikel,
``Symmetries of the gravitational scattering in the absence of peeling,''
JHEP \textbf{12} (2024), 081
doi:10.1007/JHEP12(2024)081
[arXiv:2407.07978 [gr-qc]].

\bibitem{ValienteKroon:2007bj}
J.~A.~Valiente Kroon,
``The Maxwell field on the Schwarzschild spacetime: Behaviour near spatial infinity,''
Proc. Roy. Soc. Lond. A \textbf{463} (2007), 2609-2630
doi:10.1098/rspa.2007.1896
[arXiv:gr-qc/0703147 [gr-qc]].


\bibitem{Francia:2024hja}
D.~Francia and F.~Manzoni,
``Asymptotic charges of $p-$forms and their dualities in any $D$,''
[arXiv:2411.04926 [hep-th]].


\bibitem{Romoli:2024hlc}
M.~Romoli,
``$ \mathcal{O} $(r$^{N}$) two-form asymptotic symmetries and renormalized charges,''
JHEP \textbf{12} (2024), 085
doi:10.1007/JHEP12(2024)085
[arXiv:2409.08131 [hep-th]].


\bibitem{Manzoni:2025wyr}
F.~Manzoni and M.~Romoli,
``Higher-order $p$-form asymptotic symmetries in $D = p + 2$,''
[arXiv:2503.22572 [hep-th]].


\bibitem{Henneaux:1999ct}
M.~Henneaux, B.~Julia and S.~Silva,
``Noether superpotentials in supergravities,''
Nucl. Phys. B \textbf{563} (1999), 448-460
doi:10.1016/S0550-3213(99)00536-2
[arXiv:hep-th/9904003 [hep-th]].

\bibitem{Regge:1974zd}
T.~Regge and C.~Teitelboim,
``Role of Surface Integrals in the Hamiltonian Formulation of General Relativity,''
Annals Phys. \textbf{88} (1974), 286
doi:10.1016/0003-4916(74)90404-7

\bibitem{Satishchandran:2019pyc}
G.~Satishchandran and R.~M.~Wald,
Phys. Rev. D \textbf{99} (2019) no.8, 084007
doi:10.1103/PhysRevD.99.084007
[arXiv:1901.05942 [gr-qc]].

\bibitem{Henneaux:2018hdj}
  M.~Henneaux and C.~Troessaert,
  ``Hamiltonian structure and asymptotic symmetries of the Einstein-Maxwell system at spatial infinity,''
  JHEP {\bf 1807} (2018) 171
  [arXiv:1805.11288 [gr-qc]].

\bibitem{Benguria:1976in}
R.~Benguria, P.~Cordero and C.~Teitelboim,
``Aspects of the Hamiltonian Dynamics of Interacting Gravitational Gauge and Higgs Fields with Applications to Spherical Symmetry,''
Nucl. Phys. B \textbf{122} (1977), 61-99
doi:10.1016/0550-3213(77)90426-6

\bibitem{Henneaux:2018cst}
M.~Henneaux and C.~Troessaert,
``BMS Group at Spatial Infinity: the Hamiltonian (ADM) approach,''
JHEP \textbf{03} (2018), 147
doi:10.1007/JHEP03(2018)147
[arXiv:1801.03718 [gr-qc]].

\bibitem{Tanzi:2020fmt}
R.~Tanzi and D.~Giulini,
``Asymptotic symmetries of Yang-Mills fields in Hamiltonian formulation,''
JHEP \textbf{10} (2020), 094
doi:10.1007/JHEP10(2020)094
[arXiv:2006.07268 [hep-th]].

\bibitem{Ashtekar:1978zz}
  A.~Ashtekar and R.~O.~Hansen,
  ``A unified treatment of null and spatial infinity in general relativity. I - Universal structure, asymptotic symmetries, and conserved quantities at spatial infinity,''
  J.\ Math.\ Phys.\  {\bf 19} (1978) 1542.

\bibitem{Beig:1982ifu}
R.~Beig and B.~G.~Schmidt,
``Einstein's equations near spatial infinity,''
Commun. Math. Phys. \textbf{87} (1982) no.1, 65-80
doi:10.1007/BF01211056

  \bibitem{Ultra}
 G. Szeg\H{o}, ``Orthogonal Polynomials'',  Colloquium Publications of the American Mathematical Society, Volume 23, fourth edition (Providence: 1975) (Chapter IV, in particular section 4.7)

\bibitem{Henneaux:2018mgn}
M.~Henneaux and C.~Troessaert,
``Asymptotic structure of a massless scalar field and its dual two-form field at spatial infinity,''
JHEP \textbf{05} (2019), 147
doi:10.1007/JHEP05(2019)147
[arXiv:1812.07445 [hep-th]].

\bibitem{Troessaert:2017jcm}
C.~Troessaert,
``The BMS4 algebra at spatial infinity,''
Class. Quant. Grav. \textbf{35} (2018) no.7, 074003
doi:10.1088/1361-6382/aaae22
[arXiv:1704.06223 [hep-th]].

\bibitem{Gasperin:2017apb}
E.~Gasperin and J.~A.~V.~Kroon,
``Polyhomogeneous expansions from time symmetric initial data,''
Class. Quant. Grav. \textbf{34} (2017) no.19, 195007
doi:10.1088/1361-6382/aa87bf
[arXiv:1706.04227 [gr-qc]].

\bibitem{Paetz:2018nbd}
T.~T.~Paetz,
``On the smoothness of the critical sets of the cylinder at spatial infinity in vacuum spacetimes,''
J. Math. Phys. \textbf{59} (2018) no.10, 102501
doi:10.1063/1.5037698
[arXiv:1804.05034 [gr-qc]].

\bibitem{Gasperin:2021vnm}
E.~Gasperin and J.~A.~Valiente Kroon,
``Staticity and regularity for zero rest-mass fields near spatial infinity on flat spacetime,''
Class. Quant. Grav. \textbf{39} (2022) no.1, 015014
doi:10.1088/1361-6382/ac37ce
[arXiv:2107.12875 [gr-qc]].

\bibitem{Mohamed:2021rfg}
M.~M.~A.~Mohamed and J.~A.~Valiente~Kroon,
``Asymptotic charges for spin-1 and spin-2 fields at the critical sets of null infinity,''
[arXiv:2112.03890 [gr-qc]].

\bibitem{Campiglia:2017mua}
M.~Campiglia and R.~Eyheralde,
``Asymptotic $U(1)$ charges at spatial infinity,''
JHEP \textbf{11} (2017), 168
doi:10.1007/JHEP11(2017)168
[arXiv:1703.07884 [hep-th]].

\bibitem{Campiglia:2018dyi}
M.~Campiglia and A.~Laddha,
``Asymptotic charges in massless QED revisited: A view from Spatial Infinity,''
JHEP \textbf{05} (2019), 207
doi:10.1007/JHEP05(2019)207
[arXiv:1810.04619 [hep-th]].

\bibitem{Campiglia:2017dpg}
M.~Campiglia, L.~Coito and S.~Mizera,
``Can scalars have asymptotic symmetries?,''
Phys. Rev. D \textbf{97} (2018) no.4, 046002
doi:10.1103/PhysRevD.97.046002
[arXiv:1703.07885 [hep-th]].

\bibitem{Campiglia:2018see}
M.~Campiglia, L.~Freidel, F.~Hopfmueller and R.~M.~Soni,
``Scalar Asymptotic Charges and Dual Large Gauge Transformations,''
JHEP \textbf{04} (2019), 003
doi:10.1007/JHEP04(2019)003
[arXiv:1810.04213 [hep-th]].

\bibitem{Francia:2018jtb}
D.~Francia and C.~Heissenberg,
``Two-Form Asymptotic Symmetries and Scalar Soft Theorems,''
Phys. Rev. D \textbf{98} (2018) no.10, 105003
doi:10.1103/PhysRevD.98.105003
[arXiv:1810.05634 [hep-th]].

\bibitem{Ashtekar:1987tt}
A.~Ashtekar,
``Asymptotic quantization: Based on 1984 Naples Lectures,'' Bibliopolis (Naples: 1987)

\bibitem{Strominger:2015bla}
A.~Strominger,
``Magnetic Corrections to the Soft Photon Theorem,''
Phys. Rev. Lett. \textbf{116} (2016) no.3, 031602
doi:10.1103/PhysRevLett.116.031602
[arXiv:1509.00543 [hep-th]].

\bibitem{Henneaux:2020nxi}
M.~Henneaux and C.~Troessaert,
``A note on electric-magnetic duality and soft charges,''
JHEP \textbf{06} (2020), 081
doi:10.1007/JHEP06(2020)081
[arXiv:2004.05668 [hep-th]].

\bibitem{Bieri:2013hqa}
L.~Bieri and D.~Garfinkle,
``An electromagnetic analogue of gravitational wave memory,''
Class. Quant. Grav. \textbf{30} (2013), 195009
doi:10.1088/0264-9381/30/19/195009
[arXiv:1307.5098 [gr-qc]].

\bibitem{Strominger:2014pwa}
A.~Strominger and A.~Zhiboedov,
``Gravitational Memory, BMS Supertranslations and Soft Theorems,''
JHEP \textbf{01} (2016), 086
doi:10.1007/JHEP01(2016)086
[arXiv:1411.5745 [hep-th]].

\bibitem{Geiller:2022vto}
M.~Geiller and C.~Zwikel,
``The partial Bondi gauge: Further enlarging the asymptotic structure of gravity,''
SciPost Phys. \textbf{13} (2022), 108
doi:10.21468/SciPostPhys.13.5.108
[arXiv:2205.11401 [hep-th]].

\bibitem{Fuentealba:2023huv}
O.~Fuentealba,
``Asymptotic $ \mathcal{O} $(r) gauge symmetries and gauge-invariant Poincar\'e generators in higher spacetime dimensions,''
JHEP \textbf{04} (2023), 047
doi:10.1007/JHEP04(2023)047
[arXiv:2302.13788 [hep-th]].

\bibitem{Fuentealba:2023syb}
O.~Fuentealba, M.~Henneaux and C.~Troessaert,
``Asymptotic Symmetry Algebra of Einstein Gravity and Lorentz Generators,''
Phys. Rev. Lett. \textbf{131} (2023) no.11, 111402
doi:10.1103/PhysRevLett.131.111402
[arXiv:2305.05436 [hep-th]].

\bibitem{Iwasaki}
I.~Iwasaki and K.~Katase,
``On the Spectra of Laplace Operator on $A*(S)$,"
Proc. Japan Acad. \textbf{55}, Ser. A (1979) 141

\bibitem{Copeland:1984qk}
E.~J.~Copeland and D.~J.~Toms,
``Quantized Antisymmetric Tensor Fields and Selfconsistent Dimensional Reduction in Higher Dimensional Space-times,''
Nucl. Phys. B \textbf{255} (1985), 201-230
doi:10.1016/0550-3213(85)90134-8



 


\end{thebibliography}
\end{document}